\documentclass[preprint,aps,12pt,showpacs,nofootinbib,tightenlines,amsmath,amssymb]{revtex4}
\usepackage{amsmath}
\usepackage{graphicx}
\usepackage{amssymb}
\usepackage{color}

\newcommand{\be}{\begin{eqnarray}}
\newcommand{\ee}{\end{eqnarray}}

\newcommand{\p}{\partial}

\newcommand{\nn}{\nonumber}
\newcommand{\tr}{\mathop{\rm tr}\nolimits}
\newcommand{\Tr}{\mathop{\rm Tr}\nolimits}
\newcommand{\diag}{\mathop{\rm diag}}
\newcommand{\cL}{{\mathcal L}}
\newcommand{\kL}{\mathfrak{L}}

\newcommand{\cA}{{\mathcal A}}

\newcommand{\cF}{{\mathcal F}}
\newcommand{\cQ}{{\mathbf F}}

\newcommand{\fJ}{{\mathbf J}}
\newcommand{\mJ}{{\mathsf J}}
\newcommand{\cD}{{\mathcal D}}
\newcommand{\cG}{{\mathcal G}}
\newcommand{\fG}{{\mathbf G}}
\newcommand{\cW}{{\mathcal W}}
\newcommand{\cS}{{\mathcal S}}

\newcommand{\fVh}{{\mathbf V}_h}
\newcommand{\fGh}{{\mathbf G}_h}
\newcommand{\dchi}{\delta \chi }

\newcommand{\dchiA}{\delta \chi^{\fA}}
\newcommand{\dchiB}{\delta \chi^{\fB}}

\newcommand{\1}{\mspace{1mu}}

\newcommand{\fGa}{\mathbf \Gamma}
\newcommand{\vGa}{\varGamma}
\newcommand{\dg}{\digamma}

\newcommand{\ad}{\mbox{ad}}

\newcommand{\cR}{\mathcal R}

\newcommand{\mT}{\mathbf T}
\newcommand{\sT}{\mathsf T}

\newcommand{\mM}{\mathsf M}
\newcommand{\mN}{\mathsf N}

\newcommand{\mA}{\mathsf A}
\newcommand{\mB}{\mathsf B}
\newcommand{\mC}{\mathsf C}
\newcommand{\mD}{\mathsf D}

\newcommand{\fL}{\mathbf L}
\newcommand{\fM}{\mathbf M}
\newcommand{\fN}{\mathbf N}
\newcommand{\fP}{\mathbf P}
\newcommand{\fQ}{\mathbf Q}

\newcommand{\fA}{\mathbf A}
\newcommand{\fB}{\mathbf B}
\newcommand{\fC}{\mathbf C}
\newcommand{\fD}{\mathbf D}
\newcommand{\fE}{\mathbf E}
\newcommand{\fS}{\mathbf S}

\newcommand{\mW}{\mathsf W}

\newcommand{\bc}{\bar{c}}
\newcommand{\bC}{\bar{C}}

\newcommand{\bet}{\bar{\eth}}

\newcommand{\vOm}{\varOmega }
\newcommand{\cH}{\mathbf A }

\newcommand{\chih}{\hat{\chi}}

\newcommand{\ta}{\tilde{a}}

\newcommand{\hx}{\hat{x}}

\newcommand{\tx}{\tilde{x}}
\newcommand{\mG}{\mathsf{G}}

\newcommand{\cC}{\mathcal{C}}
\newcommand{\cT}{\mathcal{T}}

\newcommand{\hPsi}{\hat{\Psi}}

\newcommand{\vPsi}{\varPsi}

\begin{document}
\def\intdk{\int\frac{d^4k}{(2\pi)^4}}
\def\sla{\hspace{-0.17cm}\slash}
\hfill


\title{Hyperunified field theory \\ and gravitational gauge-geometry duality}

\author{Yue-Liang Wu$^{a,b,c}$}\email{ylwu@itp.ac.cn; ylwu@ucas.ac.cn}
\affiliation{$^a$International Centre for Theoretical Physics Asia-Pasific(ICTP-AP), Beijing, China \\ 
$^b$Institute of Theoretical Physics, Chinese Academy of Sciences, Beijing, 100190, China\\
$^c$University of Chinese Academy of Sciences (UCAS), Beijing 100049, China }


\begin{abstract}
A hyperunified field theory is built in detail based on the postulates of gauge invariance and coordinate independence along with the conformal scaling symmetry.  All elementary particles are merged into a single hyper-spinor field and all basic forces are unified into a fundamental interaction governed by the hyper-spin gauge symmetry SP(1,$D_h$-1). The dimension $D_h$ of hyper-spacetime is conjectured to have a physical origin in correlation with the hyper-spin charge of elementary particles.  The hyper-gravifield fiber bundle structure of biframe hyper-spacetime appears naturally with the globally flat Minkowski hyper-spacetime as a base spacetime and the locally flat hyper-gravifield spacetime as a fiber that is viewed as a dynamically emerged hyper-spacetime characterized by a non-commutative geometry. The gravitational origin of gauge symmetry is revealed with the hyper-gravifield that plays an essential role as a Goldstone-like field. The gauge-gravity and gravity-geometry correspondences bring about the gravitational gauge-geometry duality. The basic properties of hyperunified field theory and the issue on the fundamental scale are analyzed within the framework of quantum field theory, which allows us to describe the laws of nature in deriving the gauge gravitational equation with the conserved current and the geometric gravitational equations of Einstein-like type and beyond.
\end{abstract}

\maketitle

\begin{widetext}
\tableofcontents
\end{widetext}

\newpage 

\section{Introduction}
 
Since Einstein established the general theory of relativity (GR)\cite{GR} in 1915, it has become a great challenge for many physicists and mathematicians to unify the then-known basic forces. Historically, the idea of unification was put forward because of the dynamical theory of the electromagnetic field formulated in 1864 by Maxwell, who combined electricity and magnetism into a unifying theory of electromagnetism, which is considered as the first successful classical unified field theory. The constancy of the speed of light in Maxwell's theory led Einstein to unify the space and time into four dimensional spacetime characterized by the global Lorentz symmetry SO(1,3), which has laid the foundation for the special theory of relativity(SR)\cite{SR}. Such a globally flat four-dimensional Minkowski spacetime holding for SR was extended by Einstein to a curved spacetime characterized by the general linear group symmetry GL(4, R), which has laid the foundation for GR\cite{GR2}. Namely, the gravitational force is characterized by a dynamic Riemannian geometry of curved spacetime. Since then, an attempt to unify the gravity and electromagnetism was pursued by many theoreticians. Some interesting progress includes the work proposed by Kaluza who extended GR to five dimensional spacetime\cite{Kaluza}, and also by Klein who proposed the fifth dimension to be curled up into an unobservable small circle\cite{Klein}. In such a Kaluza-Klein theory,  the gravitational curvature tensor corresponding to an extra spatial direction behaves as an additional force analogous to electromagnetism. Another interesting idea was proposed by Weyl, who introduced the concept of gauge field as the electromagnetic field via a local scaling transformation\cite{Weyl}. Einstein extensively set on a quest for potential unified models of the electromagnetism and gravity as a classical unified field theory; he devoted nearly all his efforts to the search for a unified field theory and spent the last two decades of his life to doing so.   

On the other hand, the Dirac spinor theory\cite{Dirac} has provided a successful unity between quantum mechanics and special relativity, which has led to the developments of relativistic quantum mechanics and quantum field theory(QFT). The framework of QFT was firstly built up in the 1940s by formulating the classical electromagnetism into the quantum electrodynamics (QED)\cite{QED1,QED2,QED02,QED3,QED03,QED003,QED4,QED04}.  QED is characterized by an Abelian gauge symmetry U(1). In 1954, Yang and Mills extended the U(1) gauge symmetry to a non-Abelian gauge symmetry for characterizing the isotopic spin symmetry SU(2)\cite{YM}. The electroweak theory with the gauge symmetry group U(1)$\times$SU(2) was developed in the 1960s\cite{EW1,EW2,EW3}, which has been a great success in unifying the electromagnetic and weak interactions. Such a theory was proven to be consistent in the sense of renormalizability\cite{RT} under the Higgs mechanism of spontaneous symmetry breaking\cite{SSB1,SSB2}. QFT has provided a successful unified description not only for the electroweak interactions, but also for the strong interaction characterized by the quantum chromodynamics(QCD)\cite{QCD1,QCD2} with the gauge symmetry group SU(3). The quantum gauge field theory governed by the symmetry group U(1)$\times$SU(2)$\times$SU(3) with spontaneous symmetry breaking is referred to as the standard model (SM) in elementary particle physics. In SM,  the leptons and quarks\cite{QK1,QK2} are regarded as the basic building blocks of nature, and three families of quarks were required to obtain a nontrivial CP-violating phase\cite{KM}. To realize the spontaneous breaking of the CP symmetry\cite{SCPV1,SCPV01}, it is necessary to go beyond the SM. For instance, the general two-Higgs doublet model as one of the simplest extensions to the SM can lead to the spontaneous breaking of CP symmetry with rich induced CP-violating sources\cite{SCPV2}.

The discovery of asymptotic freedom in QCD\cite{QCD1,QCD2} has indicated a potential unification between the strong interaction and the electroweak interactions.  As all the interactions are governed by the Abelian and non-Abelian Yang-Mills gauge symmetries, it is natural to search for unified field theories with enlarged gauge symmetries. The unity of the quark and lepton species was firstly initiated in the 1970s\cite{PS}. Some minimal grand unified theories (GUTs) for the electroweak and strong interactions were proposed based on the gauge symmetry groups SU(5)\cite{SU5} and SO(10)\cite{SO101,SO102}. The key prediction of GUTs was the instability of the proton. The current experiments have not yet observed any evidence for the proton decay, only a lower bound of the order $10^{35}$ years for its lifetime has been reached.  The enlarged gauge model SO(1,13)\cite{CW} was proposed to unify the SO(1,3) spin gauge symmetry and the SO(10) internal gauge symmetry. The SO(1,13) and SO(3,11) gauge symmetries were considered as gravity GUT models in four dimensional spacetime\cite{GGUT,GGUT0,GGUT00}.

On the other hand, the dynamical symmetry breaking mechanism of QCD at low energies\cite{DSB} reflects the color confinement of the gluons and quarks. The light scalar and pseudoscalar mesons as the bound states of the confined quarks and antiquarks were shown to behave as composite Higgs bosons\cite{DW}. Such a color confining feature forms stringlike degrees of freedom, i.e., the so-called QCD strings. Inspired by the QCD strings, a string object was motivated to be taken as a basic building block of nature as opposed to  a point-like elementary particle. A consistent string theory was found to be realized either in 26-dimensional spacetime for a bosonic string\cite{string26} or in ten-dimensional spacetime for a superstring\cite{string10a,string10b,string10c,string10c0}. Some interesting string models that are promising to realize the SM at low energies include the perturbative heterotic string models\cite{HS} and the mysterious M-theory\cite{MT}.  It was shown that the six small extra dimensions in superstring theory need to be compactified\cite{6D} on the Calabi-Yau manifold\cite{CY,CY0} in order to obtain the N=1 supersymmetry. Subsequently, string perturbation theory was found to be divergent\cite{GP}.  The theory was also demonstrated to require the inclusion of higher-dimensional objects called D-branes, which were identified with the black-hole solutions of supergravity\cite{DB}. Practically, a full holographic description of M-theory by using IIA D0 branes was formulated as a matrix theory\cite{MTD0}. Furthermore,  the anti-de Sitter/conformal field theory (AdS/CFT) correspondence was proposed to formulate string theory and study some interesting properties\cite{AC1,AC2,AC3}, which provides a new insight into the mathematical structures of string theory. Nevertheless, the basic vacuum solution of string theory remains unknown as there are $10^{500}$ possible solutions fitting the constraints of the theory\cite{SV,SV0}. Therefore, it is necessary to explore further how string theory can truly be realized as a theory of everything.

Alternatively, some gravity gauge theories were proposed to address the issue of the long-term outstanding problem about the incompatibility between GR and SM. This is because SM has successfully been described by the gauge symmetries within the framework of QFT, which motivated one to try a gauge theory description for the gravitational interaction. Numerous efforts have been made to construct gravity gauge theories, which may be found in some pioneering work\cite{GGT1,GGT2,GGT3,GGT03,GGT4,GGT5} and review articles\cite{GGTR1,GGTR2,GGTR3} in the references therein. Nevertheless, most of the gravity gauge theories were built relying on the Riemannian or non-Riemannian geometry in a curved spacetime. Some basic issues concerning the definitions of space and time as well as the quantization of gravity gauge theories remain open questions.  Recently, a quantum field theory of gravity\cite{YLWU1} was built based on the spin and scaling gauge invariances by treating the gravitational interaction on the same footing as the electroweak and strong interactions, which enables us to provide a unified description for the four basic forces within the framework of QFT. The postulates of gauge invariance and coordinate independence have been shown to be more general and fundamental than the postulate of general covariance under the general linear group GL(4,R) transformations of coordinates, so that all the basic forces are governed by gauge symmetries. The concept of {\it biframe spacetime} was found to play an essential role in such a gravitational quantum field theory. Instead of the metric field in GR, a bicovariant vector field defined in biframe spacetime is necessarily introduced as a basic {\it gravifield} to characterize the gravitational interaction. Geometrically, one frame spacetime is a globally flat coordinate Minkowski spacetime that acts as an inertial reference frame for describing the motion of the basic fields, which  enables us to derive the well-defined conservation laws and to make a physically meaningful definition for space and time in such a way that the differences of the spatial coordinates or time coordinate can be directly measured by the standard ways proposed in SR. The other frame spacetime is a locally flat non-coordinate {\it gravifield spacetime} that functions as an intrinsic interaction frame for characterizing the dynamics of basic fields, which is characterized by a non-commutative geometry and viewed as a dynamically emerging spacetime. 

Inspired by the relativistic Dirac spinor theory and the grand unified theories as well as the Einstein general theory of relativity, we are motivated to assume the hypotheses that all the spin-like charges of elementary particles should be treated on the same footing as a {\it hyper-spin charge} and the {\it hyper-spinor structures} of elementary particles are correlated with the geometric properties of {\it hyper-spacetime}. To build a reliable unified field theory within the framework of gravitational quantum field theory\cite{YLWU1}, we shall work with the postulates that the basic theory should obey the principles of gauge invariance and coordinate independence. With such hypotheses and postulates, we have presented in Ref.\cite{YLWU2} a brief description for a unified field theory of all basic forces and elementary particles in hyper-spacetime. 

In this paper, we are going to carry out a general analysis and a detailed construction for such a {\it hyperunified field theory}. The paper is organized as follows:  after Sect. 1 in which a brief outline of various attempts in exploring unified theories is presented, we then show in Sect. 2 how all the quarks and leptons as  the point-like elementary particles in SM can be merged into a column vector in the spinor representation of hyper-spacetime with a Majorana-type hyper-spinor structure. In Sect. 3, we demonstrate how all the known basic forces in nature can be unified into a fundamental interaction governed by a hyper-spin gauge symmetry SP(1,$D_h$-1) with a minimal dimension $D_h=19$. An equation of motion for the unified hyper-spinor field results characterizing a general {\it gravitational relativistic quantum theory} with a conformal scaling symmetry in hyper-spacetime.  In Sect. 4, we construct in detail a general action of {\it hyperunified field theory} in a locally flat {\it hyper-gravifield spacetime} based on the postulates of gauge invariance and coordinate independence.  By projecting into a globally flat coordinate hyper-spacetime via a bicovariant vector {\it hyper-gravifield}, we obtain in Sect. 5 the general action of hyperunified field theory within the framework of QFT. A set of equations of motion with the conserved currents are obtained describing the dynamics of all the basic fields. In Sect. 6, we derive various conservation laws in hyperunified field theory and the master equation for the dynamics of hyper-gravifield with the conserved {\it hyper-stress energy-momentum tensor}. In Sect. 7, we demonstrate the {\it gravitational origin of the gauge symmetry} and present the general action of hyperunified field theory in a {\it hidden gauge formalism}. An {\it emergent general linear group symmetry} GL($D_h$, R) is shown to characterize a Riemannian geometry of hyper-spacetime. A basic action of hyperunified field theory with a {\it general conformal scaling gauge invariance} results in Sect. 8, which enables us to demonstrate the {\it gravity-geometry correspondence} and obtain an Einstein-Hilbert type action for the gravitational interaction in hyper-spacetime, keeping the global and local conformal scaling symmetries.  In Sect. 9, we represent the basic action of hyperunifield field theory in the locally flat hyper-gravifield spacetime, which allows us to show the {\it gauge-gravity correspondence} based on the gravitational origin of gauge symmetry. In such a {\it hidden coordinate system}, we further demonstrate in Sect. 10 that the basic action of hyperunified field theory is generally characterized by a {\it non-commutative geometry} of hyper-gravifield spacetime. The {\it gravitational gauge-geometry duality} is corroborated based on various equivalent formalisms of hyperunified field theory. A complete equivalence requires to set the gauge fixing condition in a {\it flowing unitary gauge}. In Sect. 11, we present a general analysis on the basic properties of hyperunified field theory within the framework of QFT and address the issue on the fundamental mass scale relying on the conformal scaling gauge symmetry, which enables us to derive the gauge gravitational equation of the hyper-gravifield with the conserved bicovariant vector current and deduce the geometric gravitational equations of Einstein-like type and beyond, corresponding to the symmetric and antisymmetric hyper-stress energy-momentum tensor in hyper-spacetime. Our conclusions and remarks are given in the final section.

\section{Unification of elementary particles and maximal symmetry in hyper-spacetime}

The SM has been tested by ever more precise experiments including the currently running LHC. In SM, the fermionic spinors, quarks and leptons, are thought to be the basic building blocks of nature. In the electromagnetic and strong interactions, the quarks and charged leptons are the Dirac spinors. In the weak interaction,  the quarks and leptons behave as the Weyl spinors. The smallness of the neutrino masses indicates that the neutrinos are likely Majorana-type spinors. In this section, we are going to show how all the quarks and leptons as elementary particles in SM are unified into a single hyper-spinor field.

\subsection{Unity of hyper-spin charges for each family of quarks and leptons}

The Dirac equation as a combination of  quantum mechanics and special relativity has led to the successful development of relativistic quantum mechanics and quantum field theory.  Let us begin with revisiting the Poincar\'{e} covariant Dirac equation\cite{Dirac} in  four-dimensional spacetime, 
\be \label{DiracEQ}
& & \left( \gamma^{\mu} i \partial_{\mu}  - m \right) \psi = 0\, ;\; \mbox{or} \;\; (\eta^{\mu\nu} \partial_{\mu}\partial_{\nu}  + m^2 ) \psi = 0 \, , \nn \\
& & \{ \gamma_{\mu}\, \,  \gamma_{\nu} \} = \eta_{\mu\nu}\, , \quad \eta_{\mu\nu} = diag.(1, -1, -1, -1)\, ,
\ee
where $\partial_{\mu} = \partial/\partial x^{\mu}$ denotes the coordinate derivative operator, and $\gamma_{\mu}$ ($\mu =0,1,2,3$) are the 4$\times$4 $\gamma$-matrices satisfying anticommutation relations. The unity of  quantum mechanics and special relativity with the Lorentz symmetry SO(1,3) leads to a complex four-component entity $\psi^T =(\psi_1\, , \psi_2\, ,\psi_3\, , \psi_4)$ which is referred to as the Dirac spinor field. It indicates that the four-dimensional spacetime that describes the movement and rotation in coordinate systems correlates to the four-component entity that reflects the boost spin and helicity spin of the Dirac spinor field. 

The Dirac equation reveals an interesting correlation between the basic quantum numbers of fermionic spinor and the geometric features of coordinate spacetime at a more profound level. Specifically, the dimensions of spacetime are coherently related with the degrees of freedom of the Dirac spinor. A massless Dirac spinor generates new symmetries corresponding to the chirality spin and conformal scaling transformations. Recently, we have shown that treating the chirality spin on the same footing as the boost spin and helicity spin of the Dirac spinor field enables us to obtain a generalized Dirac equation in six dimensional spacetime with the Lorentz symmetry SO(1,5)\cite{YLWU3}. It has been demonstrated that the chirality spin of the Dirac spinor field does correlate to a rotation in the extra two spatial dimensions. 

Inspired by the relativistic Dirac equation, and treating all the spin-like charges of the quarks and leptons on the same footing as a {\it hyper-spin charge} $Q_h$, all the degrees of freedom in each family of quarks and leptons can be written into a column vector in a spinor representation of {\it hyper-spacetime} with the dimension $D_h= 2Q_h =14$. 

Let us first introduce a Dirac-type {\it hyper-spinor field} $\Psi$ defined in 14-dimensional {\it hyper-spacetime}
\be \label{action0}
\Psi & \equiv & \Psi(\hx)\, , \quad \hx\equiv x^{\mM}\, , \;\; \mM = 0,1,2,3,5,\ldots,14\, .
\ee
An action for a freely moving massless hyper-spinor field $\Psi(\hx)$ is simply given as follows 
\be \label{action1}
I_H = \int [d\hx] \, \frac{1}{2} [\, \bar{\Psi}(\hx) \Gamma^{\mA} \, \delta_{\mA}^{\;\; \mM} i \partial_{\mM} \Psi(\hx) + H.c. \, ]\, , 
\ee
with $\mA, \mM = 0,1,2,3,5,\ldots,14$, and $\delta_{\mA}^{\;\;\mM} $ the Kronecker symbol. $\partial_{\mM}= \partial/\partial x^{\mM}$ is the partial derivative and $\Gamma^{\mA}$ is the vector $\gamma$-matrix defined in the spinor representation of 14-dimensional spacetime. The Latin alphabet $\mA, \mB\ldots$ and the Latin alphabet starting from $\mM, \mN$ are used to distinguish vector indices in non-coordinate spacetime and coordinate spacetime, respectively.  All the Latin indices are raised and lowered by the constant metric matrices, i.e., $\eta^{\mA\mB} $ or $\eta_{\mA\mB} =$ diag.$(1,-1,\ldots,-1)$, and $\eta^{\mM\mN} $ or $\eta_{\mM\mN} =$ diag.$(1,-1,\ldots,-1)$. The system of units is chosen such that $c = \hbar = 1$. 

The action of Eq.(\ref{action1}) is invariant under global Lorentz transformations of the symmetry group SO(1,13). The coordinates $x^{\mM}$ and hyper-spinor field $\Psi(\hx)$ transform in the vector and spinor representations, respectively,
\begin{eqnarray}
x^{\mM} \to x^{'\mM} = L^{\mM}_{\; \; \; \mN}\; x^{\mN}, \quad 
\Psi(\hx) \to \Psi'(\hx') = S(L) \Psi(\hx)\, , 
\ee
with
\be
& &  L^{\mM}_{\; \;\; \mN} \in \mbox{SO}(1,13)\, , \;\; S(L) = e^{i\alpha_{\mA\mB} \Sigma^{\mA\mB}/2}  \in \mbox{SP}(1,13),     \nn \\
& &   S(L) \Gamma^{\mA} S^{-1}(L) = L^{\mA}_{\;\;\; \mB}\; \Gamma^{\mB}\, ,  \quad  \Sigma^{\mA\mB} = \frac{i}{4}[\Gamma^{\mA}, \Gamma^{\mB}] , 
\ee  
where $\Sigma^{\mA\mB}$ are the generators of the hyper-spin group SP(1,13) in the spinor representation, they satisfy the group algebra,
\be
 & & [\Sigma^{\mA\mB}, \Sigma^{\mC\mD}] =  i (\Sigma^{\mA\mD}\eta^{\mB\mC} -\Sigma^{\mB\mD}  \eta^{\mA\mC} - \Sigma^{\mA\mC} \eta^{\mB\mD} + \Sigma^{\mB\mC} \eta^{\mA\mD}) \nn \\ 
 & & [\Sigma^{\mA\mB}, \Gamma^{\mC} ]  =  i ( \Gamma^{\mA} \eta^{\mB\mC} - \Gamma^{\mB} \eta^{\mA\mC} ) \, .
\ee
The Lorentz invariance requires that the symmetry groups SP(1,13) and SO(1,13) should coincide with each other, i.e.,  SP(1,13)$\cong$ SO(1,13).

The action of Eq.(\ref{action1}) is also invariant under parallel translations of the coordinates,
\begin{eqnarray}
 x^{\mM} \to x^{'\mM} = x^{\mM} + a^{\mM}
 \end{eqnarray}
with $a^{\mM}$ a constant vector.

In general, the Dirac-type hyper-spinor field $\Psi$ in 14-dimensional spacetime contains $N_f = 2^{[D_h/2]} = 2^7 = 128$ complex degrees of freedom. Geometrically, one can explicitly construct the spinors and show how they transform under the operations of relevant symmetry groups. For instance, the massive Dirac spinors and Weyl spinors as well as Majorana spinors in four-dimensional spacetime possess the maximal Lorentz symmetry SO(1,3) that characterizes the boost spin SU$^{\ast}$(2) and the helicity spin SU(2). A massless Dirac spinor was shown to have the maximal Lorentz symmetry SO(1,5) in six dimensional spacetime\cite{YLWU3}. This is because an additional chiral symmetry emerges for the massless Dirac spinors, which reflects the degrees of freedom corresponding to the chirality spin. As a consequence, a massless Dirac spinor can be treated as a Weyl-type spinor or a Majorana-type spinor with an intrinsic W-parity in six-dimensional spacetime\cite{YLWU3}.  

To specify the typical structure of the hyper-spinor field $\Psi$ so as to characterize the internal features of the quarks and leptons, let us first identify the spinor structure for each family of the quarks and leptons. In SM, each family of the quarks and leptons has 64 independent degrees of freedom, which can be written in terms of a single field with the following hyper-spinor structure:
\be
& & \Psi_{W\1 i}^{T}  =  [ (U_i^{r}, U_i^{b}, U_i^{g}, U_i^{w}, D^{r}_{ic}, D^{b}_{ic}, D^{g}_{ic}, D^{w}_{ic}, \nn \\
& & \quad D_i^{r}, D_i^{b}, D_i^{g}, D_i^{w}, -U^{r}_{ic}, -U^{b}_{ic}, -U^{g}_{ic}, -U^{w}_{ic})_L\, ,   \nn \\
& & \quad  (U_i^{r}, U_i^{b}, U_i^{g}, U_i^{w}, D^{r}_{ic}, D^{b}_{ic}, D^{g}_{ic}, D^{w}_{ic}, \nn \\ 
& & \quad D_i^{r},  D_i^{b}, D_i^{g}, D_i^{w}, -U^{r}_{ic}, -U^{b}_{ic}, -U^{g}_{ic}, -U^{w}_{ic})_R ]^T ,
\ee
with $i=1,\ldots, n_f$ for families. The superscript $T$ denotes as the transposition of a column matrix. We have denoted the Dirac spinors of the quarks and leptons in four-dimensional spacetime $Q_i^{\alpha} = (U_i^{\alpha}, D_i^{\alpha})$ with $\alpha= (r,\, b\, , g\, , w)$ representing the trichromatic (red, blue, green) and white colors, respectively. $Q_{i c}^{\alpha} =(U_{i c}^{\alpha},  D_{i c}^{\alpha})$ are defined as follows with a charge-conjugated operation acting on the Dirac spinors:
\be \label{CC1}
Q_{i\1 c}^{\alpha} = C_4 \bar{Q}_i^T= C_4 \gamma_0 Q_i^{\ast} \, ,  \quad C_4^{\dagger} = - C_4 \, ,
\ee
with $C_4$ the charge-conjugation matrix defined in four dimensional spacetime $C_4 = i\gamma_2\gamma_0$. The subscripts ``L" and ``R" in $Q_{i L,R}^{\alpha} = (U_i^{\alpha}, D_i^{\alpha})_{L,R}$ denote the left-handed and right-handed Dirac spinors, respectively,
\be 
Q_{i L,R}^{\alpha} = \frac{1}{2}(1\mp \gamma_5) Q_i^{\alpha} \, , \quad \gamma_5   Q_{i L,R}^{\alpha}  = \mp Q_{i L,R}^{\alpha} \, .
\ee
Explicitly, the $\gamma$-matrices defined in four dimensional spacetime take the following forms,
\be
 \gamma_0 = \sigma_1\otimes \sigma_0 \, , \quad \gamma_i = i \sigma_2\otimes \sigma_i\, , \quad \gamma_5 = \sigma_3\otimes \sigma_0 \, , 
\ee
where $\sigma_i$ ($i=1,2,3$) are the Pauli matrices and $\sigma_0 \equiv I_2$ denotes the $2\times 2$ unit matrix. 

The hyper-spinor field $\Psi_{W i}$ satisfies the Majorana-Weyl type condition defined in the spinor representation of 14 spacetime dimensions
\be
\Psi_{W i}^{ \bc} = \bC_{14} \bar{\Psi}_{W i}^T = \Psi_{W i} \, , \qquad \gamma_{15} \Psi_{W i} = - \Psi_{W i}
\ee
with the explicit forms
\be
 \bC_{14} & = & \sigma_1 \otimes \sigma_2 \otimes \sigma_2 \otimes \sigma_0 \otimes \sigma_0 \otimes C_4\, ,  \nonumber \\
 \gamma_{15} & = & \sigma_3 \otimes \sigma_0 \otimes \sigma_0 \otimes \sigma_0 \otimes \sigma_0 \otimes \gamma_5   \equiv \gamma_{11} \otimes \gamma_5 \, .
 \ee

The independent degrees of freedom of  $\Psi_{W i}$ are counted as $ N_h = 2^{[D_h/2] + 1}/4 = 2^{8}/4 = 64$, which does equal those in each family of the quarks and leptons.  The spinor structure of the Dirac-type hyper-spinor field $\Psi$ can be expressed in the following form,
\be \label{SS1}
\Psi  & \equiv & \Psi_{W} + \Psi_{E} \equiv \frac{1}{\sqrt{2}}(\Psi_1 + i \Psi_2) \nn \\
 & \equiv & \Psi_{W 1} + i \Psi_{W 2} + \Psi_{E 1} + i \Psi_{E 2}  \, ,  
\ee
with 
\be \label{SS1a}
& & \Psi_{W, E} = \frac{1}{2} (1 \mp \gamma_{15}) \Psi \, , \quad \Psi_{W, E}^{\bc}  =   \frac{1}{2} (1 \mp \gamma_{15}) \Psi^{\bc}\, ,  \nonumber \\
& & \Psi_{1}  =   \frac{1}{\sqrt{2}} (\Psi+ \Psi^{ \bc} ) \, , \quad \Psi_{2} = \frac{1}{\sqrt{2}i} (\Psi- \Psi^{ \bc} ) \, , 
\ee
and  
\be  \label{SS1b}
& & \Psi_{W 1}  =  \frac{1}{2} (\Psi_{W} + \Psi_{W}^{ \bc} ) \, , \quad \Psi_{W 2} = \frac{1}{2i} (\Psi_{W} - \Psi_{W}^{ \bc} )\, , \nonumber \\
& & \Psi_{E 1}  =  \frac{1}{2} (\Psi_{E} + \Psi_{E}^{\bc} ) \, , \quad \Psi_{E 2} = \frac{1}{2i} (\Psi_{E} - \Psi_{E}^{\bc} )\, , \nn \\
& & \Psi_{1} =   \Psi_{W\1 1} + \Psi_{E\1 1} \, , \quad \Psi_{2} =   \Psi_{W\1 2} + \Psi_{E\1 2} \, ; \quad  \Psi_{i}^{\bc} = \Psi_{i} \, ,
\ee
where $\Psi_{W, E}$ are regarded as a pair of {\it mirror hyper-spinor fields} defined in 14-dimensional spacetime. In order to distinguish them from the left-handed and right-handed chiral Weyl spinors in four-dimensional spacetime, we may name a pair of mirror hyper-spinor fields $\Psi_{W}$ and $\Psi_{E}$ {\it westward} and {\it eastward } hyper-spinor fields, respectively.

With a given hyper-spinor structure, the vector $\gamma$-matrix $\Gamma^{\mA}$  should be specified to characterize the symmetry relations between the degrees of freedom in the hyper-spinor field $\Psi$. In SM, the quarks and leptons are characterized by the symmetry groups U(1)$\times$SU(2)$\times$SU(3)$\times$SP(1,3) which should be contained in the hyper-spin symmetry group SP(1,13) as subgroups. From the well-known structure of the gauge models SU(4)$\times$SU(2)$_L$$\times$SU(2)$_R$ and SO(10), we can construct the vector $\gamma$-matrix $\Gamma^{\mA}$ with the following explicit forms:
\be \label{gammamatrix}
\Gamma_{0} & = &\1 \1 \1 \sigma_{0}\otimes \sigma_{0}\otimes \sigma_0 \otimes \sigma_0 \otimes \sigma_{0}\otimes \sigma_{1}\otimes \sigma_0 \ , \nonumber \\
\Gamma_{1} & = &\1 i \sigma_{0}\otimes \sigma_{0}\otimes \sigma_0 \otimes \sigma_0 \otimes \sigma_{0}\otimes \sigma_{2}\otimes \sigma_1 \ , \nonumber \\
\Gamma_{2} & = &\1 i \sigma_{0}\otimes \sigma_{0}\otimes \sigma_0 \otimes \sigma_0 \otimes \sigma_{0}\otimes \sigma_{2}\otimes \sigma_2 \ , \nonumber \\
\Gamma_{3} & = &\1 i \sigma_{0}\otimes \sigma_{0}\otimes \sigma_0 \otimes \sigma_0 \otimes \sigma_{0}\otimes \sigma_{2}\otimes \sigma_3 \ , \nonumber \\
\Gamma_{5} & = &\1 i \sigma_{1}\otimes \sigma_0 \otimes \sigma_{1} \otimes \sigma_0 \otimes \sigma_{2}\otimes \sigma_{3}\otimes \sigma_0 \ , \nonumber \\
\Gamma_{6} & = &\1 i \sigma_{1}\otimes \sigma_0 \otimes \sigma_{2} \otimes \sigma_{3}\otimes \sigma_{2}\otimes \sigma_{3}\otimes \sigma_0\ , \nonumber \\
\Gamma_{7} & = &\1 i \sigma_{1}\otimes \sigma_0 \otimes \sigma_{1} \otimes \sigma_{2}\otimes \sigma_{3}\otimes
\sigma_{3}\otimes \sigma_0 \ , \nonumber \\
\Gamma_{8} & = &\1 i \sigma_{1}\otimes \sigma_0 \otimes  \sigma_{2}\otimes \sigma_{2}\otimes \sigma_0 \otimes \sigma_{3}\otimes \sigma_0\ , \nonumber \\
\Gamma_{9} & = &\1 i \sigma_{1}\otimes \sigma_0\otimes  \sigma_{1} \otimes \sigma_{2}\otimes \sigma_{1}\otimes \sigma_{3}\otimes \sigma_0 \ , \nonumber \\
\Gamma_{10} & = & i \sigma_{1}\otimes \sigma_0 \otimes \sigma_{2} \otimes \sigma_{1}\otimes \sigma_{2} \otimes \sigma_{3}\otimes \sigma_0\ , \nonumber \\
\Gamma_{11} & = & i \sigma_{2}\otimes \sigma_0 \otimes \sigma_0 \otimes \sigma_0 \otimes \sigma_0\otimes \sigma_{3}\otimes \sigma_0 \, , \nn \\
\Gamma_{12} & = & i \sigma_{1}\otimes \sigma_1 \otimes  \sigma_{3}\otimes \sigma_0 \otimes \sigma_0 \otimes \sigma_{3}\otimes \sigma_0 \ , \nonumber \\
\Gamma_{13} & = & i \sigma_{1}\otimes\sigma_2 \otimes  \sigma_{3}\otimes \sigma_{0} \otimes \sigma_0 \otimes \sigma_{3}\otimes \sigma_0\ , \nonumber \\
\Gamma_{14} & = & i \sigma_{1}\otimes \sigma_{3}\otimes \sigma_{3} \otimes \sigma_0 \otimes \sigma_0 \otimes \sigma_{3}\otimes \sigma_0\ , 
\ee
which enables us to write down explicitly the generators of the hyper-spin symmetry group SP(1,13), i.e.,  $\Sigma^{\mA\mB} = i [\Gamma^{\mA}\,  \Gamma^{\mB}]/4 $.

\subsection{Discrete symmetries (CPT) and hyper-spinor structure in hyper-spacetime  }

To further reveal the intrinsic properties of the hyper-spinor field and hyper-spacetime, we shall investigate the symmetry properties of the action of Eq.(\ref{action1}) under the operations of discrete symmetries of hyper-spacetime:  charge conjugation ($\mathcal{C}$), parity inversion ($\mathcal{P}$) and time reversal ($\mathcal{T}$).  

Let us first show the transformation property under the charge-conjugated operation $\bar{\cC}$ on the Majorana-type hyper-spinor field of the quarks and leptons in each family. The vector $\gamma$-matrix $\Gamma^{\mA}\equiv ( \Gamma^a\, , \Gamma^A) $  ($a = 0, 1, 2, 3,  \; A= 5,\ldots, 14$) transforms as
\be
& & \bC \Gamma^{a} \bC^{-1} = - (\Gamma^{a})^{T}\, ; \;\; \bC \Gamma^{A} \bC^{-1} =  (\Gamma^{A})^{T} ,
\ee
which implies that the ordinary four dimensional spacetime distinguishes from the other ten dimensions under the charge-conjugated operation $\bar{\cC}$. For convenience, let us express both the vector coordinate $x^{\mM}$ and the $\gamma$-matrix $\Gamma^{\mA}$ into two parts, 
\be
& & x^{\mM}  =  ( x^{\mu}, x^M)\, , \;\; \mu = 0,1,2,3\, , \; M=5,\ldots, 14 \, , \nonumber \\
& & \Gamma^{\mA} = (\Gamma^a, \Gamma^A)\, , \;\; \; a = 0,1,2,3\, , \; A=5,\ldots, 14 \, .
\ee
In order to keep the action, Eq.(\ref{action1}), invariant and nontrivial under the charge-conjugated operation $\bar{\cC}$, the hyper-spinor field has to transform as follows 
\be
\bar{\cC} \Psi(\hx) \bar{\cC}^{-1} = \bC \bar{\Psi}^T(x^{\mu}, - x^M) \, .
\ee
Unlike the ordinary four-dimensional spacetime coordinates, the signs of the ten spatial dimensions flip under the charge-conjugated operation $\bar{\cC}$. This is because if the ten spatial dimensions do not undergo a flip in sign under the charge-conjugated operation $\bar{\cC}$, the action will not be invariant due to the following identities:
\be
& & \bar{\cC}  \bar{\Psi}(\hx) \Gamma^A \Psi^(\hx) \bar{\cC}^{-1}  = \bar{\Psi}^{\bc}(\hx) \Gamma^A \Psi^{\bc}(\hx) = \bar{\Psi}(\hx) \Gamma^A \Psi(\hx)\, , \nonumber \\
& & \bar{\cC}  \bar{\Psi}(\hx) \Gamma^A  \delta^{\;\; M}_{A} i\partial_M \Psi(\hx)    \bar{\cC}^{-1}  =  \bar{\Psi}^{\bc}(\hx) \Gamma^A  \delta^{\;\; M}_{A} i\partial_M \Psi^{\bc}(\hx) \nn \\
& & \qquad \qquad \qquad \qquad = i\partial_M (\bar{\Psi}(\hx) )\, \Gamma^A   \delta^{\;\; M}_{A} \Psi(\hx)\, , \nonumber 
\ee
which has an opposite sign in comparison with the term required from the hermiticity of the action in Eq.(\ref{action1}). 

The action of Eq.(\ref{action1}) has a similar intrinsic discrete symmetry under the operation $\mathcal{W}$,
\be
& & \mathcal{W} \Psi(\hx ) \mathcal{W}^{-1} = W \Psi( x^{\mu}, -x^M )\, , \quad W= i\gamma^5 \, ,\nonumber \\
& &  W^{-1} \Gamma^{a} W = -\Gamma^{a}, \quad W^{-1} \Gamma^{A} W = \Gamma^{A} \, .
\ee
Such an operation defines an intrinsic W-parity\cite{YLWU3}. Under the combined operation $\mathcal{C} \equiv \mathcal{W\bC}$, all the coordinates in hyper-spacetime do not need to flip in sign, i.e.,  
\be
& & \Psi^{c} \equiv \mathcal{C} \Psi(\hx ) \mathcal{C}^{-1} = C_{14} \bar{\Psi}^{T} ( \hx )= C_{14} \Gamma_0 \Psi^{\ast}(\hx)\, ,  
\ee
where $C_{14}$ can simply be written as a product of the $\gamma$-matrices
\be
& & C_{14} \equiv W\bC_{14} = i \Gamma_2\Gamma_0\Gamma_6\Gamma_8\Gamma_{10}\Gamma_{12}\Gamma_{14} \, , \nn \\
& & C_{14}^{-1} \Gamma^{\mA} C_{14} = (\Gamma^{\mA})^{T}\, , 
\ee
which defines the charge-conjugation operation $\mathcal{C}$  in 14-dimensional hyper-spacetime. $C_{14}$ can be regarded as a natural extension of the charge-conjugation matrix $C_4$ in four dimensional spacetime as shown in Eq. (\ref{CC1}). Such a charge-conjugation operation $\mathcal{C}$ has the following feature: 
\be
\left(\Psi^{c}(\hx)\right)^{c} = \mathcal{C} \Psi^{c} (\hx ) \mathcal{C}^{-1} = - \Psi(\hx) \, ,
\ee
which reflects a discrete $Z_4$ property. 

Unlike the charge-conjugated operation $\bar{\cC}$, which is introduced as the Majorana-type condition to figure out the degrees of freedom in each family of the quarks and leptons, the newly defined charge-conjugation operation $\cC$ in 14-dimensional hyper-spacetime leads the Majorana-type condition of hyper-spinor field to be modified with the intrinsic W-parity,
\be
& & \cC  \Psi_{W i}(\hx) \cC^{-1} \equiv \Psi_{W i}^{c}(\hx) = C_{14} \bar{\Psi}_{W i}^T(\hx) \nn \\
& & \quad = C_{14} \Gamma_0 \Psi_{W i}^{\ast}(\hx) = \gamma_5 \Psi_{W i}(\hx) \, . 
\ee

To ensure the action of Eq.(\ref{action1}) to be invariant and nontrivial under the discrete symmetries $\mathcal{P}$ and $\mathcal{T}$  in hyper-spacetime, the hyper-spinor field should transform as follows:
\be
& & \mathcal{P} \Psi(\hx ) \mathcal{P}^{-1} = P_{14} \Psi(\tx )\, , \quad \tx = (x^0, -x^{1},\ldots, - x^{14}) \, , \nonumber \\
& &  P_{14}^{-1} \Gamma^{\mA} P_{14} = ( \Gamma^{\mA})^{ \dagger}, \qquad P_{14} = \Gamma^0 \, , 
\ee
for parity-inversion, and 
\be
& & \mathcal{T} \Psi(\hx ) \mathcal{T}^{-1} = T_{14} \Psi( -\tx ) ; \;  T_{14}^{-1} \Gamma^{\mA} T_{14} = (\Gamma^{\mA})^{T}, 
\ee
for time-reversal. The matrix $T_{14}$ is found to have an explicit form, 
\be
& & T_{14}=  \Gamma_1\Gamma_3 \Gamma_5\Gamma_7 \Gamma_9\Gamma_{11} \Gamma_{13}\gamma_{15} \, ,
\ee
which can be regarded as an extension to the time-reversal matrix $T_4= i\gamma_1\gamma_3$ defined in four dimensional spacetime. 

In general, the action Eq.(\ref{action1}) is invariant under the joint operation ${\bf \Theta} \equiv  \mathcal{CPT}$, where the hyper-spinor field transforms as
\be
& & {\bf \Theta} \Psi(\hx ) {\bf \Theta}^{-1} = \Theta \1 \bar{\Psi}^{T} ( -\hx) \, , \nonumber \\
& & \Theta^{-1}\1 \Gamma^{\mA}\1 \Theta =  (\Gamma^{\mA})^{\dagger},  \, \quad \Theta = CPT =\Gamma^0\, .
\ee

\subsection{Family and mirror hyper-spin charges with additional dimensions } 

In terms of the charge conjugation operation $\cC$, the action Eq.(\ref{action1}) can be rewritten as
\be \label{action2}
I_H & = & \int [d\hx] \, \frac{1}{2} [\, \bar{\Psi}(\hx) \Gamma^{\mA} \, \delta_{\mA}^{\;\; \mM} i \partial_{\mM} \Psi(\hx) + H.c. \, ] \nn \\
& = & \int [d\hx] \, \frac{1}{2} \bar{\hPsi}(\hx) \Gamma^{\mA} \, \delta_{\mA}^{\;\; \mM} i \partial_{\mM} \hPsi(\hx) \, , 
\ee
where we have introduced a new type of the hyper-spinor field, 
\be \label{16DMHSF}
\hPsi(\hx) \equiv \binom{\Psi(\hx) }{\Psi^{c}(\hx) }\, ,
\ee
which satisfies the Majorana-type condition,
\be
\hPsi^{c}(\hx) \equiv  \mathcal{C} \hPsi(\hx) \mathcal{C}^{-1} = C_{16}  \bar{\hPsi}^{T} ( \hx ) = \hPsi(\hx)\, .
\ee
$C_{16}$ defines a new charge-conjugation operation,
\be
C_{16} = - i \sigma_{2} \otimes C_{14} \, .
\ee
Such a charge conjugation operation in hyper-spacetime enables us to rewrite the hyper-spinor field $\Psi(\hx)$ with 128 complex degrees of freedom into the Majorana-type hyper-spinor field $\hPsi(\hx)$ with 256 real type degrees of freedom. The resulting action of Eq.(\ref{action2}) becomes self-hermitian. 

The real and imaginary parts in the decomposition of the hyper-spinor field shown in Eq.(\ref{SS1}) reflect a family hyper-spin charge between the pairs of the conjugated hyper-spinor fields $\Psi(\hx)$ and $\Psi^{c}(\hx)$, which results the action of Eq.(\ref{action2}) to have an internal SU(2) symmetry. For the massless hyper-spinor field $\hPsi(\hx)$, the action of Eq.(\ref{action2}) possesses an additional U(1) symmetry that characterizes the mirror hyper-spin charge between the westward hyper-spinor field $\Psi_W(\hx)$ and the eastward hyper-spinor field $\Psi_E(\hx)$. Such new spin-like charges are presumed to relate coherently with extra dimensions of hyper-spacetime. 

Let us now extend the action of Eq.(\ref{action2}) by requiring a maximal symmetry. Indeed, the action of Eq.(\ref{action2}) can be reconstructed to be a more general one in 18-dimensional hyper-spacetime, 
\be \label{action3}
I_H = \int [d\hx] \, \frac{1}{2} \bar{\hPsi}(\hx) \Gamma^{\fA} \, \delta_{\fA}^{\;\; \fM} i \partial_{\fM} \hPsi(\hx) \, , 
\ee
with $\fA, \fM =0,1,2,3,5, \ldots, 18$. The vector $\gamma$-matrix $\Gamma^{\fA}$ has the following explicit structure,
\be
& & \Gamma^{\fA} = \; \sigma_0 \otimes \Gamma^{\mA}\, , \nonumber \\
& & \Gamma^{15} = i \sigma_3 \otimes \gamma^{15}\, ,\nonumber \\
& & \Gamma^{16} = i \sigma_2 \otimes \gamma^{15}\, ,\nonumber \\
& & \Gamma^{17} = i \sigma_1 \otimes \gamma^{15}\, ,\nonumber \\
& & \Gamma^{18} = I_{256}  \, ,
\ee
with $\Gamma^{\mA}$ ($\mA = 0,1,2,3,5,\ldots, 14$) presented in Eq.(\ref{gammamatrix}). $I_{256}$ is the $256\times 256$ unit matrix. The action of Eq.(\ref{action3}) possesses an extended maximal symmetry,
\be
SP(1,17) \cong SO(1,17) \, ,
\ee
with the generators $ \Sigma^{\fA\fB}$ defined as
\be
& & \Sigma^{\fA\fB} = -  \Sigma^{\fB\fA} = \frac{i}{4} [\Gamma^{\fA}\, , \Gamma^{\fB}]  \, ,  \nonumber \\
& & \Sigma^{\fA18} = - \Sigma^{18\fA} = \frac{i}{2} \varGamma^{\fA}  \, , \nn \\
& & \fA, \fB = 0, 1,2,3,,5, \ldots, 17\, .
\ee

The extra four dimensions lead to the rotational symmetry SO(4)=SU(2)$\times$SU(2),  which characterizes both the family hyper-spin symmetry SU(2) and the mirror hyper-spin symmetry SU(2).

\subsection{Unification of all quarks and leptons as elementary particles in hyper-spacetime with minimal dimension and  maximal symmetry }

In SM, there are three families of quarks and leptons. The Majorana-type hyper-spinor field defined in 16-dimensional hyper-spacetime contains at most two families of quarks and leptons. To unify all the quarks and leptons in SM into a single hyper-spinor field, we need to introduce an additional family hyper-spin charge and extend the 16-dimensional hyper-spacetime to a higher-dimensional hyper-spacetime.  

Let us consider a Majorana-type hyper-spinor field $ \varPsi(\hx)$ defined in a spinor representation of 18-dimensional hyper-spacetime,
\be \label{19DMHSF}
\varPsi(\hx) =  \binom{\hPsi(\hx) }{\hPsi^{'}(\hx) } =
\begin{pmatrix}
\Psi(\hx) \\
\Psi^{c}(\hx)\\
\Psi'(\hx) \\
-\Psi^{'c}(\hx)
\end{pmatrix}\, ,
\ee
which satisfies both the Majorana-type and the Majorana-Weyl type conditions,
\be
& & \mathcal{C} \vPsi(\hx) \mathcal{C}^{-1} = \hat{C} \bar{\vPsi}^{T} ( \hx ) = \vPsi(\hx )\, , \nonumber \\
& & \mathcal{C} \vPsi_{\mp}(\hx) \mathcal{C}^{-1} = \hat{C} \bar{\vPsi}_{\mp}^{T} ( \hx ) = \vPsi_{\mp}(\hx )\, , \nn \\
& & \gamma_{19} \vPsi_{\mp}(\hx) = \mp \vPsi_{\mp}(\hx)\, , \quad \gamma_{19}^2 = 1\, , 
\ee
with 
\be
\hat{C} = -i \sigma_3\otimes \sigma_2 \otimes C_{14}\, ; \; \;  \gamma_{19} = \sigma_3\otimes \sigma_0 \otimes \gamma_{15}\, .
\ee

Taking a freely moving massless Majorana-type hyper-spinor field $ \varPsi(\hx)$ as an irreducible spinor representation in hyper-spacetime, we arrive at an action in hyper-spacetime,
\be \label{Uaction1}
I_H = \int [d\hx] \, \frac{1}{2} \bar{\vPsi}(\hx) \varGamma^{\fA} \, \delta_{\fA}^{\;\; \fM} i \partial_{\fM} \vPsi(\hx) \, ,  
\ee
with the minimal dimension $ \fA\, , \fM =0,1,2,3,5, \ldots, D_h=19$. The vector $\gamma$-matrix $\varGamma^{\fA}$ takes the following explicit form:
\be
& &  \varGamma^{\fA} \equiv (  \varGamma^{\mA},\;   \varGamma^{\ta} )\, ,  \nonumber \\
& & \varGamma^{\mA} = \; \sigma_0 \otimes \sigma_0 \otimes \Gamma^{\mA}\, , \nonumber \\
& & \varGamma^{15} = i \sigma_2 \otimes \sigma_3\otimes \gamma^{15}\, ,\nonumber \\
& & \varGamma^{16} = i \sigma_1 \otimes \sigma_0\otimes \gamma^{15}\, ,\nonumber \\
& & \varGamma^{17} = i \sigma_2 \otimes \sigma_1\otimes \gamma^{15}\, ,\nonumber \\
& & \varGamma^{18} = i \sigma_2 \otimes \sigma_2\otimes  \gamma^{15}\, ,\nonumber \\
& & \varGamma^{19} = \, I_{512} \, ,
\ee
with $\mA = 0,1,2,3,5,\ldots, 14$, $\ta = 15,\ldots, 19$. $I_{512}$ is a $512\times 512$ unit matrix. The above action possesses the maximal symmetry, 
\be
SP(1,D_h -1)\cong SO(1,D_h -1)\, , \quad D_h =19\, .
\ee
The generators of the symmetry group SP(1,$D_h$-1) are defined as follows:
\be
& & \varSigma^{\fA\fB} = -  \varSigma^{\fB\fA} = \frac{i}{4} [\varGamma^{\fA}\, , \varGamma^{\fB}]  \, , \nonumber \\
& & \varSigma^{\fA19} = - \varSigma^{19\fA} = \frac{i}{2} \varGamma^{\fA}  \, ,
\ee
with $\fA, \fB = 0, 1,2,3,5, \cdots, 18$.

The action of Eq.(\ref{Uaction1}) is invariant under the discrete symmetries $\mathcal{C}$, $\mathcal{P}$ and $\mathcal{T}$  defined in 19-dimensional hyper-spacetime. The hyper-spinor field transforms as
\be
& & \varPsi^{c}(\hx) \equiv \mathcal{C} \vPsi(\hx ) \mathcal{C}^{-1}  = C_{19} \bar{\vPsi}^{T} ( \hx )= C_{19} \varGamma_0 \vPsi^{\ast}(\hx)\, ,  \nn \\
& & C_{19}^{-1} \varGamma^{\fA} C_{19} = (\varGamma^{\fA})^{T}\, , 
\ee
for charge conjugation with $C_{19}$ given by
\be
C_{19} \equiv \hat{C} = \varGamma_2\varGamma_0\varGamma_6\varGamma_8\varGamma_{10}\varGamma_{12}\varGamma_{14}  \varGamma_{16}\varGamma_{18} \, ,
\ee
and
\be
& & \mathcal{P} \varPsi(\hx ) \mathcal{P}^{-1} = P_{19} \varPsi(\tx )\, , \nonumber \\
& &  P_{19}^{-1} \varGamma^{\fA} P_{19} = ( \varGamma^{\mA})^{ \dagger}, \qquad P_{19} = \varGamma^0 \, , 
\ee
for parity inversion with $\tx = (x^0, -x^{1},\ldots, - x^{18}, x^{19})$, as well as  
\be
\mathcal{T} \varPsi(\hx ) \mathcal{T}^{-1} = T_{19} \varPsi( -\tx ) \, ; \;  T_{19}^{-1} \varGamma^{\mA} T_{19} = (\varGamma^{\mA})^{T}\, , 
\ee
for time reversal with $T_{19}$ defined as
\be
T_{19}= i \varGamma_1\varGamma_3 \varGamma_5\varGamma_7 \varGamma_9\varGamma_{11} \varGamma_{13}\varGamma_{15} \varGamma_{17}\gamma_{19}\, .
\ee

In general, the spin-like charges as the basic quantum numbers of elementary particles determine the dimensions of hyper-spacetime with the maximal symmetry which identifies the structure of hyper-spinor field. Without considering symmetry breaking and dimension reduction, the Majorana-type hyper-spinor field in nineteen dimensional hyper-spacetime contains in general four families of vector-like quarks and leptons. Such a hyper-spinor structure allows us to define explicitly the Majorana-Weyl type hyper-spinor fields in hyper-spacetime,
\be
& & \varPsi(\hx) = \varPsi_{+}(\hx) + \varPsi_{-}(\hx) \, , \quad \varPsi_{\mp}(\hx) = \frac{1}{2} ( 1 \mp \gamma_{19} ) \varPsi(\hx)\, ,   \nonumber \\
& &  \mathcal{C} \vPsi_{\mp}(\hx ) \mathcal{C}^{-1} =  \varPsi_{\mp}(\hx) \, ;\quad 
 C_{19}^{-1} \gamma_{19}C_{19} = -\gamma_{19}^{T}\, . 
\ee

From the action Eq.(\ref{Uaction1}), we are able to obtain the Dirac-type equation of the massless hyper-spinor field, 
\be
& & \varGamma^{\fA} \, \delta_{\fA}^{\;\; \fM} i \partial_{\fM} \vPsi(\hx) = 0\, , \quad \eta^{\fM\fN}\p_{\fM}\p_{\fN} \vPsi(\hx) = 0\, , \nn \\
& & \eta^{\fM\fN} = \eta_{\fM\fN} = \diag.(1, -1, \cdots , -1) \, ,
\ee
which is regarded as an equation of motion for a generalized relativistic quantum theory in hyper-spacetime with the dimension $D_h=19$.

\section{Unification of basic forces with hyper-spin gauge symmetry and dynamics of the hyper-spinor field}

To merge the three families of quarks and leptons as the basic building blocks of nature into a single Majorana-type hyper-spinor field in the irreducible spinor representation of hyper-spacetime, one requires the minimal dimension of hyper-spacetime to be $D_h=19$. We are now going to show how the four basic forces of nature can be unified into a fundamental interaction governed by a hyper-spin gauge symmetry.

\subsection{Unification of basic forces with hyper-spin gauge symmetry} 

To ensure the action of Eq.(\ref{Uaction1}) to be invariant under the transformations of the global hyper-spin symmetry SP(1, $D_h$-1) for the freely moving massless hyper-spinor field and the global hyper-spacetime Lorentz symmetry SO(1, $D_h$-1) for the coordinates, the symmetry transformations of SP(1, $D_h$-1) and SO(1, $D_h$-1) have to coherently incorporate each other, i.e., SP(1, $D_h$-1)$\cong$SO(1, $D_h$-1). 

Based on the gauge principle of interactions, let us propose that the hyper-spin symmetry SP(1, $D_h$-1) is gauged as a local symmetry and the hyper-spacetime Lorentz symmetry SO(1, $D_h$-1) remains as a global symmetry. So that a fundamental interaction is governed by the hyper-spin gauge symmetry SP(1,$D_h$-1). In this case, the hyper-spin gauge symmetry SP(1, $D_h$-1) distinguishes from the global Lorentz symmetry SO(1, $D_h$-1) in hyper-spacetime.  To realize that, it is essential to introduce the {\it bicovaraint vector field} $\chih_{\fA}^{\;\, \fM}(\hx)$ and the {\it hyper-spin gauge field} $\cA_{\fM}(\hx)$ to preserve both the local hyper-spin gauge symmetry SP(1, $D_h$-1) and the global Lorentz symmetry SO(1, $D_h$-1). Explicitly, the kinematic term of the hyper-spinor field has to be extended as follows,
\be
\Gamma^{\fA} \delta_{\fA}^{\;\; \fM} i\partial_{\fM} \to \Gamma^{\fA}\chih_{\fA}^{\;\, \fM}(\hx) i\cD_{\fM} \, , 
\ee
with $\fA, \fM = 0,1,2,3,5,\ldots, D_h\; (D_h =19)$. The constant vector $\delta_{\fA}^{\;\; \fM}$  is replaced by the {\it bicovaraint vector field} $\chih_{\fA}^{\;\, \fM}(\hx)$, and the ordinary derivative $\partial_{\fM}$ is generalized to the covariant derivative $\cD_{\fM}$,
\be \label{HSGF0}
i\cD_{\fM} \equiv i\partial_{\fM} + \cA_{\fM}\, , \quad  \cA_{\fM}(\hx) \equiv \cA_{\fM}^{\; \fB\fC}(\hx)\, \frac{1}{2}\varSigma_{\fB\fC} \, ,
\ee
with the hyper-spin gauge field $\cA_{\fM}(\hx)$ as a hyper-spin connection in the language of differential geometry. 

We are able to generalize the action of Eq.(\ref{Uaction1}) for the freely moving massless hyper-spinor field to a gauge invariant action with the hyper-spin gauge symmetry SP(1, $D_h$-1) and the hyper-spacetime Lorentz symmetry SO(1, $D_h$-1),  
\be \label{Gaction1}
I_H = \int [d\hx] \, \frac{1}{2} \bar{\vPsi}(\hx) \varGamma^{\fA} \, \chih_{\fA}^{\;\; \fM}(\hx) i \cD_{\fM} \vPsi(\hx) \, , 
\ee
with $\fA\, , \fM =0,1,2,3,5, \ldots, D_h \; (D_h =19) $. The bold font Latin alphabet ($\fA, \fB,\ldots $) and  ($\fM, \fN,\ldots $) are adopted to distinguish the vector indices defined in the vector representations of the hyper-spin gauge group SP(1, $D_h$-1) and the hyper-spacetime Lorentz group SO(1, $D_h$-1), respectively.  

$\cA_{\fM}(\hx)$ is introduced as the {\it hyper-spin gauge field} that belongs to the adjoint representation of the hyper-spin gauge group SP(1, $D_h$-1). The action of Eq.(\ref{Gaction1}) is invariant under the global Lorentz transformations,
\be
& & x^{\fM} \to x^{'\fM} = L^{\fM}_{\; \; \; \fN}\; x^{\fN}, \quad  \vPsi(\hx) \to \vPsi'(\hx') = \vPsi(\hx)\, ,  \nonumber \\
& & \chih_{\fA}^{\;\, \fM}(\hx) \to \chih_{\fA}^{'\;\, \fM}(\hx') = L^{\fM}_{\; \; \; \fN} \chih_{\fA}^{\;\, \fN} (\hx)\, , \nn \\
& & \cA_{\fM}(\hx) \to \cA'_{\fM}(\hx') = L_{\fM}^{\; \; \; \fN}  \cA_{\fN}(\hx) \, , 
\ee
and also under the local hyper-spin gauge transformations,
\be
\cA'_{\fM}(\hx) & = &  S(\Lambda) \cA_{\fM}(\hx) S^{-1}(\Lambda)
+ S(\Lambda) i \partial_{\fM} S^{-1}(\Lambda), \nn \\
\vPsi'(\hx) & = & S(\Lambda) \vPsi(\hx)\, ,  \quad S(\Lambda) = e^{i\alpha_{\fA\fB}(\hx) \varSigma^{\fA\fB}/2}, 
\ee
with
\be
& &  S(\Lambda) \varGamma^{\fA} S^{-1}(\Lambda) = \Lambda^{\fA}_{\;\;\; \fB}(\hx)\; \varGamma^{\fB}; \;  S(\Lambda)  \in \mbox{SP}(1, D_h\mbox{-}1)\, ,     \nonumber \\
& &  L^{\fM}_{\; \;\; \fN} \in \mbox{SO}(1, D_h\mbox{-}1); \;\;  \Lambda^{\fA}_{\;\;\; \fB}(\hx) \in \mbox{SP}(1, D_h\mbox{-}1)\, ,
\ee   
where $S(\Lambda)$ and $\Lambda^{\fA}_{\;\;\; \fB}(\hx)$ are the local group elements in the spinor and vector representations of the hyper-spin gauge group SP(1,$D_h$-1), respectively. The action Eq.(\ref{Gaction1}) is also invariant under the transformations of parallel translation of coordinates  
\begin{eqnarray}
 x^{\fM} \to x^{'\fM} = x^{\fM} + a^{\fM}\, , \qquad  a^{\fM} \in P^{1, D_h-1}\, , 
 \end{eqnarray}
with $a^{\fM}$ a constant vector of the translation group $P^{1, D_h-1}$.

In general, the action, Eq.(\ref{Gaction1}), possesses a joined bimaximal symmetry 
\be \label{BMS}
G_S & \equiv &  PO(1, D_h-1)\Join SP(1, D_h-1) \, ,
\ee
where PO(1,$D_h$-1)\footnote{ A new notation PO(1,$D_h$-1) is adopted to replace the old one P(1,$D_h$-1) used before in Refs.\cite{YLWU1,YLWU2,YLWU3}.} denotes the global Poincar\'e symmetry (or non-homogeneous Lorentz symmetry) of hyper-spacetime,  
\be
 PO(1, D_h-1) \equiv P^{1, D_h-1} \ltimes SO(1, D_h-1) \, ,
 \ee
which characterizes the globally flat Minkowski hyper-spacetime. 

Note that the above global and local symmetry groups cannot be defined as a direct product group. The hyper-spinor field and the hyper-spin gauge field as well as the gauge-type hyper-gravifield belong to the spinor and  adjoint as well as vector representations of the hyper-spin gauge symmetry group SP(1, $D_h$-1), respectively.  The basic forces are characterized by the hyper-spin gauge field $\cA_{\fM}(\hx)$ with the associated bicovaraint vector field $\chih_{\fA}^{\;\, \fM}(\hx)$.

\subsection{ Equation of motion of the hyper-spinor field in a general gravitational relativistic quantum theory with a conformal scaling symmetry}

Let us further extend the action of Eq.(\ref{Gaction1}) to be invariant under the global conformal scaling transformations,
\be
& & x^{\fM} \to x^{'\fM} = \lambda^{-1} \1 x^{\fM};\;\, \cA_{\fM}(\hx) \to \cA'_{\fM}(\hx') =  \lambda\1 \cA_{\fM}(\hx), \nonumber \\
&& \vPsi(\hx) \to \vPsi'(\hx') = \lambda^{3/2}\1 \vPsi(\hx)\, , 
\ee
with $\lambda$ a constant scaling factor, and also under the local conformal scaling transformations, 
\be
& & \vPsi(\hx) \to \vPsi'(\hx) = \xi^{3/2} (\hx) \vPsi(\hx) \, , \nn \\
& & \hat{\chi}_{\fA}^{\;\; \fM}(\hx) \to  \hat{\chi}_{\fA}^{'\; \fM}(\hx) = \xi(\hx) \hat{\chi}_{\fA}^{\;\; \fM}(\hx).
\ee
with $\xi(\hx)$ a functional scaling factor. To realize that, it is useful to introduce a {\it scaling scalar field} $\phi(\hx)$ that transforms as follows, 
\be
\phi(\hx) \to \phi'(\hx) = \xi(\hx)\phi(\hx), \quad \phi(\hx) \to \phi'(\hx') = \lambda \phi(\hx).
\ee
So that we obtain a self-hermitian and {\it conformal scaling gauge invariant} action,
\be \label{Gaction2}
I_H =\int [d\hx] \, \phi^{D_h\mbox{-}4}(\hx) \chi(\hx) \1 \frac{1}{2} \bar{\vPsi}(\hx) \varGamma^{\fA} \, \chih_{\fA}^{\;\; \fM}(\hx) i \cD_{\fM} \vPsi(\hx) , \nonumber \\
\ee
where $\chi(\hx)$ is an inverse of the determinant of bicovaraint vector field $\chih_{\fA}^{\;\, \fM}(\hx)$, 
\be
 \chi(\hx) = 1/\chih(\hx)\, , \quad \chih(\hx) = \det \chih_{\fA}^{\; \; \fM}(\hx) \, .
\ee

The extended action of Eq.(\ref{Gaction2}) possesses an additional joined symmetry,
\be \label{BCSS}
G_S= S(1) \Join SG(1) \, ,
\ee
where S(1) and SG(1) denote the global {\it conformal scaling symmetry} and the local {\it conformal scaling gauge symmetry}, respectively.

From the gauge invariant action Eq.(\ref{Gaction2}), we can derive the equation of motion for the Majorana-type hyper-spinor field,
\be \label{EMSF}
& & \varGamma^{\fA} \hat{\chi}_{\fA}^{\;\; \fM}(\hx) i ( {\mathcal D} _{\fM} + {\mathsf V} _{\fM}(\hx) ) \vPsi(\hx) =   0 \, , 
\ee
where ${\mathsf V} _{\fM}$ is regarded as an {\it induced-gauge field}
\be
{\mathsf V}_{\fM} (\hx) & = &  \frac{1}{2} \partial_{\fM}\ln (\chi \phi^{D_h-3})  -\frac{1}{2}\chih_{\fB}^{\;\; \fN}\hat{\cD}_{\fN}\chi_{\fM}^{\;\; \fB} \, ,
\ee
with the covariant derivative defined as 
\be \label{COD}
& & \hat{\cD}_{\fM}\chi_{\fN}^{\;\; \fA}  = \hat{\p}_{\fM} \chi_{\fN}^{\;\; \fA} + \cA_{\fM\, \fB}^{\fA}  \chi_{\fN}^{\;\;\fB} \, , \nn \\
& &  \hat{\p}_{\fM} \equiv \partial_{\fM} + \fS_{\fM}\, , \quad \fS_{\fM}(\hx) \equiv \p_{\fM}\ln\phi(\hx)\, ,
\ee
which ensures the above equation of motion to be conformal scaling gauge invariant. $\fS_{\fM}(\hx)$ is viewed as a {\it pure conformal scaling gauge field}.

The vector field $\chi_{\fM}^{\;\; \fA}(\hx)$ is a {\it dual bicovaraint vector field} defined through the following orthonormality conditions:  
\be \label{Dual}
& & \chi_{\fM}^{\;\; \fA} (\hx)\, \chih^{\;\; \fM}_{\fB}(\hx)   =  \chi_{\fM}^{\;\; \fA} (\hx) \chih_{\fB\, \fN}(\hx) \eta^{\fM\fN} = \eta^{\;\; \fA}_{\fB} \; , \nonumber \\
& &  \chi_{\fM}^{\;\; \fA}(\hx) \chih^{\;\;\fN}_{\fA}(\hx) = \chi_{\fM\, \fA} (\hx) \chih_{\fB}^{\;\; \fN}(\hx)  \eta^{\fA\fB} = \eta_{\fM}^{\;\;\fN} \, ,
\ee
and $\chi(\hx)$ is the corresponding determinant,
\be
 \chi(\hx) = \det \chi_{\fM}^{\;\; \fA}(\hx) = \chih^{-1}(\hx)\, .
\ee

The dual bicovariant vector field  $\chi_{\fM}^{\;\; \fA}$ will characterize the gravitational interaction in hyper-spacetime. For convenience, we shall refer $\chi_{\fM}^{\;\; \fA}$ as a {\it hyper-gravifield}.  The equation of motion Eq.(\ref{EMSF}) describes the dynamics of the Majorana-type hyper-spinor field, which is viewed as the basic equation of a {\it general gravitational relativistic quantum theory} in hyper-spacetime. 

Let us deduce the quadratic derivative form for the above equation of motion Eq.(\ref{EMSF}). The explicit form is found to be,
\be \label{EMSF2}
& & \chih^{\fM\fN} (\hat{\nabla}_{\fM} + {\mathsf V} _{\fM} ) ( {\mathcal D} _{\fN} + {\mathsf V} _{\fN} )  \vPsi  =  \varSigma^{\fA\fB}\chih_{\fA}^{\;\; \fM} \chih_{\fB}^{\;\; \fN} \nn \\
& & \quad \cdot [\, \cF_{\fM\fN}  - {\cal G}_{\fM\fN}^{\fC} \chih_{\fC}^{\;\; \fP} i ( {\mathcal D} _{\fP} + {\mathsf V} _{\fP} ) + i {\cal V}_{\fM\fN}\, ]  \vPsi, 
 \ee
which is Lorentz and gauge invariant as well as global and local conformal scaling invariant.  We have used the following definitions:
 \be \label{STF1}
 & & \chih^{\fM\fN} = \chih_{\fA}^{\;\; \fM} \chih_{\fB}^{\;\; \fN} \eta^{\fA\fB}\, , \quad \chih_{\fA}^{\;\; \fM} =  \chih^{\fM\fN} \chi_{\fN}^{\;\; \fA} \, , 
 \ee
 for the symmetric tensor, and
 \be \label{CD}
 & & \hat{\nabla}_{\fM} ( {\mathcal D} _{\fN}  + {\mathsf V} _{\fN} ) \equiv {\mathcal D} _{\fM} ( {\mathcal D} _{\fN} + {\mathsf V} _{\fN} ) - \fGa_{\fM\fN}^{\fP} ( {\mathcal D} _{\fP} + {\mathsf V} _{\fP} ) ,\nn \\
 & & \fGa_{\fM\fN}^{\fP}  \equiv   \chih_{\fA}^{\;\; \fP}  \hat{\cD}_{\fM}\chi_{\fN}^{\;\;\fA} =  \chih_{\fA}^{\;\; \fP}(\hat{\p}_{\fM} \chi_{\fN}^{\;\; \fA} + \cA_{\fM\, \fB}^{\fA}  \chi_{\fN}^{\;\;\fB}\, ) \, ,
 \ee
 for the covariant derivative.

On the right-hand side of Eq.(\ref{EMSF2}), $\cF_{\fM\fN}(\hx)$ defines the field strength of the {\it hyper-spin gauge field} $\cA_{\fM}(\hx)$ 
\be
& & \cF_{\fM\fN}(\hx) \equiv  {\cal F}_{\fM\fN}^{\fA\fB}(\hx)\, \frac{1}{2}\varSigma_{\fA\fB} =  i[ \cD_{\fM}, \cD_{\fN}] \nn \\
&  & \;  = \partial_{\fM} {\cal A}_{\fN}(\hx) - \partial_{\fN} {\cal A}_{\fM}(\hx) - i [{\cal A}_{\fM}(\hx) ,  {\cal A}_{\fN}(\hx) ], 
\ee
and  ${\cal G}_{\fM\fN}^{\fA}$  defines the field strength of the gauge-type {\it hyper-gravifield} $\chi_{\fM}^{\; \fA}(\hx)$, 
 \be
 & &  {\cal G}_{\fM\fN}^{\fA}  = \hat{\cD}_{\fM} \chi_{\fN}^{\;\; \fA} - \hat{\cD}_{\fN} \chi_{\fM}^{\;\; \fA} \equiv  {\cal G}_{\fM\fN}^{\fA}(\hx) \, \frac{1}{2}\Gamma_{\fA}\, ,
 \ee
where the covariant derivative $\hat{\cD}_{\fM} \chi_{\fN}^{\;\; \fA}$ is defined in Eq.(\ref{COD}). For the field strength ${\cal V}_{\fM\fN}$ of the induced-gauge field  ${\mathsf V} _{\fM}$, we have
 \be
 {\cal V}_{\fM\fN}  = \partial_{\fM}  {\mathsf V} _{\fN} - \partial_{\fN}  {\mathsf V} _{\fM} \equiv \frac{1}{2} ( \partial_{\fN}\fGa_{\fP\fM}^{\fP} -  \partial_{\fM}\fGa_{\fP \fN}^{\fP}  ).
 \ee

The equation of motion of the hyper-spinor field $\vPsi$ is generally characterized by the hyper-spin gauge field $\cA_{\fM}^{\fA\fB}$ and the gauge-type hyper-gravifield $\chi_{\fM}^{\; \fA}$ as well as the scaling scalar field $\phi$. The dynamics of the hyper-spinor field is governed by the hyper-spin gauge field strength $\cF_{\fM\fN}^{\fA\fB}$ and the hyper-gravifield strength ${\cal G}_{\fM\fN}^{\fA}$. Note that the induced-gauge field strength ${\cal V}_{\fM\fN}$ distinguishes from other field strengths due to an imaginary factor, which can cause the hyper-spinor field to be scaled by a real scaling factor.

\section{ Fiber bundle structure of hyper-spacetime and hyperunified field theory in hyper-gravifield spacetime}

The single Majorana-type hyper-spinor field $\vPsi(\hx)$ defined in hyper-spacetime transforms in the spinor representation of the hyper-spin gauge symmetry SP(1,$D_h$-1) and the real coordinate vector $\hx\equiv x^{\fM}$ in hyper-spacetime transforms in the vector representation of the global Lorentz symmetry SO(1,$D_h$-1) ($D_h =19$). The hyper-spin gauge field associated with the gauge-type hyper-gravifield is introduced to characterize the basic forces. We are going to investigate further the structure of hyper-spacetime and construct a general action of hyperunified field theory based on the postulates of gauge invariance and coordinate independence.

\subsection{ Hyper-gravifield fiber bundle structure}

A globally flat Minkowski hyper-spacetime $M_h$ is an {\it affine spacetime}, which possesses the Poincar\'e symmetry PO(1, $D_h$-1) = P$^{1, D_h\mbox{-}1}$ $\ltimes$ SO(1, $D_h$-1). The derivative vector operator $\partial_{\fM} \equiv \partial/\partial x^{\fM}$ at point $\hx$ of $M_h$ defines a {\it tangent basis } $\{\partial_{\fM}\}\equiv \{\partial/\partial x^{\fM}\} $ for a {\it tangent hyper-spacetime} $T_h$ over $M_h$. 

Accordingly, a {\it field vector} $\chih_{\fA}(\hx)$ at point $\hx$ of $M_h$ respective to the derivative vector operator $\partial_{\fM}$ is introduced via the bicovaraint vector field $\chih_{\fA}^{\;\, \fM}(\hx)$ as follows 
\be \label{eth}
 \eth_{\fA} \equiv \chih_{\fA}(\hx) = \chih_{\fA}^{\;\, \fM}(\hx) \partial_{\fM} \, ,
\ee
which forms a non-coordinate basis $ \{\eth_{\fA} \} $ or a field basis $ \{\chih_{\fA}(\hx) \} $ for a locally flat non-coordinate hyper-spacetime over a globally flat Minkowski hyper-spacetime $M_h$. 

In the coordinate spacetime, a displacement vector $dx^{\fM}$ at point $\hx$ of $M_h$ defines a {\it dual tangent basis} $\{dx^{\fM}\}$ for {\it dual tangent hyper-spacetime} $T_h^{\ast}$ over a globally flat Minkowski hyper-spacetime $M_h$. The tangent basis and dual tangent basis satisfy the dual condition,
\be
 \langle dx^{\fM},\, \partial/\partial x^{\fN}  \rangle = \frac{\partial x^{\fM}}{\partial x^{\fN}} = \eta_{\fN}^{\; \fM}\, .
\ee

In correspondence to the displacement vector $dx^{\fM}$, a field basis $\{ \chi^{\, \fA}(\hx) \}$ at point $\hx$ of $M_h$ is introduced as a dual basis of the field basis $ \{\chih_{\fA}(\hx) \} $.  Such a dual basis $\{ \chi^{\, \fA}(\hx) \}$ is defined via the {\it dual bicovariant vector field} $\chi_{\fM}^{\;\; \fA}(\hx)$ shown in Eq.(\ref{Dual}). $\chi_{\fM}^{\;\; \fA}(\hx)$ is the inverse of $\chih_{\fA}^{\;\;\fM}(\hx)$, which exists once the determinant of $\chih_{\fA}^{\;\; \fM}(\hx)$ is nonzero, i.e., $\det \chih_{\fA}^{\;\; \fM}(\hx)\neq 0$. A {\it dual field vector} $\chi^{\, \fA}(\hx)$ is explicitly defined through the dual bicovariant vector field $\chi_{\fM}^{\;\, \fA}(\hx)$ associated with the displacement vector $dx^{\fM}$, 
\be
\delta\chi^{\fA} \equiv \chi^{\; \fA} (\hx)  = \chi_{\fM}^{\;\; \fA} (\hx) dx^{\fM} \, , 
\ee
which forms a dual non-coordinate basis $ \{\dchi_{\fA} \} $ or field basis $ \{\chi^{\fA}(\hx) \} $. The dual non-coordinate bases of field bases satisfy the dual condition,
 \be
 \langle \delta \chi^{\fA},  \eth_{\fB}\rangle & \equiv &  \langle \chi^{\, \fA}(\hx), \chih_{\fB}(\hx) \rangle = \chi_{\fM}^{\; \fA}(\hx)  \chih_{\fB}^{\;\fN} (\hx)  \langle dx^{\fM} , \partial_{\fN} \rangle \nn \\
 &  = & \chi_{\fM}^{\;\; \fA}(\hx)  \chih_{\fB}^{\;\; \fN} (\hx)  \eta_{\fN}^{\; \fM} = \eta_{\fB}^{\;\, \fA} \, .
\ee

A pair of the dual non-coordinate bases $ \{\eth_{\fA} \} $ and $ \{\dchi_{\fA} \} $, or field bases $ \{\chih_{\fA}(\hx) \} $ and $\{\chi^{\, \fA}(\hx)\}$, form a pair of dual locally flat non-coordinate hyper-spacetimes over the globally flat Minkowski hyper-spacetime $M_h$. For convenience, we shall call such dual locally flat non-coordinate hyper-spacetimes dual {\it hyper-gravifield spacetimes}, denoted $G_h$ for a tangent-like and $G_h^{\ast}$ for a dual tangent-like, respectively. $ \{\eth_{\fA} \} $ and $ \{\dchi_{\fA} \} $, or $ \{\chih_{\fA}(\hx) \} $ and $\{\chi^{\, \fA}(\hx)\}$, are referred to as a pair of {\it dual hyper-gravifield bases}. 

The hyper-gravifield $\hat{\chi}_{\fA}^{\;\;\fM}(\hx)$, which is defined on the hyper-gravifield spacetime $G_h$ and valued on the tangent Minkowski hyper-spacetime $T_h$, transforms as a bicovariant vector field under the transformations of both the hyper-spin gauge symmetry SP(1, $D_h$-1) and the global Lorentz symmetry SO(1, $D_h$-1). Such a hyper-gravifield basis does not commute and it satisfies a non-commutation relation, 
\be
& &   [ \eth_{\fA} ,\; \eth_{\fB}] = f_{\fA\fB}^{\; \fC}\,  \eth_{\fC} \, ,  \quad \mbox{or} \quad [ \chih_{\fA}(\hx) ,\; \chih_{\fB}(\hx)] = f_{\fA\fB}^{\; \fC}(\hx)\,  \chih_{\fC}(\hx)  \, , \nonumber \\
& &  f_{\fA\fB}^{\; \fC} \equiv - \chih_{\fA}^{\;\; \fM} \chih_{\fB}^{\;\; \fN} \mG_{\fM\fN}^{\; \fC} \, ; \quad \mG_{\fM\fN}^{\; \fC}(\hx) \equiv\partial_{\fM}\chi_{\fN}^{\;\; \fC}(\hx) - \partial_{\fN}\chi_{\fM}^{\;\; \fC}(\hx) \, ,
\ee 
which indicates that the locally flat hyper-gravifield spacetime $G_h$ is associated with a non-commutative geometry. Such a commutation relation generates a Lie algebra with a non-constant structure factor $f_{\fA\fB}^{\; \fC}(\hx)$, which is characterized by the hyper-gravifield strength $\mG_{\fM\fN}^{\; \fA}(\hx) $ of the hyper-gravifield $\chi_{\fM}^{\;\; \fA}(\hx)$. Such a hyper-gravifield behaves as a {\it gauge-type hyper-gravifield} that may be denoted as, 
\be
\digamma_{\fM}(\hx) \equiv \chi_{\fM}^{\;\; \fA}(\hx) \frac{1}{2} \Gamma_{\fA}\, ,
\ee
which is sided on the dual tangent Minkowski hyper-spacetime $T_h^{\ast}$ and valued on the dual hyper-gravifield spacetime $G_h^{\ast}$. 

In general, we are led to a {\it biframe hyper-spacetime} $T_h\times G_h$ with its dual {\it biframe hyper-spacetime} $T_h^{\ast}\times G_h^{\ast}$ over the coordinate hyper-spacetime $M_h$. Mathematically, the nature of globally and locally flat vector spacetimes allows for a canonical identification of vectors in the tangent Minkowski hyper-spacetime $T_h$ at points with vectors (points) in  the Minkowski hyper-spacetime itself $M_h$, and also for a canonical identification of vectors at points with its dual vectors at the same points. Physically, the globally flat Minkowski hyper-spacetime is considered as a {\it vacuum hyper-spacetime} $\fVh$.  With such a canonical identification for the vector spacetime, we arrive at a simplified structure of hyper-spacetime,
\be
 T_h \cong T_h^{\ast} \cong  M_h & \equiv & \fVh\, , \nonumber \\
G_h \cong G_h^{\ast} & \equiv & \fGh\, .
\ee

The whole structure of {\it biframe hyper-spacetime} forms a {\it hyper-gravifield fiber bundle} $\bf{E}_h$ with the {\it hyper-gravifield spacetime} as a fiber $\fGh$ and the vacuum hyper-spacetime as a base spacetime $\fVh$. The fiber bundle $\bf{E}_h$ and the product $\bf{V_h\times G_h}$ as a biframe hyper-spacetime are correlated with a continuous surjective map $\Pi_{\chi}$ which projects the bundle $\bf{E}_h$ to the base spacetime $\bf{V}_h$, i.e., $\Pi_{\chi}$: $\bf{E_h \to V_h}$. Geometrically, the {\it hyper-gravifield fiber bundle structure} of biframe hyper-spacetime is expressed as  $({\bf E_h, V_h}, \Pi_{\chi}, {\bf G_h}) $ with a trivial case,
\be
\bf{E}_h \sim \bf{V}_h \times \bf{G}_h\, .
\ee

\subsection{Hyperunified field theory with postulates of gauge invariance and coordinate independence}

It is more general to postulate that the laws of nature governed by the gauge symmetry should be independent of any choice of coordinate systems and scaling factors.  We shall work out a general action of {\it hyperunified field theory} based on the principles of {\it gauge invariance} and {\it coordinate independence} along with a {\it conformal scaling symmetry}.  

The pairs of the hyper-gravifield bases $\{\delta\chi^{\fA}\}$ and  $\{\eth_{\fA}\}$ or $\{\chi^{\fA}(\hx) \}$ and $\{\hat{\chi}_{\fA}(\hx)\}$  allow us to define an exterior differential operator in hyper-gravifield spacetime $\fGh$
\be
d_{\chi} \equiv  \delta\chi^{\fA}\wedge \eth_{\fA} =  \chi^{\fA}(\hx)\wedge \hat{\chi}_{\fA}(\hx) \, ,
\ee
which enables us to express gauge fields and their field strengths as one-form and two-form, respectively, in terms of the hyper-gravifield basis vector $\delta\chi^{\fA} $ and the exterior differential operator $d_{\chi}$. There are in general three types of gauge fields and their corresponding field strengths involved in hyperunified field theory,
\be 
& & \cA = -i \cA_{\fA}\, \dchiA;\quad  \cF = d_{\chi} \, \cA + \cA\wedge \cA =\frac{1}{2i} \cF_{\fA\fB}\, \dchiA \wedge \dchiB \; , \nonumber \\
& & \dg = -i \dg_{\fA} \, \dchiA ;\quad  {\cal G} = d_{\chi} \, \dg + \cA \wedge \dg + \mW \wedge \dg = \frac{1}{2i} {\cal G}_{\fA\fB}\, \dchiA \wedge \dchiB\; , \nonumber \\
& &  \mW = -i \mW_{\fA}\, \dchiA ; \quad {\cal W} = d_{\chi}\, \mW = \frac{1}{2i} {\cal W}_{\fA\fB}\, \dchiA \wedge \dchiB \, ,
\ee
which are defined on the locally flat hyper-gravifield spacetime $\bf{G}_h$. With the projection of the hyper-gravifield $\hat{\chi}_{\fA}^{\;\; \fM}$,  we are able to yield their relations with the usual gauge fields and field strengths defined in the globally flat vacuum hyper-spacetime $\bf{V}_h$,
\be
 \cA_{\fA} & = & \hat{\chi}_{\fA}^{\;\; \fM}\cA_{\fM}(\hx);\;   \cF_{\fA\fB} =  \hat{\chi}_{\fA}^{\;\; \fM}\,\hat{\chi}_{\fB}^{\;\; \fN} \cF_{\fM\fN}(\hx), \nonumber \\
 \dg_{\fA} & = & \hat{\chi}_{\fA}^{\;\; \fM} \dg_{\fM}(\hx); \;  \cG_{\fA\fB} =  \hat{\chi}_{\fA}^{\;\; \fM}\,\hat{\chi}_{\fB}^{\;\; \fN}\cG_{\fM\fN}(\hx).
\ee
Here the field strength of the gauge-type hyper-gravifield  $\dg_{\fM}(\hx)$ is generally defined as   
\be
{\cal G}_{\fM\fN}(\hx) & = & \hat{\cD}_{\fM}\dg_{\fN}  - \hat{\cD}_{\fN}\dg_{\fM} \nn \\
& = &  [\, \hat{\cD}_{\fM}\chi_{\fN}^{\;\; \fA} (\hx)  - \hat{\cD}_{\fN}\chi_{\fM}^{\;\; \fA}(\hx) \,] \, \frac{1}{2}\Gamma_{\fA}\, ,
\ee
with the covariant derivative given in Eq.(\ref{COD}).

To describe the dynamics of the scaling scalar field with a conformal scaling gauge invariance, it is necessary to introduce a {\it conformal scaling gauge field}  $\mW_{\fM} \equiv g_wW_{\fM}$ with a gauge coupling constant $g_w$. The conformal scaling gauge field $W_{\fM}$ transforms as an Abelian gauge field 
\be
& & \mW_{\fM}(\hx) \to \mW'_{\fM}(\hx) = \mW_{\fM}(\hx) + \partial_{\fM} \ln \xi(\hx)\, . 
\ee
The conformal scaling gauge-invariant field strength ${\cal W}_{\fM\fN} (\hx)$ is defined as, 
\be
{\cal W}_{\fM\fN} (\hx)= \partial_{\fM}\mW_{\fN}(\hx) - \partial_{\fN}\mW_{\fM}(\hx) \, .
\ee
Such a conformal scaling gauge field describes a basic force of {\it scaling gauge interaction}. The scaling gauge symmetry was proposed by Weyl\cite{Weyl} for a purpose of electromagnetic field, though it was not successful since the electromagnetic field is characterized by the U(1) gauge symmetry as is well known by now. In the hyper-gravifield basis, it is expressed as
\be
 & & \mW_{\fA} =  \hat{\chi}_{\fA}^{\;\; \fM}(\hx) \mW_{\fM}(\hx)\, , \nn \\
 & & \cW_{\fA\fB} =  \hat{\chi}_{\fA}^{\;\; \fM}(\hx)\,\hat{\chi}_{\fB}^{\;\; \fN}(\hx) \cW_{\fM\fN}(\hx) \, . 
\ee

To construct a general action of hyperunified field theory, let us express the covariant derivative as one-form in the hyper-gravifield spacetime ${\bf G}_h$ 
\be
& & {\mathcal D} = \chi^{\fA}(\hx) {\mathcal D}_{\fA}  \equiv  \delta\chi^{\fA} \left( \eth_{\fA} -i \cA_{\fA} \right)\, , 
\ee
and define the Hodge star $``\ast"$ in $D_h$-dimensional hyper-gravifield spacetime ${\bf G}_h$  
\be
\ast \cF & = & \frac{1}{2!(D_h-2)! 2i} \varepsilon_{\fA_1\fA_2\fA_3\cdots\fA_{D_h}}\, \nn \\
& & \eta^{\fA_1\fA'_1}\eta^{\fA_2\fA'_2} \cF_{\fA'_1\fA'_2}\, \dchi^{\fA_3}\wedge\cdots\wedge \dchi^{\fA_{D_h}} \, .
\ee

We are now in the position to construct a general gauge-invariant and coordinate-independent action of hyperunified field theory in hyper-gravifield spacetime ${\bf G}_h$ by applying for the exterior differential operator $ d_{\chi}$ and the dual hyper-gravifield basis vectors $\dchi^{\fA}$ and $\eth_{\fA}$. Explicitly, we arrive at the following general form, 
\be
\label{HUTaction}
I_H & = & \int \,\phi^{D_h-4}  \{\, i \bar{\vPsi}\, \dg \wedge \ast {\mathcal D} \, \vPsi -  \frac{1}{2g_w^2} \, {\cal W} \wedge \ast {\cal W}   \nn \\
& - & \frac{4}{D_h} \sum_{k=0}^{D_h-3}\alpha_{k} \Tr ({\cal F}\wedge \dg^{k}) \wedge \ast ({\cal F} \wedge \dg^{k})    \nonumber \\
& + & \frac{2}{D_h}\phi^2 [\, \sum_{k=0}^{D_h-3} \beta_{k}\Tr ({\cal G}\wedge \dg^k) \wedge \ast ({\cal G} \wedge \dg^k)  \nn \\
& - &  2 \alpha_E \,  Tr\, {\cal F} \wedge \ast (\dg \wedge \dg ) \, ]  - \frac{1}{2} d \phi \wedge \ast d \phi   \nn \\ 
& + &  \frac{4}{D_h} \1\beta_E\1 \phi^4 \, Tr\, (\dg \wedge \dg)\wedge \ast (\dg \wedge \dg )  \, \}   \, ,
\ee
where $\alpha_{k}$, $\beta_{k}$ $ ( k = 0,\ldots, D_h-3)$, $\alpha_E$ and $\beta_E$ are constant couplings. We will show that the independent numbers of couplings are actually determined by all possible independent structures of gauge interactions. 

We have used the following definitions and relations,
\be
& & \dg \equiv \eta_{\fA}^{\; \fB} \frac{1}{2}\varGamma_{\fB}\, \dchi^{\fA} \, , \quad   d \phi \equiv  (d_{\chi} -  \mW )\phi\, ,  \nn \\
& & \dg^2 \equiv (\dg \wedge \dg )  \equiv \eta_{\fA}^{\;\fA'} \eta_{\fB}^{\;\fB'}\, \frac{1}{2i} \varSigma_{\fA'\fB'} \,  \dchi^{\fA} \wedge \dchi^{\fB}\, , \nonumber \\
& & \dg^k \equiv \dg \wedge \cdots \wedge \dg 
= (\eta_{\fA_1}^{\; \fB_1} \frac{1}{2}\varGamma_{\fB_1}) \cdots (\eta_{\fA_k}^{\; \fB_k} \frac{1}{2}\varGamma_{\fB_k})\, \dchi^{\fA_1} \wedge \cdots \wedge \dchi^{\fA_k} \,  , \nn \\
& & \ast (\dg \wedge \dg)  = \frac{1}{2!(D_h-2)! 2i} \varepsilon_{\fA\fB\fA_3\cdots\fA_{D_h}}\, \eta^{\fA\fA'}\eta^{\fB\fB'} \frac{1}{2i} \varSigma_{\fA'\fB'} \, \dchi^{\fA_3}\wedge\cdots\wedge \dchi^{\fA_{D_h}}  \, , \nn \\
& & \ast d \phi =  \frac{1}{(D_h -1)!}\, (\eth_{\fA} - \mW_{\fA})\phi \, \varepsilon^{\fA}_{\;\; \fA_2\cdots \fA_{D_h }} \,  \dchi^{\fA_2} \wedge \cdots \wedge \chi^{\fA_{D_h}} \, , 
\ee
where $ \varepsilon^{\fA_1\cdots \fA_{D_h}}$ ( $\varepsilon^{01\cdots D_h} = 1$,  $ \varepsilon^{\fA_1\cdots \fA_{D_h}}= -\varepsilon_{\fA_1\cdots \fA_{D_h}}$ ) is a totally antisymmetric Levi-Civita tensor with the following general properties,
\be
& &  \varepsilon_{\fA_1\cdots \fA_n} \varepsilon^{\fB_1...\fB_n} = - n! \,\eta^{\fB_1}_{[\fA_1}\cdots\eta^{\fB_n}_{\fA_n]}\, , \quad  \varepsilon_{\fA_1\cdots \fA_n}  \varepsilon^{\fA_1\cdots \fA_n}  = - n!\, ,   \nonumber \\
 & &\varepsilon_{\fA_1\cdots \fA_k \fA_{k+1}\cdots \fA_n} \varepsilon^{\fA_1\cdots \fA_k \fB_{k+1}\cdots \fB_n} = - k! (n-k)!\, \eta^{\fB_{k+1}}_{[\fA_{k+1}}...\eta^{\fB_n}_{\fA_n]}\, , \nonumber \\
 & &  \varepsilon^{\fA_1\cdots \fA_n} \varepsilon^{\fB_1\cdots \fB_n} M_{\fA_1 \fB_1} \cdots M_{\fA_n \fB_n} 
 = n!\1 \det \mathsf{M} \, ,
 \ee
for $\mathsf{M}$ being a $n\times n$ matrix $\mathsf{M} = (M_{\fA\fB})$.

As the general action Eq.(\ref{HUTaction}) of {\it hyperunified field theory} is constructed based on the hyper-gravifield fiber bundle structure of biframe hyper-spacetime, it has a joined bimaximal global and local symmetry,
\be
G_S = PO(1, D_h\mbox{-}1)\times S(1) \Join SP(1, D_h\mbox{-}1) \times SG(1) \, .
\ee
To unify all the quarks and leptons as elementary particles in SM into a single hyper-spinor field, such a hyper-gravifield spacetime ${\bf G}_h$ as a non-coordinate hyper-spacetime requires the same minimal dimension $D_h =19$.

\section{ Hyperunified field theory within the framework of QFT and dynamics of basic fields with conserved currents }

The general action of hyperunified field theory, Eq.(\ref{HUTaction}), is obtained based on the postulates of gauge invariance and coordinate independence in the locally flat hyper-gravifield spacetime ${\bf G}_h$. In this section, we are going to reformulate such a general action of hyperunified field theory within the framework of QFT in the globally flat Minkowski hyper-spacetime and make a general analysis on the dynamics of basic fields with the conserved currents.

\subsection{Hyperunified field theory within the framework of QFT}

To show explicitly the hyper-gravifield fiber bundle structure of biframe hyper-spacetime and yield the general action of hyperunified field theory within the framework of QFT, we shall reformulate the general action of Eq.(\ref{HUTaction}) by projecting the locally flat hyper-gravifield spacetime ${\bf G}_h$ into the globally flat vacuum hyper-spacetime ${\bf V}_h$. It is simply realized by transferring the dual hyper-gravifield bases $\{\dchi^{\fA}\}$ and $\{\eth_{\fA}\}$ (or $\{\chi^{\fA}(\hx)\}$ and $\{\hat{\chi}_{\fA}(\hx)\}$) into  the corresponding dual coordinate bases $\{ dx^{\fM}\}$ and $\{ \partial_{\fM}\}$. The explicit form is found to be 
\be
\label{HUTaction1}
I_H & \equiv & \int [d\hx]\; \chi \kL = \int [d\hx]\; \chi \, \phi^{D_h-4} \{\frac{1}{2} \chih^{\fM\fN} \bar{\vPsi}(\hx) \vGa_{\fA}\chi_{\fM}^{\;\;\fA} i {\mathcal D}_{\fN} \vPsi(\hx) \nn \\
& - & \frac{1}{4}   \,[\, \hat{\chi}^{\fM\fN\fM'\fN'}_{\fA\fB\fA'\fB'} \cF_{\fM\fN}^{\fA\fB} \cF_{\fM'\fN'}^{\fA'\fB'} 
 +   \chih^{\fM\fM'} \chih^{\fN\fN'} {\cal W}_{\fM\fN} {\cal W}_{\fM'\fN'} \, ] \nonumber \\
& + &  \frac{1}{4} \phi^2 [\, \hat{\chi}^{\fM\fN\fM'\fN'}_{\fA\fA'} {\cal G}_{\fM\fN}^{\fA} {\cal G}_{\fM'\fN'}^{\fA'} - 4 \alpha_E \1 \hat{\chi}^{\;\;\fM}_{\fA} \hat{\chi}^{\;\;\fN}_{\fB}  \cF_{\fM\fN}^ {\fA \fB} \, ]   \nonumber \\
& + &  \frac{1}{2}\hat{\chi}^{\fM\fN} d_{\fM} \phi d_{\fN}\phi  -  \beta_E\1\phi^4\,   \}\, .  
\ee
Where we have introduced the following definitions and notations:
\be \label{tensor}
& & \hat{\chi}^{\fM\fN\1 \fM'\fN'}_{\fA\fB\1 \fA'\fB'} \equiv  g_1 \hat{\chi}^{\fM\fM'} \hat{\chi}^{\fN\fN'} \eta_{\fA\fA'}\eta_{\fB\fB'} + \frac{1}{2}g_2(\chih_{\fA'}^{\; \fM} \chih_{\fB'}^{\; \fN} \chih_{\fA}^{\; \fM'} \chih_{\fB}^{\; \fN'}  + \chih_{\fA'}^{\; \fN} \chih_{\fB'}^{\; \fM} \chih_{\fA}^{\; \fN'} \chih_{\fB}^{\; \fM'} ) \nn \\
& & \qquad +  \frac{1}{2}g_3 [ \eta_{\fA\fA'} ( \chih^{\fM\fM'}  \chih_{\fB'}^{\; \fN} \chih_{\fB}^{\; \fN'} +  \chih^{\fN\fN'} \chih_{\fB'}^{\; \fM} \chih_{\fB}^{\; \fM'}  ) +  (\fA\leftrightarrow \fB, \fA' \leftrightarrow \fB')  ]\nn \\
& & \qquad +  \frac{1}{2}g_4  [\eta_{\fA\fA'} (  \chih^{\fM\fM'}  \chih_{\fB}^{\; \fN} \chih_{\fB'}^{\; \fN'} + \chih^{\fN\fN'} \chih_{\fB}^{\; \fM} \chih_{\fB'}^{\; \fM'}  ) + (\fA\leftrightarrow \fB, \fA' \leftrightarrow \fB')  ] \nn \\
& &  \qquad  + \frac{1}{2}g_5 [\chih_{\fA}^{\; \fM'}\chih_{\fA'}^{\fM}  \chih_{\fB}^{\; \fN} \chih_{\fB'}^{\; \fN'} + \chih_{\fA}^{\; \fN'}\chih_{\fA'}^{\fN}  \chih_{\fB}^{\; \fM} \chih_{\fB'}^{\; \fM'} +   (\fA\leftrightarrow \fB, \fA' \leftrightarrow \fB') \, ] \nn \\
& & \qquad +  \frac{1}{2}g_6 ( \chih_{\fA}^{\; \fM} \chih_{\fB}^{\; \fN} \chih_{\fA'}^{\; \fM'} \chih_{\fB'}^{\; \fN'}  + \chih_{\fA}^{\; \fN} \chih_{\fB}^{\; \fM} \chih_{\fA'}^{\; \fN'} \chih_{\fB'}^{\; \fM'} )\, , \\ 
& &  \hat{\chi}^{\fM\fN\1 \fM'\fN'}_{\fA\fA'}  \equiv  \alpha_G  \hat{\chi}^{\fM\fM'} \hat{\chi}^{\fN\fN'} \eta_{\fA\fA'} + \beta_G     (\chih^{\fM\fM'} \chih_{\fA'}^{\; \fN} \chih_{\fA}^{\; \fN'} + \chih^{\fN\fN'} \chih_{\fA'}^{\; \fM} \chih_{\fA}^{\; \fM'} ) \nn \\ \label{tensor0}
 && \qquad \qquad \; \;  -2 \gamma_G (\chih^{\fM\fM'} \chih_{\fA}^{\; \fN} \chih_{\fA'}^{\; \fN'}  + \chih^{\fN\fN'} \chih_{\fA}^{\; \fM} \chih_{\fA'}^{\; \fM'}  ) \, , \\
& & d_{\fM} \phi = (\partial_{\fM} - g_w W_{\fM} ) \phi \, , 
\ee
where $g_i$($i=1,\ldots,6$) as the combinations of the coupling constants $\alpha_k$ ($k=0,\ldots, D_h$-3) are the general coupling constants associated with all the possible structures of the hyper-spin gauge interactions. The coupling constants $\alpha_G$, $\beta_{G}$ and $\gamma_{G}$ in relevant to $\beta_k$ ($k=0,\ldots, D_h$-3) reflect all the possible structures of the gauge-type hyper-gravifield interactions. It is noticed that the symmetric tensor field $ \hat{\chi}^{\fM\fN}(x)$ defined in Eq.(\ref{STF1}) couples to all fields.

\subsection{Equations of motion of basic fields in hyper-spacetime}

The general action of Eq.(\ref{HUTaction1}) of hyperunified field theory enables us to derive equations of motion for various basic fields including the hyper-spin gauge field, the gauge-type hyper-gravifield, the conformal scaling gauge field and the scaling scalar field. Such equations of motion are considered as the basic equations of a general gravitational relativistic quantum theory in hyper-spacetime. 

The equations of motion for the hyper-spin gauge field $\cA_{\fM}^{\fA\fB}$ and the scaling gauge field $\cW_{\fM}$ are found to be
\be   \label{EMGF}
& & \cD_{\fN} ( \phi^{D_h-4}  \chi\,  \hat{\chi}^{[\fM\fN]\fM'\fN'}_{[\fA\fB]\fA'\fB'} {\cal F}_{\fM'\fN' }^{\;\; \fA'\fB'} ) =  J_{\fA\fB}^{\; \fM}\, , \\
& &  \partial_{\fN} ( \phi^{D_h-4}  \chi   \hat{\chi}^{\fM\fM'} \hat{\chi}^{\fN\fN'}  {\cal W}_{\fM'\fN'})  =   J^{\fM}\, ,
\ee
with
\be
\hat{\chi}^{[\fM\fN]\fM'\fN'}_{[\fA\fB]\fA'\fB'} & = & \frac{1}{2} ( \hat{\chi}^{[\fM\fN]\fM'\fN'}_{\fA\fB\1 \fA'\fB'}  -  \hat{\chi}^{[\fM\fN]\fM'\fN'}_{\fB\fA\1 \fA'\fB'}  ) \, \nn \\\
 \hat{\chi}^{[\fM\fN]\fM'\fN'}_{\fA\fB\fA'\fB'} & = &  \frac{1}{2} (\hat{\chi}^{\fM\fN\1 \fM'\fN'}_{\fA\fB\1 \fA'\fB'}  -  \hat{\chi}^{\fN\fM\1 \fM'\fN'}_{\fA\fB\1 \fA'\fB'} ) .
\ee
The covariant currents have the following explicit forms:
\be \label{GCC}
J_{\fA\fB}^{\;\fM} & = & -\frac{1}{2} \chi  \phi^{D_h-4} \bar{\vPsi} \hat{\chi}_{\fC}^{\;\; \fM} \{ \varGamma^{\fC}\;\;  \frac{1}{2} \varSigma_{\fA\fB} \} \vPsi   \nn \\
& - &  \alpha_E \cD_{\fN} (\, \phi^{D_h-2}  \chi \chih^{\fM\fN}_{[\fA\fB]} \, ) +  \frac{1}{2}  \phi^{D_h-4}  \chi\1 \chih_{[\fA\fB]\fA'}^{\fM\; \fM'\fN'}  {\cal G}_{\fM'\fN'}^{\fA'}  \, , \\
J^{\fM} & = &  -g_w\chi \hat{\chi}^{\fM\fN} \phi^{D_h-3} d_{\fN} \phi \, , \label{GCC0}
\ee
with the definitions and notations,
\be \label{AST}
& & \chih_{[\fA\fB]\fA'}^{\fM\; \fM' \fN'} \equiv  \chi_{\fN\fA} \1 \hat{\chi}^{[\fM\fN]\fM'\fN'}_{\fB\fA'} -  \chi_{\fN\fB}\1  \hat{\chi}^{[\fM\fN]\fM'\fN'}_{\fA\fA'}, \nn \\
& & \hat{\chi}^{[\fM\fN]\fM'\fN'}_{\fA\fA'}  =  \frac{1}{2} (\,  \hat{\chi}^{\fM\fN\1 \fM'\fN'}_{\fA\fA'}  -  \hat{\chi}^{\fN\fM\1 \fM'\fN'}_{\fA\fA'}  \, )\, , \nn \\
& & \chih^{\fM\fN}_{[\fA\fB]}   \equiv  \chih_{\fA}^{\;\;\fM}\chih_{\fB}^{\;\;\fN }  -  \chih_{\fB}^{\;\;\fM}\chih_{\fA}^{\;\;\fN }  \, .  
\ee

The equation of motion for the gauge-type hyper-gravifield $\chi_{\fM}^{\;\; \fA}$ is  given by 
\be  \label{EMG}
& &  \bar{\cD}_{\fN}\1 {\cal G}^{\; \fM\fN}_{\fA} = J_{\fA}^{\;\; \fM} \, , 
\ee
with
\be
& &  {\cal G}^{\; \fM\fN}_{\fA}  \equiv \phi^{D_h-2}  \chi    \hat{\chi}^{[\fM\fN]\fM'\fN'}_{\fA\fA'}    {\cal G}_{\fM'\fN' }^{\fA'} \, , 
\ee
for the covariant tensor, and 
\be \label{EMGC}
& &  \bar{\cD}_{\fN}\1 {\cal G}^{\; \fM\fN}_{\fA} \equiv \bar{\p}_{\fN} {\cal G}^{\; \fM\fN}_{\fA} +  \cA_{\fN \fA}^{\;\;\; \;\fB} {\cal G}^{\; \fM\fN}_{\fB}\, , \nn \\
& & \bar{\p}_{\fN} \equiv \partial_{\fN} - \fS_{\fN}  = \partial_{\fN} - \p_{\fN}\ln\phi \, ,
\ee
for the covariant derivative. The bicovariant vector current has an explicit form
\be \label{EMGCC}
 J_{\fA}^{\;\; \fM} & = &  - \chi \hat{\chi}_{\fA}^{\;\;\fM} \kL +  \chi \phi^{D_h-4} \hat{\chi}_{\fA}^{\; \; \fP}  [\, \frac{1}{2}   \hat{\chi}_{\fA''}^{\;\; \fM}  \bar{\vPsi} \varGamma^{\fA''} i {\mathcal D}_{\fP} \vPsi -   {\cal W}_{\fP\fQ} {\cal W}^{\fM\fQ}   \nonumber \\
& - & \hat{\chi}^{[\fM\fQ]\fM'\fN'}_{\fA''\fB\; \fA'\fB'}  {\cal F}_{\fP\fQ}^{\fA''\fB} {\cal F}_{\fM'\fN'}^{\fA'\fB'}  + \phi^2\, \hat{\chi}^{[\fM\fQ]\fM'\fN'}_{\fA''\fA'}  {\cal G}_{\fP\fQ}^{\fA''} {\cal G}_{\fM'\fN'}^{\fA'}  + d_{\fP}\phi  d^{\fM} \phi  
\nn \\
& - & 2 \alpha_E \chi \phi^{2} \hat{\chi}_{\fA'}^{\;\;\fM}  \hat{\chi}_{\fB'}^{\;\; \fN'}  \cF_{\fP\fN'}^{\fA'\fB'} \, ] \, .
\ee

The equation of motion for the {\it scaling scalar field} is given by 
\be   \label{EMS}
& & d_{\fM}  ( \chi \hat{\chi}^{\fM\fN}  \phi^{D_h-4}   d_{\fN}\phi ) =  J  \, , 
\ee
with the scalar current,
\be  \label{SC}
J &= &  (D_h-4) \phi^{D_h-5} \chi [ \,  \frac{1}{2} \chih^{\fM\fN} \bar{\vPsi}(\hx) \vGa_{\fA}\chi_{\fM}^{\;\;\fA} i {\mathcal D}_{\fN} \vPsi(\hx)   + \frac{1}{2} \chih^{\fM\fN} \p_{\fM}\phi \p_{\fN}\phi   \nn \\
&  - & \frac{1}{4} (\, \hat{\chi}^{\fM\fM'} \hat{\chi}^{\fN\fN'} {\cal W}_{\fM\fN} {\cal W}_{\fM'\fN'} + \hat{\chi}^{\fM\fN\1 \fM'\fN'}_{\fA\fB\1 \fA'\fB'}{\cal F}_{\fM\fN}^{\fA\fB} {\cal F}_{\fM'\fN'}^{\fA'\fB'} \, ) \, ] \nn \\
 & + & \frac{1}{4} (D_h-2) \phi^{D_h-3} \chi [\, \hat{\chi}^{\fM\fN\1 \fM'\fN'}_{\fA\fA'}  {\cal G}_{\fM\fN}^{\fA} {\cal G}_{\fM'\fN' }^{\fA'}- 4\alpha_E g_s \hat{\chi}_{\fA}^{\;\;\fM} \hat{\chi}_{\fB}^{\;\;\fN}  {\cal F}_{\fM\fN}^{\fA\fB}\, ] \nn \\
& - & \p_{\fM}(  \phi^{D_h-2}  {\cal G}^{\; \fM\fN}_{\fA} \chi_{\fN}^{\; \fA})/\phi   - \beta_E D_h\, \chi \phi^{D_h-1}  \, .
\ee
It is noticed that all the equations of motion are conformal scaling gauge invariant.

\subsection{ Conserved currents in hyperunified field theory}

Noether's theorem\cite{NT} states that, for every differentiable symmetry generated by a local action, there exists a corresponding conserved current. We shall show that the currents generated by the hyper-spin gauge symmetry SP(1,18) and the scaling gauge symmetry SG(1) are conserved currents. 

For the hyper-spin gauge symmetry SP(1,$D_h$-1),  its conserved current can be checked from the equation of motion of the hyper-spin gauge field,
\be   \label{CC1}
\cD_{\fM}J^{\;\fM}_{ \fA\fB}  & = & \cD_{\fM} \cD_{\fN} ( \phi^{D_h-4} \chi  \hat{\chi}^{[\fM\fN]\fM'\fN'}_{[\fA\fB]\fA'\fB'} {\cal F}_{\fM'\fN' }^{\; \fA'\fB'} ) =0 \, ,
\ee
which holds due to the identity
\be
& & \cD_{\fM} \cD_{\fN} (  \phi^{D_h-4}  \chi  \hat{\chi}^{[\fM\fN]\fM'\fN'}_{[\fA\fB]\fA'\fB'}   {\cal F}_{\fM'\fN' }^{\; \fA'\fB'} ) = \partial_{\fM} \partial_{\fN} (  \phi^{D_h-4}  \chi  \hat{\chi}^{[\fM\fN]\fM'\fN'}_{[\fA\fB]\fA'\fB'}  ) {\cal F}_{\fM'\fN' }^{\; \fA'\fB'}  \nonumber \\
& & \;+ \partial_{\fM} ( \phi^{D_h-4} \chi   \hat{\chi}^{[\fM\fN]\fM'\fN'}_{[\fA\fB]\fA'\fB'} ) \cD_{\fN}  {\cal F}_{\fM'\fN' }^{\; \fA'\fB'} + \partial_{\fN} (  \phi^{D_h-4}  \chi  \hat{\chi}^{[\fM\fN]\fM'\fN'}_{[\fA\fB]\fA'\fB'}  ) \cD_{\fM} {\cal F}_{\fM'\fN' }^{\;\; \fA'\fB'}  \nonumber \\
& & \; +  \phi^{D_h-4}   \chi   \hat{\chi}^{[\fM\fN]\fM'\fN'}_{[\fA\fB]\fA'\fB'}  \cD_{\fM} \cD_{\fN} {\cal F}_{\fM'\fN' }^{\; \fA'\fB'} 
 =  \phi^{D_h-4} \chi  \hat{\chi}^{[\fM\fN]\fM'\fN'}_{[\fA\fB]\fA'\fB'}   (  {\cal F}_{\fM\fN \fC }^{\;\;  \fA'}  {\cal F}_{\fM'\fN' }^{\;\; \fC\fB'}  - {\cal F}_{\fM\fN \fC }^{\;\;  \fB'}  {\cal F}_{\fM'\fN' }^{\;\; \fC\fA'} ) = 0 , \nonumber 
\ee
where we have used the symmetric and antisymmetry properties of the tensors to show the vanishing due to a total cancellation.

When applying the conserved current equation $\cD_{\fM}J^{\fM}_{\;\; \fA\fB}  = 0$ to the explicit form of the current in Eq.(\ref{GCC}), we arrive at a correlation equation between a hyper-spin angular momentum tensor $\cS^{\fM}_{\;\; \fA\fB} $ and an antisymmetric current tensor $J_{[\fA\fB]}$,
\be \label{CC2}
  \cD_{\fM} \cS^{\fM}_{\;\; \fA\fB} =  - J_{[\fA\fB]} 
+ 2  \alpha_E  \phi^{D_h-2} \chi  (  \hat{\chi}_{\fA}^{\;\;\fM} {\cal F}_{\fM\fN \fB}^{\;\; \fC}   -  \hat{\chi}_{\fB}^{\;\;\fM} {\cal F}_{\fM\fN \fA}^{\;\; \fC}  )\hat{\chi}_{\fC}^{\;\;\fN} \, ,
\ee
with the definitions
\be 
& & \cS^{\fM}_{\;\; \fA\fB} = - \chi \, \phi^{D_h-4}  \hat{\chi}_{\fC}^{\;\; \fM} \bar{\vPsi}  \{ \varGamma^{\fC}\;\;  \frac{1}{2} \varSigma_{\fA\fB} \} \vPsi  \; , \nonumber \\
& & J_{[\fA\fB]} = J_{\fA}^{\fM} \chi_{\fM \fB} - J_{\fB}^{\fM}\chi_{\fM \fA} \, .
\ee
Such a correlation equation reflects the fact that the hyper-spin gauge invariance of the general action requires one to introduce both the hyper-spin gauge field $\cA_{\fM}^{\fA\fB}$ and the gauge-type hyper-gravifield $\chi_{\fM}^{\;\; \fA}$, so that their associated currents $\cS^{\fM}_{\fA\fB}$ and $J_{\fA}^{\; \fM}$ are correlated. 

Similarly,  the local scaling gauge symmetry leads to the conserved current,
\be   \label{CC3}
  \partial_{\fM} J^{\fM}  = \partial_{\fM} \partial_{\fN} ( \phi^{D_h-4}\, \chi   \hat{\chi}^{\fM\fM'} \hat{\chi}^{\fN\fN'}  {\cal W}_{\fM'\fN'}) \equiv 0 \, , 
\ee
which leads to the equation for a bosonic tensor when applying the explicit form of the current of Eq.(\ref{GCC0}),
\be
\p_{\fM} \{ \chi \hat{\chi}^{\fM\fN} \phi^{D_h-3} d_{\fN} \phi \, \} =0 \, .
\ee

The bicovariant vector current $J_{\fA}^{\;\;\fM}$ appearing in the equation of motion of the gauge-type hyper-gravifield $\chi_{\fM}^{\;\; \fA}$ is in general not homogeneously conserved. Explicitly, we obtain the following equation for the covariant derivative of the current,  
\be  \label{BCVC}
& & \bar{\cD}_{\fM} J_{\fA}^{\;\; \fM} =   \bar{\cD}_{\fM}   \bar{\cD}_{\fN} \cG^{\; \fM\fN}_{\fA}  =     \bar{\p}_{\fM}  [\, \phi^{D_h-2} \chi   \hat{\chi}^{[\fM\fN]\fM'\fN'}_{\fA\fA'} \, ] \hat{\cD}_{\fN} {\cal G}_{\fM'\fN'}^{\fA'} \nonumber  \\
& & \; +  \bar{\p}_{\fN}  [\, \phi^{D_h-2} \chi   \hat{\chi}^{[\fM\fN]\fM'\fN'}_{\fA\fA'} \, ] \hat{\cD}_{\fM} {\cal G}_{\fM'\fN'}^{\fA'} +  \phi^{D_h-2} \chi   \hat{\chi}^{[\fM\fN]\fM'\fN'}_{\fA\fA'}  \hat{\cD}_{\fM}   \hat{\cD}_{\fN} {\cal G}_{\fM'\fN'}^{\fA'}  \, \nonumber  \\
& & \;  = \frac{1}{2}  \phi^{D_h-2} \chi  {\cal F}_{\fM\fN \fB' }^{\; \fA'}   {\cal G}_{\fM'\fN'}^{\fB'}  \hat{\chi}^{\fM\fN\fM'\fN'}_{\fA\fA'} \, .
\ee
with  $\hat{\cD}_{\fM}$  and $\bar{\cD}_{\fM}$ defined in Eqs.(\ref{COD}) and (\ref{EMGC}), respectively.

\section{ Conservation laws and dynamics of hyper-gravifield in hyperunified field theory}

According to Noether's theorem that every differentiable symmetry of an action has a corresponding conservation law\cite{NT}, we shall investigate the conservation laws corresponding to the global Poincar\'e symmetry and scaling symmetry. 

\subsection{ Conservation law of translational invariance in hyperunified field theory }

Let us first consider the conservation law under a translational transformation of coordinates in hyper-spacetime,
\[ x^{\fM} \to x^{'\fM} = x^{\fM} + a^{\fM} \, .\]
Applying the variational principle to a translational invariant action,
\be 
\Delta I_H= \int [d\hx] \, \partial_{\fP} ( {\cal T}_{\fM}^{\;\, \fP}) a^{\fM} = 0 \, , \nonumber 
\ee
we arrive at a {\it hyper-stress energy-momentum conservation} by ignoring surface terms,
\begin{equation} \label{EMC}
\partial_{\fP} {\cal T}_{\fM}^{\;\, \fP}= 0 \, .
\end{equation}

From the general action of Eq.(\ref{HUTaction1}) of hyperunified field theory, we obtain a gauge-invariant {\it hyper-stress energy-momentum tensor},
\be \label{EMT}
{\cal T}_{\fM}^{\;\, \fP}& = & - \eta^{\; \fP}_{\fM}\chi \kL  +   \chi \phi^{D_h-4} [ \frac{1}{2}\chi 
i\bar{\vPsi} \varGamma^{\fA}{\mathcal D}_{\fM}\vPsi  \hat{\chi}^{\;\; \fP}_{\fA} - {\cal W}_{\fM \fN} {\cal W}^{\fP\fN}   \nonumber \\
& - & {\cal F}_{\fM \fN}^{\fA\fB} {\cal F}_{\fM'\fN'}^{\fA'\fB'}   \hat{\chi}^{[\fP\fN]\fM'\fN'}_{\fA\fB\1 \fA'\fB'} 
+ \phi^2 {\cal G}_{\fM \fN}^{\fA}   {\cal G}_{\fM'\fN'}^{\fA'} \hat{\chi}^{[\fP\fN]\fM'\fN'}_{\fA\fA'}   \nonumber \\
& - &   2 \alpha_E  \phi^{2}  {\cal F}_{\fM \fN}^{\; \fA\fB} \hat{\chi}_{\fA}^{\;\;\fP} \hat{\chi}_{\fB}^{\;\;\fN} +  d_{\fM}\phi  d_{\fN} \phi  \hat{\chi}^{\fN\fP} \,  ] \, .
\ee
In obtaining the above gauge invariant hyper-stress energy-momentum tensor, we have applied the equations of motion for the basic fields $\cA_{\fM}$, $W_{\fM}$, and $\chi_{\fM}^{\;\; \fA}$. 

Note that the hyper-stress energy-momentum tensor ${\cal T}_{\fM\fN} = {\cal T}_{\fM}^{\;\, \fP} \eta_{\fP\fN} $ is in general not symmetric,
\be 
{\cal T}_{\fM\fN} \neq {\cal T}_{\fN\fM} \, .
\ee

\subsection{ Conservation laws of global Lorentz and conformal scaling invariances}

We now discuss the conservation law under a global Lorentz transformation of coordinates in hyper-spacetime. Consider an infinitesimal transformation, 
\[ x'^{\fM} = x^{\fM} + \delta L^{\fM}_{\; \fN} \, x^{\fN}\, , \] 
and adopt the principle of least action, we arrive at the following conservation law for the global Lorentz transformation invariance in hyper-spacetime:
\be
\partial_{\fP} \cL^{\fP}_{\;\, \fM\fN}- {\cal T}_{[\fM\fN]} =0 \, ,
\ee
 with the definitions
 \be \label{AMC}
  \cL^{\fP}_{\;\, \fM\fN} & \equiv & {\cal T}^{\;\,\fP}_{\fM}\, x_{\fN} - {\cal T}^{\;\, \fP}_{\fN}\, x_{\fM}  \, , \nn \\
   {\cal T}_{[\fM\fN]} & \equiv & {\cal T}_{\fM}^{\; \, \fN'} \eta_{\fN'\fN} - {\cal T}_{\fN}^{\; \, \fM'} \eta_{\fM'\fM} \, .
 \ee
$\cL^{\fP}_{\;\, \fM\fN}$ defines an orbital angular momentum tensor in hyper-spacetime. As the hyper-stress energy-momentum tensor ${\cal T}_{\fM\fN}$ is in general not symmetric, the orbital angular momentum tensor is not homogeneously conserved, i.e.,  $\partial_{\fP} \cL^{\fP}_{\;\, \fM\fN}\neq 0$.
 
Let us now reformulate the conserved current equation, Eq.(\ref{CC2}), for the hyper-spin gauge invariance in hyper-spacetime as follows: 
\be \label{CSC}
 \partial_{\fP} \cS^{\fP}_{\;\; \fM\fN}  + \mT_{[\fM\fN]} & = & \cS^{\fP}_{\;\; \fA\fB} ( \chi_{\fM}^{\;\, \fA}\,  \cD_{\fP} \chi_{\fN}^{\;\, \fB} -  \chi_{\fN}^{\;\, \fA}  \cD_{\fP}\chi_{\fM}^{\;\, \fB} ) \nonumber \\
 & + & 2\alpha_E  \phi^{D_h-2} \chi \hat{\chi}_{\fA}^{\;\,\fM'} \hat{\chi}_{\fC}^{\;\,\fN'}  {\cal F}_{\fM'\fN' \fB}^{\;\;\;\;\;\,\fC} \chi_{\fM\fN}^{[\fA \fB]}  \, ,
\ee
with the definitions
\be \label{AStensor}
\cS^{\fP}_{\;\; \fM\fN}  & = & \cS^{\fP}_{\;\; \fA\fB} \chi_{\fM}^{\;\, \fA} \chi_{\fN}^{\;\, \fB} = - g_s \chi  \phi^{D_h-4} \hat{\chi}_{\fC}^{\;\; \fP} \,  \bar{\vPsi}  \{ \varGamma^{\fC}\; ,  \frac{1}{2} \varSigma_{\fA\fB} \} \vPsi \,  \chi_{\fM}^{\;\, \fA} \chi_{\fN}^{\;\, \fB}  \, , \nonumber \\
\mT_{[\fM\fN]}  & = & J_{[\fA\fB]} \chi_{\fM}^{\;\, \fA} \chi_{\fN}^{\;\, \fB} \equiv \mT_{\fM\fN} - \mT_{\fN\fM}  \, ,
\ee 
and 
\be
\chi_{\fM\fN}^{[\fA \fB]}  & = & \chi_{\fM}^{\;\; \fA}\chi_{\fN}^{\;\; \fB} - \chi_{\fM}^{\;\; \fB}\chi_{\fN}^{\;\; \fA} \, .
\ee
$\mT_{\fM\fN} $ is defined from the conformal scaling gauge-invariant hyper-stress energy-momentum tensor $\cT_{\fM}^{\;\;\fP} $, i.e., 
\be \label{STF2}
& & \mT_{\fM\fN} \equiv  {\cal T}_{\fM}^{\;\; \fP} \chi_{\fP\fN}  \, , 
\ee
with
\be
& & \chi_{\fM\fN} = \chi_{\fM}^{\;\; \fA}\chi_{\fN}^{\;\; \fB} \eta_{\fA\fB} \, ,
\ee
which is dual to the symmetric tensor field $\hat{\chi}^{\fM\fN}$ defined in Eq.(\ref{STF1}). $\chi_{\fM\fN}$ and $\chih^{\fM\fN}$ are referred to as the dual {\it hyper-gravimetric fields}.
 
 In terms of the rotational and spinning angular momentum tensors $\cL^{\fP}_{\;\; \fM\fN}$  and $\cS^{\fP}_{\;\; \fM\fN}$, respectively, let us introduce a total angular momentum tensor in hyper-spacetime
\be
 {\cal J}^{\fP}_{\;\; \fM\fN}  & \equiv & \cL^{\fP}_{\;\; \fM\fN}  + \cS^{\fP}_{\;\; \fM\fN}  \, .
\ee
From the conservation law of the Lorentz invariance Eq.(\ref{AMC}) and the conserved current of the hyper-spin gauge invariance Eq.(\ref{CSC}), we find an alternative conservation law,
\be 
 \partial_{\fP} {\cal J}^{\fP}_{\;\; \fM\fN}  & = &  \cT_{[\fM\fN]} -  \mT_{[\fM\fN]}  
 + \cS^{\fP}_{\;\; \fA\fB} ( \chi_{\fM}^{\;\, \fA}\,  \cD_{\fP} \chi_{\fN}^{\;\, \fB} -  \chi_{\fN}^{\;\, \fA} \cD_{\fP}\chi_{\fM}^{\;\, \fB} ) \nonumber \\
 & + & 2\alpha_E  \phi^{D_h-2} \chi \hat{\chi}_{\fA}^{\;\,\fM'} \hat{\chi}_{\fC}^{\;\,\fN'}  {\cal F}_{\fM'\fN' \fB}^{\;\;\;\;\;\,\fC} \chi_{\fM\fN}^{[\fA \fB]}  \, ,
\ee
which shows that such a total angular momentum tensor is not homogeneously conserved, i.e., $\partial_{\fP} {\cal J}^{\fP}_{\;\; \fM\fN} \neq 0 $, due to the presence of local gauge interactions.
 
When turning the local gauge symmetry into a global symmetry, i.e.,   
\be
 & & \cA_{\fM}^{\fA\fB} \to 0\, , \quad \hat{\chi}_{\fA}^{\;\;\fM} \to \eta_{\fA}^{\;\; \fM}\, , 
 \ee
we obtain the following relations:
 \be
 & &  \partial_{\fP} \cS^{\fP}_{\;\; \fM\fN}  = -  \cT_{[\fM\fN]}; \quad \partial_{\fP} \cL^{\fP}_{\;\; \fM\fN}  =  \cT_{[\fM\fN]} \, ,
 \ee
which reproduces the well-known conservation law for the total angular momentum tensor,
\be
\partial_{\fP} {\cal J}^{\fP}_{\;\; \fM\fN}  & = & \partial_{\fP} ( \cL^{\fP}_{\;\; \fM\fN}  + \cS^{\fP}_{\;\; \fM\fN}  )  =0 \, ,
\ee
due to the cancellation.

In conclusion, the hyper-stress energy-momentum tensor is in general asymmetric because of the existence of the hyper-spinor field as the basic building blocks of nature. Neither the angular momentum tensor nor the spinning momentum tensor is conserved homogeneously due to the antisymmetric part of the hyper-stress energy-momentum tensor. In the absence of the local gauge interactions, both the hyper-spinning momentum tensor and the hyper-angular momentum tensor are governed by the antisymmetric part of the hyper-stress energy-momentum tensor, but with an opposite sign. The total angular momentum becomes homogeneously conserved when the cancellation of the antisymmetric part of the hyper-stress energy-momentum tensor occurs in the absence of gauge interactions.

The conservation law for the global scaling invariance is found to have the following general form, 
\be
\left(x^{\fM} \frac{\partial}{\partial x^{\fM}} + D_h\right) (\chi\, \kL ) + \partial_{\fM} {\cal T}^{\fM}  - {\cal T} = 0 \, ,
\ee
with the definitions,
 \be \label{SCL}
  {\cal T}^{\fM} \equiv {\cal T}^{\;\, \fM}_{\fN}\, x^{\fN}  \, , \qquad {\cal T} \equiv {\cal T}_{\fM}^{\;\, \fN} \eta_{\fN}^{\;\, \fM}
 = {\cal T}_{\fM}^{\; \fM}  \, .
 \ee
${\cal T}^{\fM}$ is regarded as a scaling current and ${\cal T} = {\cal T}_{\fM}^{\; \fM}$ is the trace of the hyper-stress energy-momentum tensor. It is noticed that the integral $\int [d\hx] \lambda^{D_h}\,  \chi(\lambda x)\, \kL (\lambda x) $ is actually independent of $\lambda$, the differentiation with respect to $\lambda$ at $\lambda =1$ leads to the following identity:
 \be
 \int [d\hx] (x^{\fM} \frac{\partial}{\partial x^{\fM}} + D_h) (\chi \kL ) = 0 \, .\nonumber 
 \ee  
 The conservation law for the conformal scaling invariance is simply given by
 \be
 \partial_{\fM} {\cal T}^{\fM}  - {\cal T} = 0 \, .
 \ee
Only when the hyper-stress energy-momentum tensor becomes traceless, i.e., ${\cal T} = 0$, the conservation law for the conformal scaling invariance becomes homogeneous: $\partial_{\fM} {\cal T}^{\fM} = 0$ .

 \subsection{Dynamics of hyper-gravifield with conserved hyper-stress energy-momentum tensor}
 
Although the bicovariant vector current is not a homogeneously conserved current as shown in Eq.(\ref{BCVC}), it is actually correlated to the conserved hyper-stress energy-momentum tensor ${\cal T}^{\;\, \fN}_{\fM}$, Eq.(\ref{EMT}). Explicitly, there exists a simple relation,
\be
 \chi_{\fM}^{\;\; \fA} J_{\fA}^{\;\; \fN} =  {\cal T}^{\;\, \fN}_{\fM} \, .
\ee
With such a relation, the equation of motion for the hyper-gravifield, Eq.(\ref{EMG}), can be rewritten in terms of the conserved hyper-stress energy-momentum tensor as follows,
\be \label{EMGF}
 & & \hat{\nabla}_{\fP} {\cal G}^{\;\fN\fP}_{ \fM}  = {\cal T}^{\;\, \fN}_{\fM} \, ,  
 \ee
 with the covariant derivative 
 \be
 & & \hat{\nabla}_{\fP} {\cal G}^{\;\fN\fP}_{ \fM}  \equiv \partial_{\fP}\1 {\cal G}^{\;\fN\fP}_{ \fM} -  \fGa_{\fP\fM}^{\fQ} {\cal G}^{\; \fN\fP}_{\fQ} \, , 
\ee
and $\fGa_{\fP\fM}^{\fQ}$ defined in Eq.(\ref{CD}). We have introduced the covariant tensor,
\be \label{GEM2}
 & & {\cal G}^{\;\fN\fP}_{ \fM} \equiv  \phi^{D_h-2}  \chi  \hat{\chi}^{[\fN\fP]\fM'\fN'}_{\fA\fA'}  \chi_{\fM}^{\;\; \fA}\, {\cal G}_{\fM'\fN'}^{\fA'} \, , 
 \ee 
which may be regarded to as a {\it hyper-gravifield tensor}.

From the conservation law of the hyper-stress energy-momentum tensor Eq.(\ref{EMC}), we arrive at the following conserved equation:
\be \label{GFC}
& & \partial_{\fN}\1 {\cal G}^{\; \fN}_{\fM}  =  0 \, , 
\ee
with
\be
{\cal G}^{\; \fN}_{\fM} \equiv \fGa_{\fP\fM}^{\fQ} {\cal G}^{\; \fN\fP}_{\fQ} \, ,
\ee
which may be referred as a conserved {\it hyper-gravifield tensor current}.  

The equation of motion for the hyper-gravifield can be expressed as 
\be
 & &  \partial_{\fP}\1 {\cal G}^{\;\fN\fP}_{ \fM} -  {\cal G}^{\; \fN}_{\fM} = {\cal T}^{\;\, \fN}_{\fM} \, . 
 \ee

\section{Gravitational origin of gauge symmetry and hyperunified field theory in hidden gauge formalism with emergent general linear group symmetry $GL(D_h, R)$}

The general action of Eq.(\ref{HUTaction1}) formulated within the framework of QFT in the globally flat vacuum hyper-spacetime $\fVh$ enables us to derive various conserved currents and conservation laws in hyperunified field theory. As the initial action Eq.(\ref{HUTaction}) of hyperunified field theory is constructed based on the postulates of gauge invariance and coordinate independence, we shall further investigate some profound correlations between the gauge interaction and the gravitational interaction and show the gravitational origin of gauge symmetry.  The gauge-type hyper-gravifield is found to play a basic role as a Goldstone-like boson, which allows us to define a hyper-spacetime gauge field from the hyper-spin gauge field and present an equivalent action of hyperunified field theory in a hidden gauge formalism. Such a formalism relates the field dynamics of gauge-type hyper-gravifield with the geometric dynamics of hyper-spacetime in coordinate systems. We shall also demonstrate explicitly that the gravitational interaction in hyperunified field theory can equivalently be described by the Riemann geometry of hyper-spacetime with an emergent general linear group symmetry GL($D_h$, R).

\subsection{Hyper-spin gravigauge field and gravitational origin of gauge symmetry}

Let us first decompose the hyper-spin gauge field $\cA_{\fM}$ into two parts $\vOm_{\fM}$ and $\cH_{\fM}$, i.e., 
\be \label{TP1}
\cA_{\fM} = \vOm_{\fM} + \cH_{\fM} \, , 
\ee 
so that $\vOm_{\fM}$ obeys an inhomogeneous transformation of the hyper-spin gauge symmetry and $\cH_{\fM}$ transforms homogeneously under the hyper-spin gauge transformation. Namely, 
\be \label{TP2}
& & \vOm_{\fM} \to \vOm'_{\fM} = S(\Lambda) i\partial_{\fM} S^{-1}(\Lambda)  + S(\Lambda) \vOm_{\fM} S^{-1}(\Lambda) \, , \nonumber \\
& & \cH_{\fM} \to \cH'_{\fM} = S(\Lambda) \cH_{\fM} S^{-1}(\Lambda)\, , 
\ee
with $S(\Lambda) \in $ SP(1,$D_h$-1). 

For an ordinary internal gauge symmetry, one can choose a configuration in which an inhomogeneous gauge transformation part can be eliminated by a gauge transformation to only keep the independent degrees of freedom. Namely, $\vOm_{\fM}$ can generally be taken as a pure gauge field in an ordinary internal gauge field, so that its field strength becomes zero. For the hyper-spin gauge symmetry SP(1,$D_h$-1), the situation becomes distinguishable. In constructing the general action of hyperunified field theory based on the hyper-spin gauge symmetry, a gauge-type hyper-gravifield $\chi_{\fM}^{\; \fA}$ has to be introduced as an accompaniment of the hyper-spin gauge field $\cA_{\fM} $. Namely, once turning the hyper-spin gauge symmetry to coincide with the global Lorentz symmetry, both the hyper-spin gauge field $\cA_{\fM} $ and the hyper-gravifield $\chi_{\fM}^{\; \fA}$ become unnecessary. The gauge field part $\vOm_{\fM}$ with inhomogeneous transformations is presumed to correlate with the hyper-gravifield $\chi_{\fM}^{\; \fA}$. 

Indeed, $\vOm_{\fM}$ is found to be determined solely by the hyper-gravifield $\chi_{\fM}^{\; \fA}$ with the following explicit form,
\be \label{TP3}
\vOm_{\fM}^{\fA\fB} & = & \frac{1}{2}[\hat{\chi}^{\fA\fN} \mG_{\fM\fN}^{\fB} - \hat{\chi}^{\fB\fN} \mG_{\fM\fN}^{\fA} -  \hat{\chi}^{\fA\fP}  \hat{\chi}^{\fB\fQ}  \mG_{\fP\fQ}^{\fC} \chi_{\fM \fC }] \, ,
\ee
with the antisymmetric tensor $\mG_{\fM\fN}^{\fA}$ defined as
\be
\mG_{\fM\fN}^{\fA} & \equiv & \p_{\fM} \chi_{\fN}^{\; \fA} -  \p_{\fN} \chi_{\fM}^{\;\fA} \, ,
\ee
where $\mG_{\fM\fN}^{\fA}$ is viewed as the field strength of the gauge-type hyper-gravifield $\chi_{\fM}^{\; \fA}$. When the hyper-gravifield $\chi_{\fM}^{\; \fA}$ transforms in a vector representation of the hyper-spin gauge symmetry SP(1,$D_h$-1),  $\vOm_{\fM}^{\fA\fB}$ does transform as a gauge field in an adjoint representation of the hyper-spin gauge symmetry SP(1,$D_h$-1). Explicitly, we have the following transformation properties under the hyper-spin gauge transformations, 
\be
 \vOm_{\fM}^{'\fA\fB} & = & \Lambda^{\fA}_{\; \fC}  \Lambda^{\fB}_{\; \fD} \vOm_{\fM}^{\fC\fD}  + \frac{i}{2} ( \Lambda^{\fA}_{\; \fC}  \partial_{\fM}  \Lambda^{\fB\fC}- \Lambda^{\fB}_{\; \fC}  \partial_{\fM}  \Lambda^{\fA\fC} )  \nn \\
\chi_{\fM}^{'\; \fA} & = &  \Lambda^{\fA}_{\;\; \fC} \chi_{\fM}^{\; \fC}; \quad   \Lambda^{\fA}_{\; \fC}  \in \mbox{SP(1,$D_h$-1)}\, ,
\ee
where $\vOm_{\fM}^{\fA\fB}$ is completely governed by the hyper-gravifield $\chi_{\fM}^{\; \fA}$. 

From such a decomposition of the hyper-spin gauge field $\cA_{\fM}$, it involves not any extra independent degrees of freedom. For convenience, we may refer  by $\vOm_{\fM}^{\fA\fB}$ to the {\it hyper-spin gravigauge field} and by $\cH_{\fM}^{\fA\fB}$ to the {\it hyper-spin homogauge field}. Unlike the usual internal gauge field, there exists not any gauge transformation that can make the hyper-spin gravigauge field $\vOm_{\fM}^{\fA\fB}$ to vanish. In general, the hyper-spin gauge field strength consists of two parts,
\be
\cF_{\fM\fN}^{\fA\fB} &\equiv & \cR_{\fM\fN}^{\fA\fB} + \cQ_{\fM\fN}^{\fA\fB} \, , 
\ee
with
\be \label{FS1}
\cR_{\fM\fN}^{\fA\fB} & = & \partial_{\fM} \vOm_{\fN}^{\fA\fB} - \partial_{\fN} \vOm_{\fM}^{\fA\fB} + \vOm_{\fM \fC}^{\fA} \vOm_{\fN}^{\fC \fB} -  \vOm_{\fN \fC}^{\fA} \vOm_{\fM}^{\fC \fB},  \\  
\cQ_{\fM\fN}^{\fA\fB} & = & \cD_{\fM} \cH_{\fN}^{\fA\fB} - \cD_{\fN} \cH_{\fM}^{\fA\fB} +  \cH_{\fM \fC}^{\fA} \cH_{\fN}^{\fC \fB} -  \cH_{\fN \fC}^{\fA} \cH_{\fM}^{\fC \fB} \, , 
\ee
where the covariant derivative is defined as 
\be
\cD_{\fM} \cH_{\fN}^{\fA\fB} = \partial_{\fM}  \cH_{\fN}^{\fA\fB}  +  \vOm_{\fM \fC}^{\fA} \cH_{\fN}^{\fC \fB} + \vOm_{\fM \fC}^{\fB} \cH_{\fN}^{\fA \fC}  \, .
\ee

As the hyper-spin gauge symmetry SP(1,$D_h$-1) is essentially characterized by the gauge-type hyper-gravified $\chi_{\fM}^{\; \fA}$, which displays the {\it gravitational origin of gauge symmetry}.

\subsection{ Hyper-spacetime gauge field and Goldstone-like hyper-gravifield }

The hyper-gravifield $\chi_{\fM}^{\; \fA}$ allows us to project tensors defined in the locally flat hyper-gravifield spacetime into those defined in the globally flat vacuum hyper-spacetime and to formulate the general action of hyperunified field theory in a hidden gauge formalism. So that the hyper-spin gauge symmetry is transmuted into a hidden gauge symmetry. 

The hidden gauge invariant {\it hyper-spacetime gauge field} can be constructed from the hyper-spin gauge field and the hyper-gravifield. Explicitly, we make the following definition for the hyper-spacetime gauge field,
\be \label{STGF1}
 & & \cA_{\fM\fQ}^{\fP}   \equiv  \chih_{\fA}^{\;\; \fP} \partial_{\fM} \chi_{\fQ}^{\;\;\fA} + \chih_{\fA}^{\;\; \fP}   \cA_{\fM\1 \fB}^{\fA} \chi_{\fQ}^{\;\;\fB}\, .
\ee
As the hyper-spin gauge field $ \cA_{\fM}^{\fA\fB}$ consists of two parts as shown in Eqs.(\ref{TP1})-(\ref{TP3}),  the {\it hyper-spacetime gauge field} $\cA_{\fM\fQ}^{\fP}$ is decomposed, correspondingly, into two parts,
\be
 \cA_{\fM\fQ}^{\fP}   \equiv   \vGa_{\fM\fQ}^{\fP}  +  \cH_{\fM\fQ}^{\fP}  \, ,
\ee
where $\vGa_{\fM\fQ}^{\fP}$ defines the {\it hyper-spacetime gravigauge field}  and  $ \cH_{\fM\fQ}^{\fP} $ represents the {\it hyper-spacetime homogauge field}. Their explicit forms are found to be 
\be \label{HSGF}
\vGa_{\fM\fQ}^{\fP}  & \equiv &  \chih_{\fA}^{\;\; \fP} \p_{\fM} \chi_{\fQ}^{\;\;\fA} +  \chih_{\fA}^{\;\; \fP}   \vOm_{\fM\1 \fB}^{\fA} \chi_{\fQ}^{\;\;\fB} \, \nn \\
& = & \frac{1}{2} \hat{\chi}^{\fP\fL} (\, \p_{\fM} \chi_{\fQ \fL} + \p_{\fQ} \chi_{\fM \fL} - \p_{\fL}\chi_{\fM\fQ} \, ) \, , \\
 \cH_{\fM\fQ}^{\fP} & \equiv &  \chih_{\fA}^{\;\; \fP} \cH_{\fM}^{\fA \fB} \chi_{\fQ \fB} \, .
 \ee
Where the hyper-gravifield $\chi_{\fM}^{\;\;\fA}$ plays an essential role as a {\it Goldstone-like field} of gauge symmetry. The dual tensor fields $\chih^{\fM\fN}$ and $\chi_{\fM\fN}$ defined in Eqs.(\ref{STF1}) and (\ref{STF2}) as the composite fields of the hyper-gravifield with a hidden gauge symmetry are regarded as the dual {\it Goldstone-like hyper-gravimetric fields}. The symmetric {\it hyper-spacetime gravigauge field} $\vGa_{\fM\fQ}^{\fP} = \vGa_{\fQ\fM}^{\fP}$ is described by the dual Goldstone-like hyper-gravimetric fields $\chi_{\fM\fN}$ and $\chih^{\fM\fN}$.  Note that both $ \vGa_{\fM\fQ}^{\fP}$ and  $\cH_{\fM\fQ}^{\fP}$ are not conformal scaling gauge invariant.

We can apply the dual Goldstone-like hyper-gravimetric fields $\chi_{\fM\fN}$ and $\chih^{\fM\fN}$ to make redefinitions for the {\it hyper-spacetime homogauge field} $\cH_{\fM\fQ}^{\fP} $,  
\be
& & \cH_{\fM \fP \fQ} \equiv \cH_{\fM\fQ}^{\fP'}  \chi_{\fP'\fP} =   \cH_{\fM}^{\fA \fB}  \chi_{\fP\fA}\chi_{\fQ\fB} \, , \nn \\
 & & \cH_{\fM}^{\; \fP \fQ} \equiv \cH_{\fM\fN}^{\fP}  \chih^{\fN\fQ} =   \cH_{\fM}^{\fA \fB}  \chih_{\fA}^{\;\; \fP}\chih_{\fB}^{\;\; \fQ} \, , 
\ee
so that the hyper-spacetime homogauge field shows the explicit antisymmetry property,
 \be
 \cH_{\fM \fP \fQ} = - \cH_{\fM \fQ \fP}  \, , \qquad \cH_{\fM}^{\; \fP \fQ} = - \cH_{\fM}^{\; \fQ \fP}  \, .
 \ee 
In such definitions, the hyper-gravimetric fields $\chi_{\fM\fN}$ and $\chih^{\fM\fN}$ have been adopted to lower and raise the indices of the hyper-spacetime tensors, respectively, in a hidden gauge formalism.

As the dual Goldstone-like hyper-gravimetric field $\chi_{\fM\fN}$ (or $\chih^{\fM\fN}$) is symmetric and the {\it hyper-spacetime homogauge field} $\cH_{\fM \fP \fQ}$ (or $\cH_{\fM}^{ \fP \fQ} $) is antisymmetric, the total independent degrees of freedom in a hidden gauge formalism are reduced by amounts of $D_h(D_h-1)/2$ in bosonic gauge interactions. This reflects the fact that the hyper-spin gauge symmetry SP(1,$D_h$-1) is transmuted, through the {\it Goldstone-like hyper-gravifield} $\chi_{\fM}^{\; \fA}$, into a hidden gauge symmetry.

\subsection{ Field strength of hyper-spacetime gauge field in hidden gauge formalism}

From the hyper-spacetime gauge field, we can define the corresponding hyper-spacetime field strength. Explicitly, we have,
\be
 \cR_{\fM\fN\fQ}^{\;\fP}  = \partial_{\fM} \vGa_{\fN\fQ}^{\fP} - \partial_{\fN} \vGa_{\fM\fQ}^{\fP}  + \vGa_{\fM\fL}^{\fP} \vGa_{\fN\fQ}^{\fL}  - \vGa_{\fN\fL}^{\fP} \vGa_{\fM\fQ}^{\fL},
\ee
for the hyper-spacetime gravigauge field strength, and 
\be
& & \cQ_{\fM\fN\fQ}^{\;\fP}  = \nabla_{\fM} \cH_{\fN\fQ}^{\fP} - \nabla_{\fN} \cH_{\fM\fQ}^{\fP} + \cH_{\fM\fL}^{\fP} \cH_{\fN\fQ}^{\fL}   - \cH_{\fN\fL}^{\fP} \cH_{\fM\fQ}^{\fL}  \, , \nonumber \\
& & \nabla_{\fM} \cH_{\fN\fQ}^{\fP} = \partial_{\fM} \cH_{\fN\fQ}^{\fP} - \vGa_{\fM\fQ}^{\fL} \cH_{\fN\fL}^{\fP} + \vGa_{\fM\fL}^{\fP}  \cH_{\fN\fQ}^{\fL} \, .
\ee
for the hyper-spacetime homogauge field strength. 

In terms of the antisymmetric hyper-spacetime homogauge field, the corresponding field strength is given by  
\be
& & \cQ_{\fM\fN}^{\, \fP\fQ} = \nabla_{\fM} \cH_{\fN}^{\fP\fQ} - \nabla_{\fN} \cH_{\fM}^{\fP\fQ} + \cH_{\fM\fL}^{\fP} \cH_{\fN}^{\fL\fQ} - \cH_{\fN\fL}^{\fP} \cH_{\fM}^{\fL\fQ}  \, , \nonumber \\
& & \nabla_{\fM} \cH_{\fN}^{\fP\fQ} = \p_{\fM} \cH_{\fN}^{\fP\fQ} + \vGa_{\fM\fL}^{\fP} \cH_{\fN}^{\fL\fQ} + \vGa_{\fM\fL}^{\fQ}  \cH_{\fN}^{\fP\fL} \, , 
\ee
or
\be \label{HFS}
& & \cQ_{\fM\fN\fP\fQ} = \nabla_{\fM} \cH_{\fN\fP\fQ} - \nabla_{\fN} \cH_{\fM\fP\fQ} - \cH_{\fM\fQ}^{\fL}\cH_{\fN\fP\fL}   + \cH_{\fN\fQ}^{\fL} \cH_{\fM\fP\fL}  \, , \nonumber \\
& & \nabla_{\fM} \cH_{\fN \fP\fQ} = \p_{\fM} \cH_{\fN \fP\fQ} - \vGa_{\fM\fP}^{\fL} \cH_{\fN\fL\fQ} - \vGa_{\fM\fQ}^{\fL}  \cH_{\fN\fP\fL} \, . 
\ee
Their symmetryproperties lead to the following relations
\be
 & & \cQ_{\fM\fN}^{\, \fP\fQ} \equiv \cQ_{\fM\fN\fQ'}^{\;\fP}\chih^{\fQ'\fQ} =  - \cQ_{\fM\fN}^{\, \fQ\fP} \, , \nn \\
 & & \cQ_{\fM\fN \fP\fQ} \equiv \cQ_{\fM\fN\fQ}^{\;\fP'}\chi_{\fP'\fP} =  - \cQ_{\fM\fN\fQ\fP}  \, .
\ee
Similarly, we can apply the hyper-gravimetric fields $\chi_{\fM\fN}$ and $\chih^{\fM\fN}$ to lower and raise the indices of the hyper-spacetime gravigauge field strength,
\be
 & & \cR_{\fM\fN}^{\, \fP\fQ} \equiv \cR_{\fM\fN\fQ'}^{\;\fP}\chih^{\fQ'\fQ} =  - \cR_{\fM\fN}^{\, \fQ\fP} \, , \nn \\
 & &  \cR_{\fM\fN \fP\fQ} \equiv \cR_{\fM\fN\fQ}^{\;\fP'}\chi_{\fP'\fP} =  - \cR_{\fM\fN\fQ\fP}  \, .
\ee

In such a hidden gauge formalism, the gauge covariant field strength of the hyper-gravifield is given by the hyper-spacetime homogauge field and the scalar field,
\be
\cG_{\fM\fN}^{\fP} & \equiv& \cH_{[\fM\fN]}^{\fP} + \fS_{[\fM\fN]}^{\fP} \equiv \tilde{\cH}_{[\fM\fN]}^{\fP} \, , \nn \\
\cH_{[\fM\fN]}^{\fP} & = & \cH_{\fM\fN}^{\fP} - \cH_{\fN\fM}^{\fP}\, , \quad  \fS_{[\fM\fN]}^{\fP} =  \fS_{\fM\fN}^{\fP} - \fS_{\fN\fM}^{\fP} \, , \nn \\
\fS_{\fM\fN}^{\fP} & = & \chi_{\fM\fN}\chih^{\fP\fQ}\fS_{\fQ} - \eta_{\fM}^{\; \fP} \fS_{\fN};\; \fS_{\fM} \equiv \p_{\fM}\ln\phi,
\ee
or
\be
& & \cG_{\fM\fN\fP} \equiv -\cH_{[\fM\fN]\fP} - \fS_{[\fM\fN]\fP} \equiv - \tilde{\cH}_{[\fM\fN]\fP} \, , \nn \\
& &  \fS_{\fM\fN\fP}  =  \chi_{\fM\fP} \fS_{\fN} -  \chi_{\fM\fN} \fS_{\fP} \, , \nn \\
& &  \fS_{[\fM\fN]\fP} =  - \fS_{\fM}\chi_{\fN\fP} + \fS_{\fN}\chi_{\fM\fP}  \, ,
\ee
which leads to a general relation between the hyper-spin homogauge field $\cH_{\fM\fP\fQ}$ and the gauge covariant hyper-gravifield strength $\cG_{\fM\fP\fQ}$,
\be \label{Hrelation2}
\cH_{\fM \fP\fQ} & = & \frac{1}{2} ( \cH_{[\fM \fP]\fQ} - \cH_{[\fM\fQ]\fP} - \cH_{[\fP\fQ]\fM} ) \nn \\
\tilde{\cH}_{\fM \fP\fQ} & = & \frac{1}{2} ( \tilde{\cH}_{[\fM \fP]\fQ} - \tilde{\cH}_{[\fM\fQ]\fP} - \tilde{\cH}_{[\fP\fQ]\fM} ) \nn \\
& = & \cH_{\fM \fP\fQ}  + \frac{1}{2} ( \fS_{[\fM \fP]\fQ} - \fS_{[\fM\fQ]\fP} - \fS_{[\fP\fQ]\fM} ) \nn \\
& = & -\frac{1}{2} (\cG_{\fM \fP\fQ} - \cG_{\fM\fQ\fP} - \cG_{\fP\fQ\fM} )\, . 
\ee

The hyper-spacetime gravigauge field strength $\cR_{\fM\fN\fQ}^{\;\fP}$ is characterized by the Goldstone-like hyper-gravimetric field $\chi_{\fM\fN}$. Thus $\chi_{\fM\fN}$ is considered as a basic field for the hyper-spin gravigauge interaction in a hidden gauge formalism. In general, the field strength of the hyper-spin gauge field is related to the field strength of the hyper-spacetime gauge field via the following relations,
\be
& & \cF_{\fM\fN}^{\fA\fB} \equiv  \cR_{\fM\fN}^{\fA\fB} + \cQ_{\fM\fN}^{\fA\fB} = \frac{1}{2} \cF_{\fM\fN}^{\fP\fQ} \, \chi_{ \fP\fQ}^{[\fA\fB]}  \, , 
\ee
and
\be
& &  \cR_{\fM\fN}^{\, \fA\fB}   =  \frac{1}{2} \cR_{\fM\fN\fL}^{\fP}\, \chi_{ \fP\fQ}^{[\fA\fB]} \chih^{\fL\fQ} = \frac{1}{2} \cR_{\fM\fN}^{\fP\fQ}\, \chi_{ \fP\fQ}^{[\fA\fB]} \, , \nn \\
& &  \cQ_{\fM\fN}^{\, \fA\fB}  =  \frac{1}{2} \cQ_{\fM\fN\fL}^{\fP} \, \chi_{ \fP\fQ}^{[\fA\fB]} \chih^{\fL\fQ} = \frac{1}{2} \cQ_{\fM\fN}^{\fP\fQ} \, \chi_{ \fP\fQ}^{[\fA\fB]} \, , 
\ee 
with $\chi_{ \fP\fQ}^{[\fA\fB]} $ defined in Eq.(\ref{AStensor}). 

In obtaining the above relations, we have used the following identities,
\be \label{IDS}
& & \nabla_{\fM}\chi_{\fN}^{\; \fA} = \p_{\fM} \chi_{\fN}^{\; \fA} + \vOm_{\fM \fB}^{\fA} \chi_{\fN}^{\; \fB}  
- \vGa_{\fM\fN}^{\fP} \chi_{\fP}^{\; \fA} =0\, , \nonumber \\
& & \nabla_{\fM} \chih_{\fB}^{\;\, \fP} =  \p_{\fM} \chih_{\fB}^{ \;\, \fP} 
- \vOm_{\fM \fB}^{\fA} \chih_{\fA}^{\;\,\fP} + \vGa_{\fM\fN}^{\fP} \chih_{\fA}^{\; \fN}=0\, , \nn \\
& & \nabla_{\fM} \chi_{\fP\fQ} = \p_{\fM} \chi_{\fP\fQ}  - \vGa_{\fM\fP}^{\fN} \chi_{\fN\fQ} - \vGa_{\fM\fQ}^{\fN} \chi_{\fP\fN} = 0\, ,\nn \\
& & \nabla_{\fM} \chih^{\fP\fQ} = \p_{\fM} \chih^{\fP\fQ}  + \vGa_{\fM\fN}^{\fP} \chih^{\fN\fQ} + \vGa_{\fM\fN}^{\fQ}\chih^{\fP\fN} = 0 \, .
\ee

\subsection{General covariance and Riemannian geometry of hyper-spacetime}

Geometrically, considering the gauge invariant tensor field $\chi_{\fM\fN}(\hx)$ as the hyper-gravimetric field of hyper-spacetime, the {\it hyper-spacetime gravigauge field} $\vGa_{\fM\fQ}^{\fP}(\hx)$ given in Eq.(\ref{HSGF}) is regarded as the {\it Levi-Civita connection} or the {\it Christoffel symbols} in hyper-spacetime. The Christoffel symbols of the first kind is obtained by lowering the index of the second kind $\vGa_{\fM\fQ}^{\fP}(\hx)$ through the hyper-gravimetric field $\chi_{\fM\fN}(\hx)$,
\be \label{HSGGF}
\vGa_{\fM\fP\fQ} = \vGa_{\fM\fQ}^{\fP'} \chi_{\fP'\fP} = \frac{1}{2} (\, \p_{\fM} \chi_{\fQ \fP} + \p_{\fQ} \chi_{\fM \fP} - \p_{\fP}\chi_{\fM\fQ} \, ) = \vGa_{\fQ\fP\fM}\, . 
 \ee
There are general contracting relations for the Christoffel symbols,  
\be
& & \vGa_{\fM\fP}^{\fP} = \eta_{\fP}^{\fQ}\vGa_{\fM\fQ}^{\fP} = \p_{\fM}\ln \chi \, , \nn \\
& & \chih^{\fM\fQ}\vGa_{\fM\fQ}^{\fP}  = - \chih\1 \p_{\fM} (\chi \chih^{\fM\fP})\, .
\ee

Such Christoffel symbols in hyper-spacetime define a principal connection of general linear symmetry group GL($D_h$, R), which allows us to extend the global non-homogeneous Lorentz transformations of the Poincar\'e symmetry PO(1,$D_h$-1) in the globally flat Minkowski hyper-spacetime to the general coordinate transformations of the local symmetry GL($D_h$, R) in the Riemannian curved hyper-spacetime. A general coordinate transformation is defined as an arbitrary reparametrization of coordinate systems, i.e., $\hx' \equiv \hx'(\hx)$, which is a local transformation for describing a distinct reparametrization at every point in hyper-spacetime. It obeys the following transformation laws,
\be
 \p_{\fM} & \to & \p'_{\fM} = T_{\fM}^{\;\, \fN}\, \p_{\fN} \, ,\quad  T_{\fM}^{\;\; \fN} \equiv \frac{\p x^{\fN}}{\p x^{'\fM}} \, ; \nn \\
dx^{\fM} & \to &  dx^{'\fM}= T^{\fM}_{\;\; \; \fN}\, dx^{\fN}\, , \quad  T^{\fM}_{\;\; \; \fN}  \equiv \frac{\p x^{'\fM}}{\p x^{\fN}}\, ,
\ee
which enables us to provide a general definition for a covariant vector field $A_{\fM}(\hx)$, with lower index, and  a contravariant vector field $V^{\fM}(\hx)$, with upper index, from the following transformation properties,
\be \label{TensorT1}
A'_{\fM}(\hx') = T_{\fM}^{\;\; \fN}\, A_{\fN}(\hx); \quad  V^{'\fM}(\hx') = T^{\fM}_{\;\; \; \fN}\, V^{\fN}(\hx) \, .
\ee
The hyper-gravifield $\chi_{\fM}^{\;\fA}(\hx)$ is a covariant vector field and its dual vector $\chih_{\fA}^{\; \fM}(\hx)$ is a contravariant vector field in hyper-spacetime. The hyper-gravimetric field $\chi_{\fM\fN}(\hx)$ is a covariant tensor field, its transformation law is given by, 
\be
\chi'_{\fM\fN}(\hx')  = (\frac{\p x^{\fP}}{\p x^{'\fM}} \frac{\p x^{\fQ}}{\p x^{'\fN}} ) \, \chi_{\fP\fQ}(\hx) =  T_{\fM}^{\;\; \fP}  T_{\fN}^{\;\; \fQ}\, \chi_{\fP\fQ}(\hx)  \, .
\ee
With such transformation laws, the Christoffel symbols in hyper-spacetime transform as follows,
\be
\vGa_{\fM\fN}^{'\fP} (\hx') & = &  ( \frac{\p x^{\fM'}}{\p x^{'\fM}}\frac{\p x^{\fN'}}{\p x^{'\fN}} ) \left(\, \frac{\p x^{'\fP}}{\p x^{\fP'}} \vGa_{\fM'\fN'}^{\fP'} (\hx) - \frac{\p^2 x^{'\fP}}{\p x^{\fM'}\p x^{\fN'}} \, \right)\nn \\
& \equiv & T_{\fM}^{\;\; \fM'}  T_{\fN}^{\;\; \fN'} T^{\fP}_{\;\; \; \fP'}  \left(\, \vGa_{\fM'\fN'}^{\fP'} (\hx) -  T^{\fP'}_{\;\; \, \fQ} \p_{\fM'} T^{\fQ}_{\;\; \, \fN'}  \, . \right)
\ee

The covariant derivatives for both the covariant vector field $A_{\fN}(\hx)$ and the contravariant vector field $V^{\fN}(\hx)$ are defined as 
\be
& & \nabla_{\fM}A_{\fN} = \p_{\fM} A_{\fN} - \vGa_{\fM\fN}^{\fP} A_{\fP} \, , \nn \\
& & \nabla_{\fM}V^{\fN} = \p_{\fM} V^{\fN} + \vGa_{\fM\fP}^{\fN} V^{\fP} \, . 
\ee
Their transformation laws are given by 
\be  \label{TensorT5}
& & (\nabla_{\fM}A_{\fN})^{'} =  ( \frac{\p x^{\fP}}{\p x^{'\fM}}\frac{\p x^{\fQ}}{\p x^{'\fN}} ) (\nabla_{\fP}A_{\fQ}) \equiv T_{\fM}^{\;\; \fP}  T_{\fN}^{\;\; \fQ} (\nabla_{\fP}A_{\fQ}) \, , \nn \\
& & (\nabla_{\fM}V^{\fN})^{'} =   ( \frac{\p x^{\fP}}{\p x^{'\fM}})  ( \frac{\p x^{'\fQ}}{\p x^{\fN}} ) (\nabla_{\fP}V^{\fQ}) 
\equiv T_{\fM}^{\;\; \fP}  T^{\fN}_{\;\; \fQ}  (\nabla_{\fP}V^{\fQ})  \, ,
\ee
which generates the general covariance under the general coordinate transformations.

The field strength $\cR_{\fM\fN\fQ}^{\;\fP}$ of the hyper-spacetime gravigauge field defines the Riemann curvature tensor in hyper-spacetime. Lowering indices with the hyper-gravimetric field, i.e., $\cR_{\fM\fN\fP\fQ} = \chi_{\fP\fP'}\cR_{\fM\fN\fQ}^{\;\fP'}$, we have,
\be
\cR_{\fM\fN\fP\fQ} & = &  \frac{1}{2} (\p_{\fM}\p_{\fQ} \chi_{\fN\fP} + \p_{\fN}\p_{\fP} \chi_{\fM\fQ} - \p_{\fN}\p_{\fQ} \chi_{\fM\fP} -  \p_{\fM}\p_{\fP} \chi_{\fN\fQ} )  \nn \\ 
& + & \chi_{\fL\fL'}( \vGa_{\fM\fQ}^{\fL}\vGa_{\fN\fP}^{\fL'} -  \vGa_{\fN\fQ}^{\fL}\vGa_{\fM\fP}^{\fL'} )\, ,
\ee
which has symmetry properties,
\be \label{SP1}
& & \cR_{\fM\fN\fP\fQ} = -\cR_{\fN\fM\fP\fQ} \, , \quad \cR_{\fM\fN\fP\fQ} = -\cR_{\fM\fN\fQ\fP}\, , \nn \\
& &  \cR_{\fM\fN\fP\fQ} = \cR_{\fP\fQ\fM\fN}\, ,
\ee
and satisfies two Bianchi identities,
\be
& & \cR_{\fM\fN\fQ\fP} + \cR_{\fN\fQ\fM\fP} + \cR_{\fQ\fM\fN\fP} = 0 \, , \nn \\
& & \nabla_{\fL}\cR_{\fM\fN\fQ}^{\;\fP} + \nabla_{\fM}\cR_{\fN\fL\fQ}^{\;\fP} + \nabla_{\fN}\cR_{\fL\fM\fQ}^{\;\fP} = 0\, .
\ee

 A unique nontrivial way of contracting the Riemann tensor results in the Ricci curvature tensor, 
 \be
\cR_{\fM\fN} = \cR_{\fM\fQ\fN}^{\;\fP} \eta_{\fP}^{\, \fQ} = - \cR_{\fQ\fM\fN}^{\;\fP} \eta_{\fP}^{\, \fQ} = - \cR_{\fP\fM\fN}^{\fP} \, , 
\ee 
which is a symmetric tensor with the following explicit form,
\be \label{SP2}
 \cR_{\fM\fN} = \cR_{\fN\fM} = \nabla_{\fM} \p_{\fN}\ln \chi  - \partial_{\fP} \vGa_{\fM\fN}^{\fP}  + \vGa_{\fM\fL}^{\fP} \vGa_{\fP\fN}^{\fL}   \, .
\ee
The trace of the Ricci curvature tensor yields the Ricci scalar curvature in hyper-spacetime
\be \label{RSC}
\cR = \chih^{\fM\fN} \cR_{\fM\fN} = \chih^{\fM\fN}  \cR_{\fM\fQ\fN}^{\;\fP} \eta_{\fP}^{\, \fQ} = - \chih^{\fM\fN} \chih^{\fP\fQ} \cR_{\fM\fQ\fN\fP}\, .
\ee

\subsection{Hyperunified field theory in hidden gauge formalism and emergent general linear group symmetry GL($D_h$, R) }

With the above analysis, we are able to rewrite the general action Eq.(\ref{HUTaction1}) in a hidden gauge  formalism,
\be
\label{HUTaction3}
I_H & \equiv & \int [d\hx]\; \chi  \phi^{D_h-4}\,  \{ \bar{\vPsi}\dg^{\fM} [ i \p_{\fM}  +  (\vGa_{\fM\fP\fQ} + \cH_{\fM\fP\fQ} +  \varXi_{\fM\fP\fQ} ) \frac{1}{2}\varSigma^{\fP\fQ} ] \vPsi  \nonumber \\
& - &  \frac{1}{4} [\, (\, \chih^{\fM\fN\1 \fM'\fN'\1\fP\fQ\1\fP'\fQ'}   \cQ_{\fM\fN\fP\fQ} \cQ_{\fM'\fN'\fP'\fQ'} + {\cal W}_{\fM\fN} {\cal W}^{\fM\fN} \, ) \nn \\
& + &  (g_1 + g_2 + g_3 ) (\, \cR^{\fM\fN\fP\fQ} \cR_{\fM\fN\fP\fQ} + 2 \cR^{\fM\fN\fP\fQ} \cQ_{\fM\fN\fP\fQ} )    \nonumber \\
& + &   (g_4 + g_5) (\cR^{\fM\fN}\cR_{\fM\fN} + 2\cR^{\fM\fN}\cQ_{\fM\fN} ) + g_6 (\cR^2 + 2\cR\cQ) \, ] \nn \\
& + & \frac{1}{4} \phi^2\,  [\, (\alpha_G -\beta_G)\tilde{\cH}_{[\fM\fQ]\fP} \tilde{\cH}^{[\fM\fQ]\fP} + 2(\alpha_E- \beta_G) \tilde{\cH}_{\fM\fP\fQ}\tilde{\cH}^{[\fP\fQ]\fM} \, ]  \nn \\
& + & \phi^{2} (\alpha_E - \gamma_G)  (\eta^{\,\fM}_{\fP}\tilde{\cH}_{\fM\fN}^{\fP} ) (\eta_{\fP'}^{\,\fM'}\tilde{\cH}_{\fM'\fN'}^{\fP'} ) \chih^{\fN\fN'}  \nonumber \\
& + &   \alpha_E    [\, \phi^2\cR   - (D_h-1)(D_h-2) \p_{\fM}\phi \p^{\fM}\phi  \nn \\
& - &  \phi^2 \chih^{\fM\fP}\chih^{\fN\fQ}\hat{\nabla}_{[\fM}\cH_{\fN]\fP\fQ} \, ]  +  \frac{1}{2} d_{\fM} \phi d^{\fM}\phi  - \beta_E\1 \phi^4 \, \}\, ,
\ee
with the definitions,
 \be
& & \dg^{\fM} \equiv \chih^{\;\fM}_{\fC} \frac{1}{2}\vGa^{\fC}\, , \quad \varSigma^{\fP\fQ} \equiv \varSigma^{\fA\fB} \chih_{\fA}^{\; \fP}\chih_{\fB}^{\; \fQ} \, , \\
& & \varXi_{\fM\fP\fQ}\equiv \frac{1}{2} [ (\p_{\fM}\chi_{\fP}^{\;\fC}) \chi_{\fQ\fC} - (\p_{\fM}\chi_{\fQ}^{\;\fC}) \chi_{\fP\fC} ] \, , 
\ee
and
\be
& & \hat{\nabla}_{\fM}\cH_{\fN\fP\fQ}  = \nabla_{\fM}\cH_{\fN\fP\fQ}  + \fS_{\fM\fQ}^{\fL} \cH_{\fN\fP\fL} -  \fS_{\fM\fP}^{\fL} \cH_{\fN\fQ\fL}, \nn \\
& &  \hat{\nabla}_{[\fM}\cH_{\fN]\fP\fQ}  \equiv \hat{\nabla}_{\fM}\cH_{\fN\fP\fQ} - \hat{\nabla}_{\fN}\cH_{\fM\fP\fQ}\, .
\ee

The general tensor $\chih^{\fM\fN\1 \fM'\fN'\1 \fP\fQ\1\fP'\fQ'}$ is defined as 
\be \label{tensor2}
& & \chih^{\fM\fN\1 \fM'\fN'\1 \fP\fQ\1\fP'\fQ'} \equiv  g_1 \hat{\chi}^{\fM\fM'} \hat{\chi}^{\fN\fN'} \chih^{\fP\fP'}\chih^{\fQ\fQ'}  \nn \\
& & + \frac{1}{2}g_2 (\chih^{\fM\fP'} \chih^{\fN\fQ'} \chih^{\fM'\fP} \chih^{\fN'\fQ}  + \chih^{\fN\fP'} \chih^{\fM\fQ'} \chih^{\fN'\fP} \chih^{\fM'\fQ} )  \nn \\
& & +  \frac{1}{2}g_3 [ \chih^{\fP\fP'} ( \chih^{\fM\fM'}  \chih^{\fN\fQ'} \chih^{\fN'\fQ} +  \chih^{\fN\fN'} \chih^{\fM\fQ'} \chih^{\fM'\fQ}  ) +  \chih^{(\fP,\fP'\leftrightarrow \fQ,\fQ' )}\, ] \nn \\
& & + \frac{1}{2}g_4 [\chih^{\fP\fP'} (  \chih^{\fM\fM'}  \chih^{\fN\fQ} \chih^{\fN'\fQ'} + \chih^{\fN\fN'} \chih^{\fM\fQ} \chih^{\fM'\fQ'}  ) + \chih^{(\fP,\fP'\leftrightarrow \fQ,\fQ' )}  ] \nn \\
& &  + \frac{1}{2}g_5 [\chih^{\fM'\fP}\chih^{\fM\fP'}  \chih^{\fN\fQ} \chih^{\fN'\fQ'} + \chih^{\fN'\fP}\chih^{\fN\fP'}  \chih^{\fM\fQ} \chih^{\fM'\fQ'} + \chih^{(\fP,\fP'\leftrightarrow \fQ,\fQ' )}  \,  ] \nn \\
& & + \frac{1}{2}g_6 ( \chih^{\fM\fP} \chih^{\fN\fQ} \chih^{\fM'\fP'} \chih^{\fN'\fQ'}  + \chih^{\fN\fP} \chih^{\fM\fQ} \chih^{\fN'\fP'} \chih^{\fM'\fQ'} )\, .
 \ee
In obtaining the above action, we have used the general properties of the Riemann curvature tensors shown in Eqs.(\ref{SP1})-(\ref{SP2}) and the following identities,
\be
& & \cR^{\fM\fN\fP\fQ} \cR_{\fM\fQ\fP\fN} =\frac{1}{2} \cR^{\fM\fN\fP\fQ} \cR_{\fM\fN\fP\fQ}  \, , \nn \\
& & \cR^{\fM\fN\fP\fQ} \cQ_{\fM\fQ\fP\fN} =\frac{1}{2} \cR^{\fM\fN\fP\fQ} \cQ_{\fM\fN\fP\fQ}\, ,
\ee
as well as the symmetry properties of the hyper-spacetime homogauge field with the following relevant identity,
\be
\tilde{\cH}_{[\fM\fP]\fQ}\tilde{\cH}^{[\fM\fQ]\fP} = - \frac{1}{2} \tilde{\cH}_{[\fM\fP]\fQ}\tilde{\cH}^{[\fM\fP]\fQ}  
-  \tilde{\cH}_{\fM\fP\fQ}\tilde{\cH}^{[\fP\fQ]\fM} \, .
\ee

The general action of Eq.(\ref{HUTaction}) of hyperunified field theory is constructed based on the postulates of gauge invariance and coordinate independence in a locally flat hyper-gravifield spacetime ${\bf G}_h$. By projecting it into the globally flat vacuum hyper-spacetime ${\bf V}_h$ through the dual bicovariant hyper-gravifield $\chi_{\fM}^{\;\fA}(\hx)$ ($ \hat{\chi}_{\fA}^{\; \fM}(\hx)$), we find the general action of Eq.(\ref{HUTaction1}) within the framework of QFT in the globally flat Minkowski hyper-spacetime. As the postulate of coordinate independence is considered to be more general and fundamental in the construction of hyperunified field theory, the resulting action is expected to hold in any frame of coordinates. By taking the dual bicovariant vector hyper-gravifield $\chi_{\fM}^{\;\fA}(\hx)$ as the Goldstone-like field of gauge symmetry, we arrive at the general action Eq.(\ref{HUTaction3}) in a hidden gauge formalism. 

Such a general action of Eq.(\ref{HUTaction3}) can be shown to be invariant under the general coordinate transformation that is defined as an arbitrary reparametrization in coordinate systems, i.e., 
\be
 \hx' & \equiv & \hx'(\hx); \;\; dx^{\fM} \to  dx^{'\fM}= T^{\fM}_{\;\;\, \fN}\, dx^{\fN} \, , \nn \\
  \p'_{\fM} & = & T_{\fM}^{\;\, \fN}\, \p_{\fN}; \; \; T_{\fM}^{\;\, \fN}, \, T^{\fM}_{\;\;\, \fN} \in GL( D_h, R).
\ee
The tensor transformation laws, Eqs.(\ref{TensorT1})-(\ref{TensorT5}), lead the scalar product of all the tensors to be invariant under the general coordinate transformations. So that the general action of Eq.(\ref{HUTaction3}) generates an emergent general linear group symmetry, 
\be
G_S = \mbox{GL(}D_h, \mbox{R)},
\ee
which is considered to be a natural deduction of the postulate of coordinate independence in the construction of hyperunified field theory.

\section{ Hyperunified field theory with general conformal scaling gauge invariance and  Einstein-Hilbert type action with essential gauge massless condition and gravity-geometry correspondence}

The general action of Eq.(\ref{HUTaction3}) of hyperunified field theory in a hidden gauge formalism shows an emergent general coordinate transformation invariance though such a hyperunified field theory is initially built based on the postulate of gauge invariance in a hyper-gravifield spacetime. The gravitational interaction is described by the Riemann and Ricci curvature tensors of hyper-spacetime. Here we are going to explore some intriguing properties by imposing a {\it general conformal scaling gauge invariance} and requiring an {\it essential gauge massless condition} for gauge fields. In particular, we shall demonstrate how the gravitational interaction in hyperunified field theory is described solely by a {\it conformal scaling gauge invariant Einstein-Hilbert type action}. Consequently, the basic action of hyperunified field theory in a hidden gauge formalism enables us to reveal the {\it gravity-geometry correspondence}.

\subsection{ Hyperunified field theory with general conformal scaling gauge invariance and essential gauge massless condition}

The general action Eq.(\ref{HUTaction1}) of hyperunified field theory formulated in the hyper-gravifield fiber bundle has the explicit conformal scaling gauge invariance with the introduction of the scaling gauge field and scalar field. In the general action Eq.(\ref{HUTaction3}) expressed in a hidden gauge formalism, the gauge interactions are described by the field strength $\cR_{\fM\fN\fP\fQ}$ of the hyper-spacetime gravigauge field $\vGa_{\fM\fN}^{\fP}$ and the field strength $\cQ_{\fM\fN\fP\fQ}$ of the hyper-spacetime homogauge field $\cH_{\fM}^{\fP\fQ}$. Each term of the gauge interactions is in general not conformal scaling gauge invariant though the sum of those interaction terms as a whole is conformal scaling gauge invariant. This is because both the hyper-spacetime gravigauge field $\vGa_{\fM\fN}^{\fP}$ as the Christoffel symbols and the hyper-spacetime homogauge field $\cH_{\fM\fN}^{\fP}$ are no longer invariant though the hyper-spin gauge field $\cA_{\fM}^{\fA\fB}$ carries no charge under the conformal scaling gauge transformations. Explicitly, they transform as follows,
\be \label{CSG}
\vGa_{\fM\fN}^{\fP} & \to & \vGa_{\fM\fN}^{\fP} - \eta_{\fM}^{\; \fP} \p_{\fN}\ln \xi - \eta_{\fN}^{\; \fP} \p_{\fM}\ln \xi   + 
\chi_{\fM\fN} \p^{\fP}\ln \xi , \nn \\
\cH_{\fM\fN}^{\fP} & \to & \cH_{\fM\fN}^{\fP} + (\eta_{\fM}^{\; \fP} \eta_{\fN}^{\;\fQ} - \chi_{\fM\fN}\chih^{\fP\fQ} )\p_{\fQ} \ln \xi \, .
\ee

The combination of the Riemann curvature tensor and the Ricci curvature tensor allows us to obtain the conformal scaling gauge invariant Weyl curvature tensor,
\be
\cC_{\fM\fN\fP\fQ} & = & \cR_{\fM\fN\fP\fQ} + \frac{1}{(D_h-1)(D_h-2)}(\chi_{\fM\fP} \chi_{\fN\fQ} - \chi_{\fN\fP} \chi_{\fM\fQ} )\1  \cR     \nn \\
& + & \frac{1}{D_h -2}\1 (  \chi_{\fM\fQ} \cR_{\fN\fP} - \chi_{\fN\fQ} \cR_{\fM\fP}   +\chi_{\fN\fP}  \cR_{\fM\fQ} - \chi_{\fM\fP}  \cR_{\fN\fQ}) \, ,
\ee
which leads to the following conformal scaling gauge-invariant interaction:
\be
\cC_{\fM\fN\fP\fQ} \cC^{\fM\fN\fP\fQ} = \cR_{\fM\fN\fP\fQ} \cR^{\fM\fN\fP\fQ} - \frac{4}{D_h -2} \1 \cR_{\fM\fN} \cR^{\fM\fN} 
+ \frac{2}{(D_h - 1)(D_h -2)} \1 \cR^2 \, .
\ee 

When the couplings satisfy the relations, 
\be
& & g_4 + g_5 = - \frac{4}{D_h-2} \1 g_C \, , \nn \\
& & g_6 = \frac{2}{(D_h-1)(D_h-2)} \1 g_C\, , \nn \\
& & g_C \equiv g_1 + g_2 + g_3 \, , 
\ee 
we are able to simplify the gauge interaction terms in Eq.(\ref{HUTaction3}) to the following form,
\be
& & (g_1 + g_2 + g_3 ) (\, \cR^{\fM\fN\fP\fQ} \cR_{\fM\fN\fP\fQ} + 2 \cR^{\fM\fN\fP\fQ} \cQ_{\fM\fN\fP\fQ} ) 
 \nn \\  
 & & + (g_4 + g_5) (\cR^{\fM\fN}\cR_{\fM\fN} + 2\cR^{\fM\fN}\cQ_{\fM\fN} ) + g_6 (\cR^2 + 2 \cR\cQ)  \nn \\
& & = g_C (\, \cC^{\fM\fN\fP\fQ} \cC_{\fM\fN\fP\fQ} + 2 \cC^{\fM\fN\fP\fQ} \cQ_{\fM\fN\fP\fQ} ) \, ,
\ee
where the first term on the right-hand side of the equality is conformal scaling gauge invariant, while the second term is not. Only when taking $g_C =0$, we are able to obtain the general conformal scaling gauge-invariant interactions that concern solely the field strength of the hyper-spacetime homogauge field. 

In general, when the coupling constants satisfy the relations,
\be \label{CCrelation1}
g_1 + g_2 + g_3=0\, , \quad g_4 + g_5 = 0\, , \quad g_6 = 0\, ,
\ee
we arrive at a general conformal scaling gauge-invariant action of hyperunified field theory. We may refer to such relations as  {\it general conformal scaling invariance conditions}.

The combined hyper-spacetime homogauge field $\tilde{\cH}_{\fM\fN}^{\fP} = \cH_{\fM\fN}^{\fP} + \fS_{\fM\fN}^{\fP} $ is conformal scaling gauge invariant, it leads to the following gravitational ``mass-like" terms:  
\be
& & \tilde{\cH}_{[\fM\fQ]\fP} \tilde{\cH}^{[\fM\fQ]\fP} = \chih^{\fM\fM'}\chih^{\fQ\fQ'} \chih^{\fP\fP'}  \tilde{\cH}_{[\fM\fQ]\fP} \tilde{\cH}_{[\fM'\fQ']\fP'} \, , \nn \\
& & \tilde{\cH}_{\fM\fP\fQ} \tilde{\cH}^{[\fP\fQ]\fM} = \chih^{\fM\fM'}\chih^{\fQ\fQ'} \chih^{\fP\fP'}  \tilde{\cH}_{\fM\fP\fQ} \tilde{\cH}_{[\fP'\fQ']\fM'}\, , \nn \\
& & \tilde{\cH}_{\fQ}^{\fQ\fM} \tilde{\cH}_{\fP\fM}^{\fP} = \chih^{\fM\fM'}\chih^{\fP\fQ} \chih^{\fP'\fQ'} \tilde{\cH}_{\fP\fQ\fM} \tilde{\cH}_{\fP'\fQ'\fM'}\, . \nn 
\ee
They satisfy a general relation,
\be
\tilde{\cH}_{[\fP\fQ]\fM} \tilde{\cH}^{[\fP\fQ]\fM} = 2 \tilde{\cH}_{\fM\fP\fQ} \tilde{\cH}^{\fM\fP\fQ}  +  \tilde{\cH}_{\fM\fP\fQ} \tilde{\cH}^{[\fP\fQ]\fM}\, .
\ee
With such a relation, we come to discuss several cases,
\be \label{cases}
& & (\alpha_G -\beta_G)\tilde{\cH}_{[\fP\fQ]\fM} \tilde{\cH}^{[\fP\fQ]\fM} + 2(\alpha_E-\beta_G) \tilde{\cH}_{\fM\fP\fQ}\tilde{\cH}^{[\fP\fQ]\fM} \nn \\
& & = \begin{cases}
 2 (\alpha_G -\beta_G)\tilde{\cH}_{\fP\fQ\fM} \tilde{\cH}^{\fP\fQ\fM}\, , & \mbox{if} \quad 3\beta_G - \alpha_G = 2\alpha_E \\
 6 (\alpha_G -\beta_G)\cH_{[\fP\fQ\fM]} \cH^{[\fP\fQ\fM]}\, , & \mbox{if} \quad  \alpha_G + \beta_G = 2\alpha_E  \\
 0 \, , & \mbox{if} \quad \alpha_G = \beta_G = \alpha_E\, , 
 \end{cases}
\ee
with $\cH_{[\fP\fQ\fM]}$ a totally antisymmetric hyper-spacetime homogauge field defined as
\be
& & \cH_{[\fP\fQ\fM]} = \frac{1}{3} ( \cH_{\fP\fQ\fM} + \cH_{\fQ\fM\fP} + \cH_{\fM\fP\fQ}  ) \, \nn \\
& & \tilde{\cH}_{[\fP\fQ\fM]}  =  \cH_{[\fP\fQ\fM]} \, , \quad \fS_{[\fP\fQ\fM]} = 0  \, .
\ee

It is seen that, for the case $3\beta_G - \alpha_G = 2\alpha_E$, the hyper-spacetime homogauge field $\cH_{\fM}^{\fP\fQ}$ has an ordinary gravitational mass-like term $\cH_{\fM\fP\fQ} \cH^{\fM\fP\fQ}$ in proportion to the combined coupling $(\alpha_G -\beta_G)$.  For the case $\alpha_G + \beta_G = 2\alpha_E$, the totally antisymmetric hyper-spacetime homogauge field $\cH_{[\fM\fP\fQ]}$ gets the gravitational mass-like term $\cH_{[\fM\fP\fQ]} \cH^{[\fM\fP\fQ]}$ in proportion to the combined coupling $(\alpha_G -\beta_G)$. Once the following relations hold:
\be \label{CCrelation2}
\alpha_G = \beta_G = \gamma_G= \alpha_E\, ,
\ee
all the gravitational ``mass-like" terms concerning the hyper-spacetime homogauge field $\cH_{\fM}^{\fP\fQ}$ disappear. For convenience, we may refer to such relations by an essential {\it gauge massless condition} for the hyper-spin gauge field. In fact, it provides a {\it general hyper-spin gauge-invariance condition} for the dynamics of the hyper-gravifield.

When the {\it general conformal scaling invariance condition} given in Eq.(\ref{CCrelation1}) holds, we obtain the simplified action of hyperunified field theory in the hidden gauge formalism,
\be
\label{HUTaction04}
& & I_H =  \int [d\hx]\1 \chi   \phi^{D_h-4} \1 \{ \bar{\vPsi}\dg^{\fM} [ i \p_{\fM}  +  (\varXi_{\fM}^{\fP\fQ} + g_h\cH_{\fM}^{\fP\fQ} ) \frac{1}{2}\varSigma_{\fP\fQ} ] \vPsi  \nonumber \\
& & - \frac{1}{4}\tilde{\chi}^{\fM\fN\1 \fM'\fN'\1\fP\fQ\1\fP'\fQ'}  \cQ_{\fM\fN\fP\fQ} \cQ_{\fM'\fN'\fP'\fQ'}  
 + \cL_{\fA} \nn \\
 && + \alpha_E  (\, \phi^2 \cR   - (D_h-1)(D_h-2)\p_{\fM}\phi \p^{\fM}\phi \, ) - \beta_E\1\phi^4  \nn \\
& &+  \frac{1}{2}d_{\fM} \phi d^{\fM}\phi -\frac{1}{4} {\cal W}_{\fM\fN} {\cal W}^{\fM\fN}   \}  + 2 \alpha_Eg_h\p_{\fM} (\chi \phi^{D_h-2}\cH_{\fN}^{\fN\fM} ) ,
\ee
with
\be \label{mass}
\cL_{\fA} & = & 
\begin{cases}
0 &  \alpha_G = \beta_G = \gamma_G = \alpha_E\, ,  \\
\frac{3}{2} \bar{\alpha}\, \phi^2 \cH_{[\fM\fP\fQ]} \cH^{[\fM\fP\fQ]} & \alpha_G + \beta_G = 2\alpha_E ,  
\gamma_G = \alpha_E , \\
 \frac{1}{2}\bar{\alpha}\, \phi^2 \tilde{\cH}_{\fM\fP\fQ} \tilde{\cH}^{\fM\fP\fQ} + \bar{\alpha}_E \phi^2  \tilde{\cH}_{\fM\fN}^{\fM}  \tilde{\cH}_{\fM'}^{\fM'\fN}    &  3\beta_G - \alpha_G = 2\alpha_E.
\end{cases}
\ee
and with $\bar{\alpha} \equiv \alpha_G -\beta_G$ and $\bar{\alpha}_E \equiv \alpha_E - \gamma_G$. The field strength $\cQ_{\fM\fN\fP\fQ}$ is defined in Eq.(\ref{HFS}) and the Ricci scalar curvature $\cR$ is given in Eq.(\ref{RSC}). 
In obtaining the above action, we have used the identity,
\be
  & & - \chih^{\fM\fP}\chih^{\fN\fQ} \hat{\nabla}_{\fM}\cH_{\fN\fP\fQ} =  \hat{\nabla}_{\fM}\cH_{\fN}^{\fN\fM} \nn \\
  & & \equiv    \nabla_{\fM}\cH_{\fN}^{\fN\fM}  + (D_h-2)\p_{\fM}(\ln\phi)  \cH_{\fN}^{\fN\fM}  \nn \\
  & & =  (\chi \phi^{D_h-2})^{-1}  \p_{\fM} (\chi \phi^{D_h-2}\cH_{\fN}^{\fN\fM})  \, , 
\ee
with the definition $\cH_{\fN}^{\fN\fM} = -\cH_{\fN\fP\fQ}  \chih^{\fN\fQ}\chih^{\fP\fM}$.

The tensor factor $\tilde{\chi}^{\fM\fN\1 \fM'\fN'\1 \fP\fQ\1\fP'\fQ'}$ has the following form:
\be \label{tensor3}
& & \tilde{\chi}^{\fM\fN\1 \fM'\fN'\1 \fP\fQ\1\fP'\fQ'} \equiv   \nn \\
& &\quad  \frac{1}{4} \{ [ \hat{\chi}^{\fM\fM'} \chih^{\fP\fP'} (\hat{\chi}^{\fN\fN'} \chih^{\fQ\fQ'} - 2  \chih^{\fN\fQ'} \chih^{\fN'\fQ}) + \chih^{(\fM,\fM'\leftrightarrow\fN, \fN' )} ]+  \chih^{(\fP,\fP'\leftrightarrow \fQ,\fQ' ) }    \} \nn \\
& & \; + \frac{1}{4}\alpha_W\{ [ (\chih^{\fM\fP'} \chih^{\fM'\fP} - 2\chih^{\fM\fM'} \chih^{\fP\fP'}  )  \chih^{\fN\fQ'}  \chih^{\fN'\fQ} 
+  \chih^{(\fM,\fM'\leftrightarrow\fN, \fN' )} ] +  \chih^{(\fP,\fP'\leftrightarrow \fQ,\fQ' ) }    \}    \nn \\
& & \;  + \frac{1}{2}\beta_W \{ [ (\chih^{\fP\fP'}  \chih^{\fM\fM'} - \chih^{\fM'\fP}\chih^{\fM\fP'}) \chih^{\fN\fQ} \chih^{\fN'\fQ'} + 
 \chih^{(\fM,\fM'\leftrightarrow\fN, \fN' )} ]+  \chih^{(\fP,\fP'\leftrightarrow \fQ,\fQ' )} \}   \, ,
\ee
where we have normalized the kinetic term of the hyper-spacetime homogauge field interactions by taking the following conventions: 
\be \label{ICC}
& & \cH_{\fM}^{\fP\fQ}\to g_h\1 \cH_{\fM}^{\fP\fQ} \, , \qquad  g_1 \equiv g_h^{-2} \, , \nn \\
& & g_2 \equiv g_h^{-2}\1 \alpha_W\, , \quad \qquad g_4 \equiv g_h^{-2}\1 \beta_W\, ,
\ee
which shows that to preserve the general conformal scaling gauge invariance in the gauge interactions of the hyper-spacetime homogauge field $ \cH_{\fM}^{\fP\fQ}$, there exist in general three independent coupling constants $g_h$, $\alpha_W$ and $\beta_W$. Note that the last term of the action of Eq.(\ref{HUTaction4}) reflects a surface effect.

In SM, both the electromagnetic field and the gluon fields are massless, and the weak gauge bosons receive masses through the Higgs mechanism of spontaneous symmetry breaking. To match such phenomena in SM, it is natural to postulate that {\it the hyper-spin gauge field as a basic field should be massless}. By applying the essential {\it gauge massless condition} given in Eq.(\ref{CCrelation2}), we arrive at the basic action of hyperunified field theory in the hidden gauge formalism,
\be
\label{HUTaction4}
& & I_H =  \int [d\hx]\1 \chi  \phi^{D_h-4} \1 \{ \bar{\vPsi}\dg^{\fM} [ i \p_{\fM}  +  (\varXi_{\fM}^{\fP\fQ} + g_h\cH_{\fM}^{\fP\fQ} ) \frac{1}{2}\varSigma_{\fP\fQ} ] \vPsi  \nonumber \\
& & - \frac{1}{4}\tilde{\chi}^{\fM\fN\1 \fM'\fN'\1\fP\fQ\1\fP'\fQ'}  \cQ_{\fM\fN\fP\fQ} \cQ_{\fM'\fN'\fP'\fQ'}  \nn \\
 && + \alpha_E  (\, \phi^2 \cR   - (D_h-1)(D_h-2)\p_{\fM}\phi \p^{\fM}\phi \, ) - \beta_E\1\phi^4  \nn \\
& &+  \frac{1}{2}d_{\fM} \phi d^{\fM}\phi -\frac{1}{4} {\cal W}_{\fM\fN} {\cal W}^{\fM\fN}  \, \}  + 2 \alpha_Eg_h\p_{\fM} (\chi \phi^{D_h-2}\cH_{\fN}^{\fN\fM} ) .
\ee

It is noticed that all the quadratic Riemann tensor and Ricci tensor terms disappear due to the general conformal scaling invariance condition Eq.(\ref{CCrelation1}). Namely, the higher derivative gravitational interactions in hyper-spacetime are absent and the dynamics of the hyper-gravimetric field is described solely by an Einstein-Hilbert type action with the global and local conformal scaling symmetries, which is ensured by introducing the dynamical scaling scalar field $\phi(\hx)$ associated with the scaling gauge field $W_{\fM}(\hx)$. The hyper-spacetime gravigauge field $ \vGa_{\fM\fN}^{\fP}$ as the Christoffel symbols decouples from the hyper-spinor interactions due to the self-hermiticity of the Majorana-type hyper-spinor field interactions and the symmetric property of the Christoffel symbols  $\vGa_{\fM\fN}^{\fP} =\vGa_{\fN\fM}^{\fP}$.

\subsection{Einstein-Hilbert type action with gravity-geometry correspondence in hyper-spacetime and symmetric Goldstone-like hyper-gravifield with unitary gauge}

The basic action given in Eq.(\ref{HUTaction4}) shows that the bosonic gauge interactions are described by the symmetric {\it hyper-spacetime gravigauge field} $\vGa_{\fM\fN}^{\fP}$ and the antisymmetric {\it hyper-spacetime homogauge field} $\cH_{\fM}^{ \fP \fQ}$. When the {\it general conformal scaling invariance condition} Eq.(\ref{CCrelation1}) holds for the coupling constants, the gravitational gauge interactions of  the {\it hyper-spacetime gravigauge field} $\vGa_{\fM\fN}^{\fP}$ as the Christoffel symbols are solely characterized by the scaling gauge-invariant Einstein-Hilbert type action that is governed by the Ricci scalar tensor of hyper-spacetime and the scalar field interactions, i.e.,  
\be
I_{GG} \equiv  \int [d\hx]\; \chi  \phi^{D_h-4} \alpha_E [ \phi^2 \cR - (D_h-1)(D_h-2) \p_{\fM}\phi\p^{\fM}\phi  ]  + 2 \alpha_Eg_h\p_{\fM} (\chi \phi^{D_h-2}\cH_{\fN}^{\fN\fM} ). \nn
\ee
The absence of terms in quadratic Riemann tensor and Ricci tensor indicates that such a hyperunified field theory should get rid of the so-called unitarity problem caused by the higher derivative gravitational interactions.  

The hyper-spacetime gravigauge field $\vGa_{\fM\fN}^{\fP}$ as the Christoffel symbols is characterized by the Goldstone-like hyper-gravimetric field $\chi_{\fM\fN}$, which contains $N_h =D_h(D_h+1)/2$ degrees of freedom. The gravitational interactions of the hyper-spinor field are governed by the Goldstone-like hyper-gravifield  $\chi_{\fM}^{\;\fA}$ (or $\chih_{\fA}^{\; \fM}$) through the gauge-type field $\varXi_{\fM}^{\fP\fQ}$ and the $\gamma$-matrix terms $\dg^{\fM}$ and  $\varSigma^{\fP\fQ}$. In general, the basic gravitational field remains the Goldstone-like hyper-gravifield $\chi_{\fM}^{\;\fA}$, which concerns $N_h = D_h\times D_h$ degrees of freedom. The extra degrees of freedom in $\chi_{\fM}^{\;\fA}$ are accounted as $\Delta N_h = D_h^2 - D_h(D_h+1)/2=D_h(D_h -1)/2$, which reflects the equivalence classes of the hyper-spin gauge symmetry SP(1, $D_h$-1) that involves the same degrees of freedom $D_h(D_h -1)/2$. Therefore, we can always make a specific hyper-spin gauge transformation $\bar{\Lambda}(\hx)$ to set a gauge fixing condition, so that the gauge transformed hyper-gravifield $\chi'_{\fM\fA}(\hx)$ becomes symmetric, i.e., 
\be
\chi_{\fM \fA}(\hx) \to  \chi'_{\fM\fA}(\hx) = \chi_{\fM\fB}(\hx) \bar{\Lambda}^{\fB}_{\; \, \fA}(\hx)  =  \chi'_{\fA\fM}(\hx)\, .
\ee 
We may refer to such a gauge fixing condition b ay {\it unitary gauge}. 

In such a {\it unitary gauge}, both the Goldstone-like hyper-gravifield $\chi_{\fM\fA}(\hx)$ (we shall omit the prime for simplicity) and the Goldstone-like hyper-gravimetric field $\chi_{\fM\fN}(\hx)$ are symmetric. They involve the same degrees of freedom and are correlated with the following relation: 
\be
\chi_{\fM\fN} = \chi_{\fM\fA} \chi_{\fN\fB} \eta^{\fA\fB} = \chi_{\fM\fA}\eta^{\fA\fB} \chi_{\fB\fN}  \equiv  (\chi_{\fM\fA})^2 \, .
\ee

To be more explicit, it is useful to take a nonlinearly realized exponential representation, 
\be
& & \chi_{\fM\fA} = (e^{G})_{\fM\fA}\, , \quad  G_{\fM\fA} = G_{\fA\fM} \, , \nn \\
& & \chi_{\fM\fN} =(e^{G} e^{G} )_{\;\fM\fN}= (e^{2G})_{\fM\fN} \, ,
\ee 
where $G_{\fM\fA}=G_{\fA\fM}$ is regarded as a nonlinearly realized {\it symmetric Goldstone-like hyper-gravifield}. One can make use of the general properties of matrix exponential, such as,
\be
& &  \p_{\fM} (e^{G}) =  e^{G}\, \frac{1- e^{- \ad_{G} }}{\ad_{G}}\,\p_{\fM}G \, ; \quad \det (e^{G}) = e^{\tr (G)}\, , \nn \\
& &  \frac{1- e^{- \ad_{G} }}{\ad_{G}} = \sum_{k=0}^{k=\infty} \frac{(-1)^k}{(k+1)!} (\ad_{G})^k\, , \quad \ad_{G}(\p_{\fM}G) \equiv [G\, , \p_{\fM}G]\, .
\ee

It is clear that in the basic action of hyperunified field theory formulated in terms of hidden gauge formalism, it is reliable to set the gauge fixing condition to be the unitary gauge. So that the basic gravitational field is chosen to be the symmetric Goldstone-like hyper-gravifield $\chi_{\fM\fA}(\hx)= \chi_{\fA\fM}(\hx)$ (or nonlinearly realized symmetric Goldstone-like hyper-gravifield $G_{\fM\fA}(\hx)=G_{\fA\fM}(\hx)$) and the basic gauge field is taken as the antisymmetric hyper-spacetime homogauge field $\cH_{\fM}^{\fP\fQ}(\hx)$. In this case, the gravitational interactions of both the hyper-spinor and the bosonic fields are described by the symmetric Goldstone-like hyper-gravifield $\chi_{\fM\fA}(\hx)$. Meanwhile, the geometry and dynamics of hyper-spacetime are essentially characterized by the symmetric Goldstone-like hyper-gravifield $\chi_{\fM\fA}(\hx)$ which determines the hyper-gravimetric field  $\chi_{\fM\fN}(\hx) = ( \chi_{\fM\fA}(\hx) )^2$ and the hyper-spacetime gravigauge field $\vGa_{\fM\fN}^{\fP}(\hx)$ as the Christoffel symbols. In this sense, we arrive at the general {\it gravity-geometry correspondence} in hyper-spacetime. 

To keep the Goldstone-like hyper-gravifield $\chi_{\fM\fA}(\hx)$ be symmetric in the unitary gauge, both the hyper-spin symmetry SP(1,$D_h$-1) of the hyper-spinor field and the Lorentz symmetry SO(1,$D_h$-1) of coordinates in hyper-spacetime have to be the global symmetries and coincide with each other, i.e.,  

\[ SP(1,D_h-1)\cong SO(1,D_h-1). \]

In conclusion, for a given unitary gauge with the symmetric hyper-gravifield $\chi_{\fM\fA}(\hx)=\chi_{\fA\fM}(\hx)$, the basic action of hyperunified field theory in the hidden gauge formalism has only the global Poincar\'e symmetry $\mbox{PO(1},D_h\mbox{-1)})$ as well as the global and local scaling symmetries S(1) and SG(1). Namely, when making the gauge fixing condition to be the unitary gauge, the basic action of Eq.(\ref{HUTaction4}) of hyperunified field theory in the hidden gauge formalism possesses only a minimized maximal symmetry, 
\be
G_S = PO(1,D_h-1) \times S(1)\Join SG(1)\, ,
\ee
which plays a role as a fundamental symmetry in hyperunified field theory.

\section{Hyperunified field theory in hidden coordinate formalism and gauge-gravity correspondence }

Alternatively, we shall investigate a correlation between the gauge interaction and the gravitational effect in a hidden coordinate formalism. As the general action of hyperunified field theory given in Eq.(\ref{HUTaction}) is constructed based on the postulates of gauge-invariance and coordinate independence in the locally flat hyper-gravifield spacetime ${\bf G}_h$, we shall explicitly show how the gravitational interaction can be described by the gauge interaction in a hidden coordinate formalism and demonstrate the general gauge-gravity correspondence relying on the gravitational origin of gauge symmetry. 

\subsection{Hyper-spin gauge field and field strength in a hidden coordinate formalism}

To express the basic action of hyperunified field theory in terms of hidden coordinate formalism, it is useful to make the redefinitions for the hyper-spin gauge field and the corresponding hyper-spin gravigauge and homogauge fields defined in Eqs.(\ref{TP1})-(\ref{TP3}) in the locally flat hyper-gravifield spacetime ${\bf G}_h$. 

Let us begin with the redefinition for the hyper-spin gauge field in the hyper-gravifield spacetime ${\bf G}_h$,
\be
 \cA_{\fC}^{\fA\fB} \equiv \chih_{\fC}^{\;\, \fM} \cA_{\fM}^{\fA\fB}\, .
\ee
Correspondingly, the gravitational origin of the hyper-spin gauge symmetry shown in Eqs.(\ref{TP1})-(\ref{TP3}) enables us to rewrite the hyper-spin gauge field into two parts, 
\be
\cA_{\fC}^{\fA\fB} =  \vOm_{\fC}^{\fA\fB} + \cH_{\fC}^{\fA\fB} \, ,
\ee
with $\vOm_{\fC}^{\fA\fB}$ and $\cH_{\fC}^{\fA\fB}$ the hyper-spin gravigauge and homogauge fields, respectively,  
\be
& & \vOm_{\fC}^{\fA\fB} \equiv \chih_{\fC}^{\;\, \fM} \vOm_{\fM}^{\fA\fB}\, , \quad \cH_{\fC}^{\fA\fB} \equiv \chih_{\fC}^{\;\, \fM} \cH_{\fM}^{\fA\fB}  \, .
\ee
The hyper-spin gravigauge field $\vOm_{\fC}^{\fA\fB}$ is characterized by the gauge-type hyper-gravifield, 
\be \label{HSGGF1}
\vOm_{\fC}^{\fA\fB}  & = &  \frac{1}{2} [\, \chih^{\fA\fM} \eth_{\fC}\chi_{\fM}^{\;\, \fB} - \chih^{\fB\fM} \eth_{\fC}\chi_{\fM}^{\;\, \fA} - \chih_{\fC}^{\;\,\fM} ( \eth^{\fA}\chi_{\fM}^{\;\, \fB} \nn \\
& - &  \eth^{\fB}\chi_{\fM}^{\;\, \fA} ) + ( \eth^{\fA}\chih^{\fB\fM} -  \eth^{\fB}\chih^{\fA\fM} ) \chi_{\fM\fC}], 
\ee
with $\eth_{\fC}  \equiv \chih_{\fC}^{\;\, \fM} \p_{\fM}$ defined in Eq.(\ref{eth}).

Similarly, the field strength of the hyper-spin gauge field consists of two parts, 
\be
\cF_{\fC\fD}^{\fA\fB} = \cR_{\fC\fD}^{\fA\fB} + \cQ_{\fC\fD}^{\fA\fB} \, , 
\ee
with explicit forms given by
\be
& & \cF_{\fC\fD}^{\fA\fB} = \tilde{\cD}_{\fC} \cA_{\fD}^{\fA\fB} - \tilde{\cD}_{\fD} \cA_{\fC}^{\fA\fB} + ( \cA_{\fC \fE}^{\fA} \cA_{\fD}^{\fE \fB} -  \cA_{\fD \fE}^{\fA} \cA_{\fC}^{\fE \fB} )  \, , \nn \\
& &  \tilde{\cD}_{\fC} \cA_{\fD}^{\fA\fB}  =  \eth_{\fC} \cA_{\fD}^{\fA\fB}  - \vOm_{\fC \fD}^{\fE} \cA_{\fE}^{\fA \fB} \, , 
\ee
for the hyper-spin gauge field strength, and 
\be \label{GGFS}
& & \cR_{\fC\fD}^{\fA\fB} = \tilde{\cD}_{\fC} \vOm_{\fD}^{\fA\fB} - \tilde{\cD}_{\fD} \vOm_{\fC}^{\fA\fB} + \vOm_{\fC \fE}^{\fA} \vOm_{\fD}^{\fE \fB} -  \vOm_{\fD \fE}^{\fA} \vOm_{\fC}^{\fE \fB}   \, , \nn \\
& &  \tilde{\cD}_{\fC} \vOm_{\fD}^{\fA\fB}  =  \eth_{\fC} \vOm_{\fD}^{\fA\fB}  - \vOm_{\fC \fD}^{\fE} \vOm_{\fE}^{\fA \fB} \, , 
\ee
for the hyper-spin gravigauge field strength, as well as 
\be \label{HGFS}
 & & \cQ_{\fC\fD}^{\fA\fB} = \cD_{\fC} \cH_{\fD}^{\fA\fB} - \cD_{\fD} \cH_{\fC}^{\fA\fB}  +  \cH_{\fC \fE}^{\fA} \cH_{\fD}^{\fE \fB} -  \cH_{\fD \fE}^{\fA} \cH_{\fC}^{\fE \fB}  \,  ,\nn \\
 & & \cD_{\fC} \cH_{\fD}^{\fA\fB} =  \tilde{\cD}_{\fC}  \cH_{\fD}^{\fA\fB} + \vOm_{\fC \fE}^{\fA} \cH_{\fD}^{\fE \fB} + \vOm_{\fC \fE}^{\fB} \cH_{\fD}^{\fA \fE}   \, , \nn \\
& & \tilde{\cD}_{\fC}  \cH_{\fD}^{\fA\fB} = \eth_{\fC}  \cH_{\fD}^{\fA\fB}  -  \vOm_{\fC \fD}^{\fE} \cH_{\fE}^{\fA\fB} \, , 
\ee
for the hyper-spin homogauge field strength. 

The conformal scaling gauge field strength is given by  
\be
& & {\cal W}_{\fC\fD} = \tilde{\cD}_{\fC}\mW_{\fD} - \tilde{\cD}_{\fD}\mW_{\fC} \, , \quad  \mW_{\fC} = \chih_{\fC}^{\;\, \fM} \mW_{\fM}\, ,
 \nn \\
 & & \tilde{\cD}_{\fC}\mW_{\fD} = \eth_{\fC}\mW_{\fD} - \vOm_{\fC \fD}^{\fE} \mW_{\fE}\, . 
\ee

For the gauge-type hyper-gravifield strength, it is determined by the hyper-spin homogauge field $\cH_{\fC }^{\fA\fB}$ and the pure gauge field $\fS_{\fC}$ of the scalar field,
\be
& & \cG_{\fC\fD}^{\fA} \equiv \cH_{[\fC\fD]}^{\fA} + \fS_{[\fC\fD]}^{\fA} \equiv \tilde{\cH}_{[\fC\fD]}^{\fA} \, , \nn \\
& &  \cH_{[\fC\fD]}^{\fA} = \cH_{\fC\fD}^{\fA} - \cH_{\fD\fC}^{\fA}\, , \quad  \fS_{[\fC\fD]}^{\fA} =  \fS_{\fC\fD}^{\fA} - \fS_{\fD\fC}^{\fA} \, , \nn \\
& & \fS_{\fC\fD}^{\fA} = \eta_{\fC\fD}\eta^{\fA\fB}\fS_{\fB} - \eta_{\fC}^{\; \fA} \fS_{\fD}, \; \; \fS_{\fC} \equiv \eth_{\fC}\ln \phi,
\ee
and 
\be
& & \cG_{\fC\fD\fA} \equiv -\cH_{[\fC\fD]\fA} - \fS_{[\fC\fD]\fA} \equiv - \tilde{\cH}_{[\fC\fD]\fA} \, , \nn \\
& &  \fS_{\fC\fD\fA}  =  \eta_{\fC\fA} \fS_{\fD} -  \eta_{\fC\fD} \fS_{\fA} \, , \nn \\
& &  \fS_{[\fC\fD]\fA} =  - \fS_{\fC}\eta_{\fD\fA} + \fS_{\fD}\eta_{\fC\fA}  \, .
\ee

\subsection{ Riemann-like and Ricci-like tensors in locally flat hyper-gravifield spacetime and symmetry properties of field strengths }

The field strengths in the locally flat hyper-gravifield spacetime relate to those in the hyper-gravifield fiber bundle as follows,
 \be
& & \cF_{\fC\fD}^{\fA\fB} =\cF_{\fM\fN}^{\fA\fB}  \chih^{\;\fM}_{\fC} \chih^{\;\fN}_{\fD}; \; \; \cF_{\fC\fD\fA\fB} =\cF_{\fM\fN\fA\fB}  \chih^{\;\fM}_{\fC} \chih^{\;\fN}_{\fD} \nn \\
& & \cR_{\fC\fD}^{\fA\fB} = \cR_{\fM\fN}^{\fA\fB}  \chih^{\;\fM}_{\fC} \chih^{\;\fN}_{\fD}; \;  \; \cR_{\fC\fD \fA\fB} = \cR_{\fM\fN \fA\fB}  \chih^{\;\fM}_{\fC} \chih^{\;\fN}_{\fD}\, , \nn \\
& & \cQ_{\fC\fD}^{\fA\fB}  =\cQ_{\fM\fN}^{\fA\fB} \chih^{\;\fM}_{\fC} \chih^{\;\fN}_{\fD}; \; \;  \cQ_{\fC\fD\fA\fB} =\cQ_{\fM\fN\fA\fB}  \chih^{\;\fM}_{\fC} \chih^{\;\fN}_{\fD}\, ,
 \ee
and 
\be
& & \cG_{\fC\fD}^{\fA}= \cG_{\fM\fN}^{\fA}\chih^{\;\fM}_{\fC} \chih^{\;\fN}_{\fD}; \; \; \cG_{\fC\fD\fA}= \cG_{\fM\fN \fA}\chih^{\;\fM}_{\fC} \chih^{\;\fN}_{\fD} \, , \nn \\
& & \cW_{\fC\fD}  = \cW_{\fM\fN}\chih^{\;\fM}_{\fC} \chih^{\;\fN}_{\fD} \, .  
\ee

The field strength of the hyper-spin gravigauge field in the hidden coordinate formalism is related to the one of the hyper-spacetime gravigauge field in the hidden gauge formalism:
\be
 \cR_{\fC\fD \fA\fB} =  \cR_{\fM\fN \fA\fB}  \chih^{\;\fM}_{\fC} \chih^{\;\fN}_{\fD} =\cR_{\fM\fN \fP\fQ}  \chih^{\;\fM}_{\fC} \chih^{\;\fN}_{\fD}  \chih^{\;\fP}_{\fA} \chih^{\;\fQ}_{\fB}\, ,
\ee
which defines the Riemann-like tensor in the locally flat hyper-gravifield spacetime. Such relations indicate both the {\it gauge-gravity and gravity-geometry correspondences}. 

Applying for the symmetry properties of the Riemann tensors shown in Eqs.(\ref{SP1})$-$(\ref{SP2}) and the identities Eq.(\ref{IDS}), we obtain similar symmetry properties,
\be \label{SP3}
& & \cR_{\fC\fD\fA\fB} = -\cR_{\fD\fC\fA\fB} \, , \quad \cR_{\fC\fD\fA\fB} = -\cR_{\fC\fD\fB\fA}\, , \nn \\
& &  \cR_{\fC\fD\fA\fB} = \cR_{\fA\fB\fC\fD}\, ,
\ee
and the analogous Bianchi identities,
\be
& & \cR_{\fC\fD\fB\fA} + \cR_{\fD\fB\fC\fA} + \cR_{\fB\fC\fD\fA} = 0 \, , \nn \\
& & \cD_{\fE}\cR_{\fC\fD\fB}^{\;\fA} + \cD_{\fC}\cR_{\fD\fE\fB}^{\;\fA} + \cD_{\fD}\cR_{\fE\fC\fB}^{\;\fA} = 0\, .
\ee

We can define the symmetric Ricci-like curvature tensor by contracting the Riemann-like tensor in the locally flat hyper-gravifield spacetime,
 \be \label{RLT}
\cR_{\fC\fD} = \cR_{\fC\fB\fD}^{\;\fA} \eta_{\fA}^{\, \fB} = - \cR_{\fB\fC\fD}^{\;\fA} \eta_{\fA}^{\, \fB} = - \cR_{\fA\fC\fD}^{\fA} = \cR_{\fD\fC}\, . 
\ee 
The trace of the Ricci-like curvature tensor defines the scalar curvature 
\be  \label{RLC}
\cR \equiv \eta^{\fC\fD} \cR_{\fC\fD} = \eta^{\fC\fD}  \cR_{\fC\fB\fD}^{\;\fA} \eta_{\fA}^{\, \fB} = - \eta^{\fC\fD} \eta^{\fA\fB} \cR_{\fC\fB\fD\fA}\, .
\ee

Based on the above properties of the Riemann-like and Ricci-like tensors in the locally flat hyper-gravifield spacetime, we arrive at the following identities:
\be
& & \cR^{\fC\fD\fA\fB} \cR_{\fC\fB\fA\fD} =\frac{1}{2} \cR^{\fC\fD\fA\fB} \cR_{\fC\fD\fA\fB}  \, , \nn \\
& & \cR^{\fC\fD\fA\fB} \cQ_{\fC\fB\fA\fD} =\frac{1}{2} \cR^{\fC\fD\fA\fB} \cQ_{\fC\fD\fA\fB}\, .
\ee

In terms of the hidden coordinate formalism, there exists a general relation between the hyper-spin homogauge field $\cH_{\fC\fA\fB}$ and the hyper-gravifield strength $\cG_{\fC\fA\fB}$ as well as the gauge-type field strength $\fS_{\fC\fA\fB}$ of the scalar field,
\be \label{Hrelation3}
\cH_{\fC \fA\fB} & \equiv & \frac{1}{2} ( \cH_{[\fC \fA]\fB} - \cH_{[\fC\fB]\fA} - \cH_{[\fA\fB]\fC} ) \nn \\
\tilde{\cH}_{\fC \fA\fB} & \equiv & \frac{1}{2} ( \tilde{\cH}_{[\fC \fA]\fB} - \tilde{\cH}_{[\fC\fB]\fA} - \tilde{\cH}_{[\fA\fB]\fC} ) \nn \\
& = & \cH_{\fC \fA\fB}  + \frac{1}{2} ( \fS_{[\fC \fA]\fB} - \fS_{[\fC\fB]\fA} - \fS_{[\fA\fB]\fC} ) \nn \\
& = & -\frac{1}{2} (\cG_{\fC \fA\fB} - \cG_{\fC\fB\fA} - \cG_{\fA\fB\fC} )  \, ,
\ee
which results in the following identity,
\be
\tilde{\cH}_{[\fC\fA]\fB}\tilde{\cH}^{[\fC\fB]\fA} = - \frac{1}{2} \tilde{\cH}_{[\fC\fA]\fB}\tilde{\cH}^{[\fC\fA]\fB}  
-  \tilde{\cH}_{\fC\fA\fB}\tilde{\cH}^{[\fA\fB]\fC} \, .
\ee

\subsection{Hyperunified field theory in locally flat hyper-gravifield spacetime and gauge-gravity correspondence}

With the above analysis, the general action of hyperunified field theory in the locally flat hyper-gravifield spacetime is found to be, 
\be
\label{HUTactionGG1}
I_H & = & \int [\dchi]\,  \phi^{D_h-4} \{  \frac{1}{2}\bar{\vPsi}\vGa^{\fC} [ i \eth_{\fC}  + ( \vOm_{\fC\fA\fB} + \cH_{\fC\fA\fB}) \frac{1}{2}\varSigma^{\fA\fB} ] \vPsi  \nonumber \\
& - &  \frac{1}{4}  [\, ( \tilde{\eta}^{\fC\fD\1 \fC'\fD'\1\fA\fB\1\fA'\fB'}   \cQ_{\fC\fD\fA\fB} \cQ_{\fC'\fD'\fA'\fB'}  + {\cal W}_{\fC\fD} {\cal W}^{\fC\fD}\, )  \nn \\
& + &    \, (g_1 + g_2 + g_3 ) (\, \cR^{\fC\fD\fA\fB} \cR_{\fC\fD\fA\fB} + 2 \cR^{\fC\fD\fA\fB} \cQ_{\fC\fD\fA\fB} )    \nonumber \\
& + &   (g_4 + g_5) (\cR^{\fC\fD}\cR_{\fC\fD} + 2\cR^{\fC\fD}\cQ_{\fC\fD} ) + g_6 (\cR^2 + 2\cR\cQ)  \, ] \nn \\
& + & \frac{1}{4} \phi^2 \,  [\, (\alpha_G -\beta_G)\tilde{\cH}_{[\fC\fB]\fA} \tilde{\cH}^{[\fC\fB]\fA} + 2(\alpha_E- \beta_G) \tilde{\cH}_{\fC\fA\fB}\tilde{\cH}^{[\fA\fB]\fC} \, ]  \nn \\
& + & \phi^2 (\alpha_E - \gamma_G)  (\eta^{\,\fC}_{\fA}\tilde{\cH}_{\fC\fD}^{\fA} ) (\eta_{\fA'}^{\,\fC'}\tilde{\cH}_{\fC'\fD'}^{\fA'} ) \eta^{\fD\fD'}  \nonumber \\
& + & \alpha_E  [\, \phi^2 \eta^{\fC\fD}\cR_{\fC\fD} - (D_h-1)(D_h-2) \eth_{\fC}\phi\eth^{\fC}\phi  + 2(D_h-2) \phi \eth_{\fC}\phi \cH_{\fD}^{\fD\fC} \, ]  \nn \\
& + &  \frac{1}{2}\eta^{\fC\fD}\bet_{\fC} \phi \bet_{\fD}\phi  - \beta_E\1 \phi^4  + 2\alpha_E\phi^2 \cD_{\fC}\cH_{\fD}^{\fD\fC}   \, \}\, , 
\ee
which has a similar form as the general action of Eq.(\ref{HUTaction3}) in the hidden gauge formalism. 

In an analogous manner, by combining the Riemann-like curvature tensor and the Ricci-like curvature tensor, we can define a conformal scaling gauge invariant Weyl-like curvature tensor in the locally flat hyper-gravifield spacetime
\be
\cC_{\fC\fD\fA\fB} & = & \cR_{\fC\fD\fA\fP} +  \frac{1}{(D_h-1)(D_h-2)}(\eta_{\fC\fA} \eta_{\fD\fB} - \eta_{\fD\fA} \eta_{\fC\fB} )\1  \cR  \nn \\
& + & \frac{1}{D_h -2}\1 (  \eta_{\fC\fB} \cR_{\fD\fA} - \eta_{\fD\fB} \cR_{\fC\fA}   +\eta_{\fD\fA}  \cR_{\fC\fB} - \eta_{\fC\fA}  \cR_{\fD\fB}) \, ,
\ee
which leads to the conformal scaling gauge-invariant interaction,
\be
\cC_{\fC\fD\fA\fB} \cC^{\fC\fD\fA\fB}  =  \cR_{\fC\fD\fA\fB} \cR^{\fC\fD\fA\fB}  - \frac{4}{D_h -2} \1 \cR_{\fC\fD} \cR^{\fC\fD} + \frac{2}{(D_h - 1)(D_h -2)} \1 \cR^2 \, .
\ee 

When taking the {\it general conformal scaling invariance condition} given in Eq. (\ref{CCrelation1}) for the coupling constants $g_i$ ($i=1,\ldots, 6$) and the conventional redefinitions shown in Eq. (\ref{ICC}), all the quadratic terms of the Riemann-like tensor and the Ricci-like tensor as well as their cross terms with the field strength of the hyper-spin homogauge field cancel each other. Once the coupling constants $\alpha_G$, $\beta_G$, $\gamma_G$ and $\alpha_E$ satisfy the essential {\it gauge massless condition} given in Eq.(\ref{CCrelation2}),  all the gravitational ``mass-like" terms  $\tilde{\cH}_{[\fC\fB]\fA} \tilde{\cH}^{[\fC\fB]\fA}$, $\tilde{\cH}_{\fC\fA\fB} \tilde{\cH}^{[\fA\fB]\fC}$ and $\tilde{\cH}_{\fB}^{\fB\fC} \tilde{\cH}_{\fA\fC}^{\fA}$ disappear and the hyper-spin gauge field becomes ``massless". In this case, the general action of Eq.(\ref{HUTactionGG1}) is simplified, and we obtain the basic action of hyperunified field theory in a locally flat hyper-gravifield spacetime,  
\be
\label{HUTactionGG2}
& & I_H =  \int [\dchi]\,  \phi^{D_h-4}  \{  \frac{1}{2} \bar{\vPsi} \vGa^{\fC} [ i \eth_{\fC}  + g_h (\vOm_{\fC}^{\fA\fB} + \cH_{\fC}^{\fA\fB} ) \frac{1}{2}\varSigma_{\fA\fB} ] \vPsi  \nonumber \\
& & - \frac{1}{4}  \tilde{\eta}^{\fC\fD\1 \fC'\fD'\1\fA\fB\1\fA'\fB'}  \cQ_{\fC\fD\fA\fB} \cQ_{\fC'\fD'\fA'\fB'}  - \beta_E\1 \phi^4 \nonumber \\
& & +  \alpha_E (\phi^2 g_h \eta^{\fC\fD}\cR_{\fC\fD}  - (D_h-1)(D_h-2) \eth_{\fC}\phi\eth^{\fC}\phi)  \nn \\
& & +   \frac{1}{2} \bet_{\fC} \phi \bet^{\fC}\phi   -\frac{1}{4}  {\cal W}_{\fC\fD} {\cal W}^{\fC\fD}  
 + 2g_h \alpha_E  \phi^2 \hat{\cD}_{\fC}\cH_{\fD}^{\fD\fC} \, \}  ,
\ee
where the field strength $\cQ_{\fC\fD\fA\fB}$ is given in Eq.(\ref{HGFS}) and the Ricci-like tensor $\eta^{\fC\fD}\cR_{\fC\fD}$ is defined in Eq.(\ref{RLT}).  

We have introduced the following definitions,
\be \label{tensor5}
& & \tilde{\eta}^{\fC\fD\1 \fC'\fD'\1 \fA\fB\1\fA'\fB'} \equiv   \chi_{\fM}^{\; \fC}  \chi_{\fN}^{\; \fD}  \chi_{\fM'}^{\; \fC'}  \chi_{\fN'}^{\; \fD'}  \chi_{\fP}^{\; \fA}  \chi_{\fQ}^{\; \fB}  \chi_{\fP'}^{\; \fA'}  \chi_{\fQ'}^{\; \fB'}\, \tilde{\chi}^{\fM\fN\1 \fM'\fN'\1 \fP\fQ\1\fP'\fQ'}   \nn \\
& &\quad = \frac{1}{4} \{ [ \eta^{\fC\fC'} \eta^{\fA\fA'} (\eta^{\fD\fD'} \eta^{\fB\fB'} - 2  \eta^{\fD\fB'} \eta^{\fD'\fB}) + \eta^{(\fC,\fC'\leftrightarrow\fD, \fD' )} ]+  \eta^{(\fA,\fA'\leftrightarrow \fB,\fB' ) }    \} \nn \\
& & \quad + \frac{1}{4}\alpha_W\{ [ (\eta^{\fC\fA'} \eta^{\fC'\fA} - 2\eta^{\fC\fC'} \eta^{\fA\fA'}  )  \eta^{\fD\fB'}  \eta^{\fD'\fB} 
+  \eta^{(\fC,\fC'\leftrightarrow\fD, \fD' )} ] +  \eta^{(\fA,\fA'\leftrightarrow \fB,\fB' ) }    \}    \nn \\
& & \quad + \frac{1}{2}\beta_W \{ [ (\eta^{\fA\fA'}  \eta^{\fC\fC'} - \eta^{\fC'\fA}\eta^{\fC\fA'}) \eta^{\fD\fB} \eta^{\fD'\fB'} + 
 \eta^{(\fC,\fC'\leftrightarrow\fD, \fD' )} ]+  \eta^{(\fA,\fA'\leftrightarrow \fB,\fB' )} \}  \, ,
\ee
for the general tensor factor resulting from the {\it general conformal scaling gauge invariance}, and 
\be
\hat{\cD}_{\fC} \cH_{\fD}^{\fD\fC} & \equiv & \cD_{\fC} \cH_{\fD}^{\fD\fC} + (D_h-2)\fS_{\fC} \cH_{\fD}^{\fD\fC}  \nn \\
& = &  \eth_{\fC}  \cH_{\fD}^{\fD\fC} + g_h\vOm_{\fC\fA}^{\fC} \cH_{\fD}^{\fD\fA} + (D_h-2)\fS_{\fC} \cH_{\fD}^{\fD\fC}  \, , \nn \\ 
g_h \vOm_{\fC\fA}^{\fC} & = & (\p_{\fM}\chi_{\fN}^{\;\fC} - \p_{\fN}\chi_{\fM}^{\;\fC} )\chih_{\fA}^{\;\fM}\chih_{\fC}^{\;\fN} ; \; \; \bet_{\fC} \phi = (\eth_{\fC} - \mW_{\fC} ) \phi \, , 
\ee
for the covariant derivative.

In such a hidden coordinate formalism, the fundamental gauge interactions are described by the hyper-spin gravigauge field $\vOm_{\fC}^{\fA\fB}$ and the hyper-spin homogauge field $\cH_{\fC}^{\fA\fB}$.  The hyper-spin gravigauge field $\vOm_{\fC}^{\fA\fB}$ is determined by the gauge-type hyper-gravifield $\chi_{\fM}^{\; \fA}$ as shown in Eq.(\ref{HSGGF1}). The dynamics of $\vOm_{\fC}^{\fA\fB}$ is described by the Ricci-like scalar tensor $\cR \equiv \eta^{\fC\fD} \cR_{\fC\fD} = -\eta^{\fC\fD} \eta_{\fA}^{\, \fB} \cR_{\fC\fB\fD}^{\;\fA} $. Therefore, the gravitational interaction in the hidden coordinate formalism is characterized by an analogous conformal scaling gauge invariant Einstein-Hilbert type action governed by the hyper-spin gauge symmetry. Namely, in the locally flat hyper-gravifield spacetime, the basic gravitational interaction is formulated as the gauge interaction of the hyper-spin gravigauge field $\vOm_{\fC}^{\fA\fB}$, which corroborates the general {\it gauge-gravity correspondence}.  Such a gauge-gravity correspondence distinguishes the so-called gauge-gravity duality associated with the holographic idea\cite{AC1,AC2,AC3}. Namely a gravity theory defined on a bulk D-dimensional spacetime can be equivalent to a gauge theory defined on a (D-1)-dimensional spacetime that forms the bulk boundary. The original conjecture made in\cite{AC1} concerns an anti-de Sitter spacetime being equivalent to a conformal field theory on its boundary, i.e., the so-called AdS/CFT conjecture. More precisely, it was conjectured in Ref.\cite{AC1} that the type IIB string theory on a $AdS_5\times S^5$ background is equivalent to $N=4$ four-dimensional super-Yang Mills field theory.

\section{ Gravitational gauge-geometry duality and hyperunified field theory with non-commutative geometry in locally flat hyper-gravifield spacetime }

We have shown the gravity-geometry correspondence in the hidden gauge formalism and the gauge-gravity correspondence in the hidden coordinate formalism. The analogy between two basic actions of Eqs.(\ref{HUTaction3}) and (\ref{HUTactionGG1}) or  Eqs.(\ref{HUTaction4}) and (\ref{HUTactionGG2}) displays the gauge-geometry correspondence. We shall further investigate their correlations and reveal the {\it gravitational gauge-geometry duality}. In general, the locally flat hyper-gravifield spacetime is viewed as a dynamically emerged spacetime, which is characterized by a non-commutative geometry.

\subsection{Gravitational gauge-geometry duality with flowing unitary gauge}

To make an explicit comparison, let us put together the basic action Eq.(\ref{HUTaction4}) expressed in the hidden gauge formalism and the basic action of Eq.(\ref{HUTactionGG2}) formulated in the hidden coordinate formalism,
\be
\label{HUTactionGGG}
& & I_H \equiv  \int [d\hx]\; \chi  \phi^{D_h-4} \,  \{ \bar{\vPsi}\dg^{\fM} [ i \p_{\fM}  +  (\varXi_{\fM}^{\fP\fQ} + g_h\cH_{\fM}^{\fP\fQ} ) \frac{1}{2}\varSigma_{\fP\fQ} ] \vPsi  \nonumber \\
& &-\frac{1}{4} \tilde{\chi}^{\fM\fN\1 \fM'\fN'\1\fP\fQ\1\fP'\fQ'}  \cQ_{\fM\fN\fP\fQ} \cQ_{\fM'\fN'\fP'\fQ'} - \beta_E\1 \phi^4 \nn \\
& &+   \alpha_E  (\,\phi^2\chih^{\fM\fN}\cR_{\fM\fN} - (D_h-1)(D_h-2)\p_{\fM}\phi\p^{\fM}\phi )   \nn \\
& &+   \frac{1}{2} d_{\fM} \phi d^{\fM}\phi -\frac{1}{4} {\cal W}_{\fM\fN} {\cal W}^{\fM\fN}  + 2g_h\alpha_E\phi^2\hat{\nabla}_{\fM}(\cH_{\fN}^{\fN\fM} ) \}\,  \nn \\
& &\equiv  \int [\dchi]  \phi^{D_h-4}\, \{  \frac{1}{2} \bar{\vPsi} \vGa^{\fC} [ i \eth_{\fC}  +  g_h (\vOm_{\fC}^{\fA\fB} + \cH_{\fC}^{\fA\fB} ) \frac{1}{2}\varSigma_{\fA\fB} ] \vPsi  \nonumber \\
& & - \frac{1}{4} \tilde{\eta}^{\fC\fD\1 \fC'\fD'\1\fA\fB\1\fA'\fB'}  \cQ_{\fC\fD\fA\fB} \cQ_{\fC'\fD'\fA'\fB'}  - \beta_E\1 \phi^4  \nonumber \\
& & +  \alpha_E (\phi^2g_h\eta^{\fC\fD}\cR_{\fC\fD} - (D_h-1)(D_h-2) \eth_{\fC}\phi\eth^{\fC}\phi )  \nn \\
& & +  \frac{1}{2}\bet_{\fC} \phi \bet^{\fC}\phi  - \frac{1}{4} {\cal W}_{\fC\fD} {\cal W}^{\fC\fD} + 2g_h\alpha_E\phi^2 \hat{\cD}_{\fC}\cH_{\fD}^{\fD\fC} \, \} \, .  \nn
\ee

In terms of the hidden gauge formalism,  the basic action of hyperunified field theory emerges a general linear group symmetry GL($D_h$,R) under the general coordinate transformations and the hyper-spin gauge symmetry SP(1,$D_h$-1) becomes a hidden symmetry. The gravitational interactions are characterized by the dynamics of the Riemannian geometry in a curved hyper-spacetime and all the bosonic fields couple to the Goldstone-like hyper-gravimetric field $\chi_{\fM\fN}$, but the hyper-spinor field only interacts with the Goldstone-like hyper-gravifield $\chi_{\fM}^{\;\fA}$. To ensure the gravitational interactions for all the basic fields be described by a common basic gravitational field, it requires us to take the gauge fixing condition to be the unitary gauge, so that the gauge-type hyper-gravifield $\chi_{\fM\fA}(\hx)$ is made to be a symmetric Goldstone-like field, i.e., $\chi_{\fM\fA}(\hx)=\chi_{\fA\fM}(\hx)$. In such a unitary gauge, the symmetric Goldstone-like hyper-gravimetric field can be expressed as a square of the symmetric Goldstone-like hyper-gravifield, i.e., $\chi_{\fM\fN} \equiv (\chi_{\fM\fA})^2 = \chi_{\fM\fA}\eta^{\fA\fB} \chi_{\fB\fN} $. 

In terms of the hidden coordinate formalism, the basic action of hyperunified field theory possesses the explicit hyper-spin gauge symmetry SP(1,$D_h$-1) and the general linear group symmetry GL($D_h$,R) is a hidden symmetry. The gravitational interactions are described by the hyper-spin gravigauge field $\vOm_{\fC}^{\fA\fB}$, which is characterized by the gauge-type hyper-gravifield $\chi_{\fM}^{\; \fA}$. 

The similarity of two formalisms reflects the {\it gauge-geometry correspondence}. In the hidden gauge formalism, we shall further show how to make such a gauge fixing procedure hold at any point in hyper-spacetime. Namely, for every general coordinate transformation that is defined as an arbitrary reparametrization of coordinate systems, $\hx' \equiv \hx'(\hx)$, we can always carry out a specific hyper-spin gauge transformation associated with a distinct local reparametrization at every point in hyper-spacetime. Such a gauge fixing procedure can be realized with two-step transformations. Namely, considering a general coordinate transformation,
\be
 & & dx^{\fM} \to dx^{'\fM} = T^{\fM}_{\;\; \, \fN} \, dx^{\fN}\, , \quad \p_{\fM} \to \p'_{\fM} =  T_{\fM}^{\;\; \fN} \p_{\fN}\, , \nn \\
 & & T_{\fM}^{\;\, \fN}\, ,  \; T^{\fM}_{\; \; \, \fN} \in GL(D_h, R) \, , 
 \ee
as the first step, and making a hyper-spin gauge transformation at point $\hx' \equiv \hx'(\hx)$,
 \be
& &\vGa^{\fA} \to S'(\Lambda') \vGa^{\fA} S^{'-1}(\Lambda') =  \Lambda^{'\fA}_{\; \; \fB}(\hx') \vGa^{\fB}\, , \nn \\
& & \Lambda^{'\fA}_{\; \; \fB}(\hx') \in SP(1,D_h\mbox{-}1)\, , 
\ee
as the second step. So that the hyper-gravifield at every point in hyper-spacetime is made to be symmetric,
\be
& & \chi_{\fM\fA}(\hx) = \chi_{\fA\fM}(\hx) \to \chi'_{\fM\fA}(\hx') = T_{\fM}^{\;\; \fN} \, \chi_{\fN\fA}(\hx)  \, , \nn \\
& & \chi'_{\fM\fA}(\hx') \neq \chi'_{\fA\fM}(\hx')  \to \nn \\
& & \chi^{''}_{\fM\fA}(\hx') = \Lambda_{\fA}^{'\; \fB}(\hx')\, \chi'_{\fM\fB}(\hx') = \chi^{''}_{\fA\fM}(\hx') \, .
\ee

As the symmetry groups GL($D_h$, R) and SP(1,$D_h$-1) do not act as a direct product group, the gauge fixing has to associate with an arbitrary reparametrization of coordinate systems to make the specific hyper-spin gauge transformation in order to obtain a symmetric hyper-gravifield. Therefore, in the hyper-gravifield fiber bundle structure with considering general transformations under the symmetry groups GL($D_h$, R) and SP(1,$D_h$-1), such a gauge fixing procedure leads a running gauge fixing condition to ensure the symmetric hyper-gravifield at arbitrary point of coordinate systems in hyper-spacetime, which may be referred to by the {\it flowing unitary gauge}. 

Let us now discuss and outline some general properties of hyperunified field theory based on various formalisms. 

Geometrically, the general action of hyperunified field theory should hold in any coordinate system as it is constructed with the postulate of coordinate independence. The basic action Eq.(\ref{HUTaction4}) of hyperunified field theory formulated in the hidden gauge formalism displays an emergent general coordinate transformation invariance. The gravitational interaction in hyper-spacetime of coordinates is described by the dynamics of the Riemannian geometry in a curved hyper-spacetime, which is characterized by the Christoffel symbols $\vGa_{\fM\fN}^{\fP}$ defined through the Goldstone-like hyper-gravimetric field $\chi_{\fM\fN}$. Such a gravitational interaction in hyper-spacetime is governed by the general linear group symmetry, 
\be
G_S =  \mbox{GL(}D_h,\mbox{R)} \, , \nn
\ee
which is a real Lie group of dimension $N_D= D_h^2 $ and consists of matrices that have non-zero determinant. Such a general linear group symmetry lays the foundation of general relativity in four dimensional spacetime.  

Locally, at every point of hyper-spacetime, the general action of hyperunified field theory is invariant under the hyper-spin gauge transformations as it is constructed with the postulate of gauge-invariance in the locally flat hyper-gravifield spacetime. The hyper-spin homogauge field $\cH_{\fM}^{\fA\fB}$ and the hyper-spin gravigauge field $\vOm_{\fM}^{\fA\fB}$ characterized by the gauge-type hyper-gravifield $\chi_{\fM}^{\; \fA}$ are adopted to describe the fundamental gauge interactions governed by the hyper-spin gauge symmetry, 
\be
G_S =  \mbox{SP(1},D_h\mbox{-1)} \, , \nn
\ee
which is known as the gauge principle; it is applied to build gauge theories within the framework of QFT. 

In general, there appears to have a bimaximal local symmetry in the hyper-gravifield fiber bundle structure of biframe hyper-spacetime, 
\be
G_S =  \mbox{GL(}D_h, \mbox{R)} \Join \mbox{SP(1},D_h\mbox{-1)} \, ,
\ee
which is viewed as a joined Lie group as two local symmetries cannot be operated as a direct product group. Such a joined bimaximal symmetry can be utilized as a symmetry principle to construct a general action of hyperunified field theory. One is the maximal general linear group symmetry GL($D_h$, R) operating on the coordinates in the curved Riemannian hyper-spacetime, and the other is the maximal hyper-spin gauge symmetry SP(1,$D_h$-1) operating on the hyper-spinor field and gauge field in the locally flat hyper-gravifield spacetime.

In practice, as the general action of hyperunified field theory is built based on the postulates of gauge invariance and coordinate independence in the locally flat hyper-gravifield spacetime, the resulting action should hold for all systems of coordinates in hyper-spacetime. Namely, the postulates of gauge invariance and coordinate independence are more general and fundamental, we can always choose the globally flat Minkowski hyper-spacetime as a base spacetime in the hyper-gravifield fiber bundle structure, so that the general action of hyperunified field theory possesses actually the maximal global Poincar\'e symmetry PO(1,$D_h$-1) and the maximal hyper-spin gauge symmetry SP(1,$D_h$-1), 
\be
G_S =  \mbox{PO(1},D_h\mbox{-1)}\Join \mbox{SP(1},D_h\mbox{-1)}\, .
\ee
Such a joined bimaximal symmetry group has a total dimension $N_D = 2\times D_h(D_h-1)/2 + D_h = D_h^2$, which has the same dimension as the symmetry group GL($D_h$,R). The Lorentz group SO(1,$D_h$-1) is a subgroup of GL($D_h$, R), i.e., SO(1,$D_h$-1)$\in$GL($D_h$, R). 

Physically, either the hyper-spin gauge symmetry group SP(1,$D_h$-1) that governs the dynamics of the hyper-spinor field and gauge fields or the emergent general linear symmetry group GL($D_h$, R) that describes the dynamics of the Riemannian geometry of hyper-spacetime, it provides a symmetry principle to construct the general action of hyperunified field theory in hyper-spacetime. Such a local gauge symmetry represents each physically distinct configuration of the system as an equivalence class of detailed local field configurations. Namely, for any two detailed configurations that are in the same equivalence class, they are related by a gauge transformation. As physical observables should be gauge independent, they can only be obtained with a reasonable prescription for removing those unphysical degrees of freedom. Thus a gauge fixing is required to provide a mathematical procedure for dealing with the redundant degrees of freedom in field variables. 

Gravitationally, the {\it flowing unitary gauge} enables us to take the symmetric Goldstone-like hyper-gravifield as a basic gravitational field at every point in hyper-spacetime to characterize gravitational interactions for all basic fields.  The concept of the {\it flowing unitary gauge} allows us to take all points in hyper-spacetime to be equivalent for making the gauge fixing be the unitary gauge. Therefore, in a physically meaningful unitary gauge, the basic action of hyperunified field theory in the hidden gauge formalism possesses only the global Poincar\'e symmetry ,
\be
G_S =  PO(1,D_h-1) \equiv \mbox{P}^{1,D_h\mbox{-}1} \ltimes \mbox{SO(1},D_h\mbox{-1)}; \; \quad 
\mbox{SP(1},D_h\mbox{-1)} \cong \mbox{SO(1},D_h\mbox{-1)} \, .
\ee   

From the above analysis on the general properties of hyperunified field theory, we should conclude that the dynamics of  gravitational interactions can be described either as a dynamical gauge interaction of the gauge-type hyper-gravifield or equivalently as a dynamical Riemannian geometry of the symmetric Goldstone-like hyper-gravifield in a general flowing unitary gauge. Such an equivalent description reveals the {\it gravitational gauge-geometry duality}, which is attributed to the gravitational origin of gauge symmetry in hyper-spacetime.

It becomes clear that the hyper-spacetime Poincar\'e symmetry plays a basic role as a fundamental symmetry in hyperunified field theory, which essentially demands the existence of the globally flat Minkowski hyper-spacetime to define the inertial reference frames and act as a vacuum base spacetime for describing the kinematics and dynamics of all basic fields and deriving the conservation laws of all symmetries.

 \subsection{Hyperunified field theory with non-commutative geometry in locally flat hyper-gravifield spacetime }
 
Geometrically, the locally flat hyper-gravifield spacetime ${\bf G}_h$ is associated with a non-commutative geometry due to the nontrival commutation relation of the non-coordinate field basis $\{\eth_{\fC}\} $,
 \be \label{NCG}
 & & [\eth_{\fC}\, , \eth_{\fD} ] = f_{\fC\fD}^{\fA}\1 \eth_{\fA}  \, , \quad f_{\fC\fD}^{\fA} = g_h( \vOm_{\fC\1 \fD}^{\fA} - \vOm_{\fD\1 \fC}^{\fA}) \equiv g_h\vOm_{[\fC \fD]}^{\fA} \, ,
 \ee
which generates a special Lie algebra. The structure factor $f_{\fC\fD}^{\fA}$ is no longer a constant and it is characterized by the {\it hyper-spin gravigauge field} $\vOm_{\fC }^{\fA\fB}$. Therefore, to explore the geometric properties of the locally flat hyper-gravifield spacetime ${\bf G}_h$, one needs to investigate the dynamics of the hyper-spin gravigauge field. 

In the locally flat hyper-gravifield spacetime ${\bf G}_h$, the gauge interactions are characterized by both the hyper-spin gravigauge field $\vOm_{\fC}^{\fA\fB}$ and the hyper-spin homogauge field $\cH_{\fC}^{\fA\fB}$.  The fundamental interaction of the hyper-spinor field are described solely by the hyper-spin gauge field $\cA_{\fC}^{\fA\fB} \equiv \vOm_{\fC}^{\fA\fB} + \cH_{\fC}^{\fA\fB}$ due to the fact that the gravity is treated on the same footing as the other forces. It is natural to express the basic action of hyperunified field theory in terms of the hyper-spin gauge field $\cA_{\fC}^{\fA\fB}$ and the hyper-spin gravigauge field $\vOm_{\fC}^{\fA\fB}$. 

The basic action of hyperunified field theory Eq.(\ref{HUTactionGG2}) can be rewritten into the following form, 
\be
\label{HUTactionGG3}
I_H & \equiv & \int [\dchi ]\; \kL = \int [\dchi] \,  \phi^{D_h-4} \{\frac{1}{2} \bar{\vPsi} \varGamma^{\fC} ( i \eth_{\fC} + g_h\cA_{\fC} )  \vPsi \nonumber \\
& - & \frac{1}{4}  \tilde{\eta}^{\fC\fD\1 \fC'\fD'\1 \fA\fB\1 \fA'\fB' } \cF_{\fC\fD\fA\fB}\cF_{\fC'\fD' \fA'\fB'}   \nn \\
& + & \alpha_E  \phi^2 [g_h^2 (\,\vOm_{\fC\fA\fB}\vOm^{\fC\fA\fB} -\frac{1}{2} \vOm_{[\fC\fD]\fA}\vOm^{[\fC\fD]\fA} - \vOm_{\fC}^{\fC\fB}\vOm_{\fD\fB}^{\fD} )  \nn \\
& - & 2(D_h-2) \vOm_{\fD}^{\fD\fC} \eth_{\fC}\ln\phi  - (D_h-1)(D_h-2) \eth_{\fC}\ln\phi\1 \eth^{\fC}\ln\phi 
 \, ]   \nn \\
 & + &  \frac{1}{2}\bet_{\fC} \phi \bet^{\fC}\phi   - \frac{1}{4}{\cal W}_{\fC\fD} {\cal W}^{\fC\fD} - \beta_E\1 \phi^4  +  2 g_h\alpha_E\phi^2 \hat{\cD}_{\fC}\cA_{\fD}^{\fD\fC} \, \}\, ,
\ee
with the definitions
\be
\cF_{\fC\fD}^{\fA\fB} & = & \tilde{\cD}_{\fC} \cA_{\fD}^{\fA\fB} - \tilde{\cD}_{\fD} \cA_{\fC}^{\fA\fB} + g_h ( \cA_{\fC \fE}^{\fA} \cA_{\fD}^{\fE \fB} -  \cA_{\fD \fE}^{\fA} \cA_{\fC}^{\fE \fB} ) \, , \nn \\
\tilde{\cD}_{\fC} \cA_{\fD}^{\fA\fB} & = & \eth_{\fC} \cA_{\fD}^{\fA\fB}  - g_h\vOm_{\fC \fD}^{\fE} \cA_{\fE}^{\fA \fB} \, , \nn \\
\hat{\cD}_{\fC}\cA_{\fD}^{\fD\fC}  & \equiv&  \cD_{\fC} \cA_{\fD}^{\fD\fC} + (D_h-2)\fS_{\fC}  \cA_{\fD}^{\fD\fC} \nn \\
& = & \eth_{\fC}  \cA_{\fD}^{\fD\fC} + g_h\vOm_{\fC\fA}^{\fC} \cA_{\fD}^{\fD\fA} + (D_h-2)\fS_{\fC}  \cA_{\fD}^{\fD\fC}  \, .
\ee
In obtaining the above expression, we have used the equality,
\be
& & \cR_{\fC\fD}^ {\fA\fB}  +  \cD_{[\fC}\cH_{\fD]}^{\fA\fB}  = g_h\vOm_{[\fC}^{\fA\fA'} \vOm_{\fD]}^{\fB\fB'} \eta_{\fA'\fB'} + \cD_{[\fC}\cA_{\fD]}^{\fA\fB} \, , \nn \\
& & \vOm_{[\fC}^{\fA\fA'} \vOm_{\fD]}^{\fB\fB'} \equiv \vOm_{\fC}^{\fA\fA'} \vOm_{\fD}^{\fB\fB'}  - \vOm_{\fD}^{\fA\fA'} \vOm_{\fC}^{\fB\fB'}  \, .
\ee
In the first equality, each term on the left-hand side is gauge covariant, on the right-hand side, only the combination of two terms becomes gauge covariant.  

In the above basic action of hyperunified field theory,  the hyper-spin gauge field $\cA_{\fC}^{\fA\fB} \equiv \vOm_{\fC}^{\fA\fB} + \cH_{\fC}^{\fA\fB}$ has the same gauge transformation property as the hyper-spin gravigauge field $\vOm_{\fC}^{\fA\fB}$ which characterizes a non-commutative geometry of the locally flat hyper-gravifield spacetime. In general, the hyper-spin gauge symmetry and the general linear group symmetry cannot be treated as a direct product group\cite{PHO} as they operate on two vector spacetimes in the hyper-gravifield fiber bundle structure of biframe hyper-spacetime. It is helpful to adopt the concept of biframe spacetime and treat two spacetimes in parallel to describe the gauge and gravitational interactions in the light of a non-commutative geometry\cite{NCG}.

\section{ Basic properties of hyperunified field theory and gravitational equations of Einstein-like and beyond with fundamental mass scale }

We have formulated the basic action of hyperunified field theory in both the hidden gauge formalism and the hidden coordinate formalism as shown in Eqs.(\ref{HUTaction4}) and (\ref{HUTactionGG2}), respectively. Such formalisms enable us to demonstrate both the {\it gravity-geometry} and {\it gauge-gravity} correspondences and to corroborate the {\it gravitational gauge-geometry duality}. We shall further analyze the basic properties of hyperunified field theory and address the issue on the fundamental mass scale in hyperunified field theory based on the conformal scaling gauge symmetry.  We then derive the gauge gravitational equation with the conserved current and present the geometric gravitational equations of Einstein-like and beyond that correspond to the symmetric and antisymmetric hyper-stress energy momentum tensors, respectively.

\subsection{Basic properties of hyperunified field theory within the framework of QFT }

Relying on the basic action of hyperunified field theory Eq.(\ref{HUTactionGG3}) which results from the gauge symmetry principle in the locally flat hyper-gravifield spacetime,  we are able to obtain the basic action of hyperunified field theory within the framework of QFT by projecting Eq.(\ref{HUTactionGG3}) into the coordinate hyper-spacetime by applying for the Goldstone-like hyper-gravifield,
\be \label{HUTactionQFT}
I_H  & \equiv & \int [d\hx ]\chi \1 \kL  =    \int  [d\hx ] \chi  \phi^{D_h-4} \{  \chih^{\fM\fN} \frac{1}{2} \bar{\vPsi} \varGamma_{\fA} \chi^{\;\fA}_{\fM} ( i \p_{\fN} + g_h\1 \cA_{\fN} )  \vPsi \nonumber \\
&  -&   \frac{1}{4} (\tilde{\chi}^{\fM\fN\1 \fM'\fN'}_{\fA\fB\1 \fA'\fB' } \cF_{\fM\fN}^{\fA\fB}\cF_{\fM'\fN'}^{ \fA'\fB'}  + \chih^{\fM\fM'}\chih^{\fN\fN'}{\cal W}_{\fM\fN} {\cal W}_{\fM'\fN'} )\nn \\
&  + & \alpha_E \phi^2  \frac{1}{4} \tilde{\chi}^{\fM\fN\fM'\fN'}_{\fA\fA'} \fG_{\fM\fN}^{\fA}\fG_{\fM'\fN'}^{\fA'}    + \frac{1}{2}\chih^{\fM\fN} d_{\fM} \phi d_{\fN}\phi - \beta_E\phi^4  \}  \nn \\
& + & 2\alpha_E g_h \p_{\fM}(\chi\phi^{D_h-2} \cA_{\fN}^{\fN\fM} ) \, ,
\ee
where we have used the following conformal scaling gauge invariant definitions,
\be
 \cF_{\fM\fN}^{\fA\fB} & = & \p_{\fM} \cA_{\fN}^{\fA\fB} - \p_{\fN}\cA_{\fM}^{\fA\fB} + g_h( \cA_{\fM \fC}^{\fA} \cA_{\fN}^{\fC \fB} -  \cA_{\fN \fC}^{\fA} \cA_{\fM}^{\fC \fB} ),\nn \\
 \fG_{\fM\fN}^{\fA} & = & \hat{\p}_{\fM} \chi_{\fN}^{\; \fA} - \hat{\p}_{\fN}\chi_{\fM}^{\; \fA}; \;\; \hat{\p}_{\fM}\equiv  \p_{\fM} +  \p_{\fM}\ln \phi,   \nn \\
 d_{\fM}\phi & = & (\p_{\fM} - g_wW_{\fM})\phi  ;\;\; \cA_{\fN}^{\fN\fM}  = - \chih_{\fA}^{\fM} \chih_{\fB}^{\; \fN} \cA_{\fN}^{\fA\fB}\, .
\ee

The tensor factors are defined as 
\be \label{tensors}
& & \tilde{\chi}^{\fM\fN\fM'\fN'}_{\fA\fB\fA'\fB'} \equiv \chih_{\fC}^{\;\fM}\chih_{\fD}^{\;\fN} \chih_{\fC'}^{\;\fM'}\chih_{\fD'}^{\;\fN'}\tilde{\eta}^{\fC\fD\fC'\fD'}_{\fA\fB\fA'\fB'}\, , \nn \\
& & \tilde{\chi}^{\fM\fN\fM'\fN'}_{\fA\fA'} \equiv \chih_{\fC}^{\;\fM}\chih_{\fD}^{\;\fN} \chih_{\fC'}^{\;\fM'}\chih_{\fD'}^{\;\fN'}\tilde{\eta}^{\fC\fD\fC'\fD'}_{\fA\fA'}\, , 
\ee
where the constant tensor factor $\tilde{\eta}^{\fC\fD\1 \fC'\fD'}_{ \fA\fB\1\fA'\fB'}$ is given by 
\be \label{tensor6}
\tilde{\eta}^{\fC\fD\1 \fC'\fD'}_{ \fA\fB\1\fA'\fB'} & \equiv & \frac{1}{4}\1 \{\, [\, \eta^{\fC\fC'} \eta_{\fA\fA'} (\eta^{\fD\fD'} \eta_{\fB\fB'} - 2  \eta^{\fD}_{\fB'} \eta^{\fD'}_{\fB}) + \eta^{(\fC,\fC'\leftrightarrow\fD, \fD' )}\, ]+  \eta_{(\fA,\fA'\leftrightarrow \fB,\fB' ) }  \, \} \nn \\
&  + & \frac{1}{4}\alpha_W\, \{\, [\, (\eta^{\fC}_{\fA'} \eta^{\fC'}_{\fA} - 2\eta^{\fC\fC'} \eta_{\fA\fA'}  )  \eta^{\fD}_{\fB'}  \eta^{\fD'}_{\fB} 
+  \eta^{(\fC,\fC'\leftrightarrow\fD, \fD' )} \, ] +  \eta_{(\fA,\fA'\leftrightarrow \fB,\fB' ) }  \,  \}    \nn \\
& + & \frac{1}{2}\beta_W\, \{\, [\, (\eta_{\fA\fA'}  \eta^{\fC\fC'} - \eta^{\fC'}_{\fA}\eta^{\fC}_{\fA'}) \eta^{\fD}_{\fB} \eta^{\fD'}_{\fB'} + \eta^{(\fC,\fC'\leftrightarrow\fD, \fD' )}\, ]+  \eta_{(\fA,\fA'\leftrightarrow \fB,\fB' )} \, \}  \, ,  
\ee
which may be referred to by a {\it general conformal invariance  tensor factor}. The constant tensor $\tilde{\eta}^{\fC\fD\fC'\fD'}_{\fA\fA'} $ is defined as  
\be \label{tensor7}
 \tilde{\eta}^{\fC\fD\fC'\fD'}_{\fA\fA'}  & \equiv &    \eta^{\fC\fC'} \eta^{\fD\fD'} \eta_{\fA\fA'}  
+  \eta^{\fC\fC'} ( \eta_{\fA'}^{\fD} \eta_{\fA}^{\fD'}  -  2\eta_{\fA}^{\fD} \eta_{\fA'}^{\fD'}  ) \\
& + & \eta^{\fD\fD'} ( \eta_{\fA'}^{\fC} \eta_{\fA}^{\fC'} -2 \eta_{\fA}^{\fC} \eta_{\fA'}^{\fC'} ) \, , 
\ee
which may be called a {\it general hyper-spin gauge invariance tensor factor}. This is because such a tensor factor ensures the dynamic term $\fG_{\fM\fN}^{\fA}\fG_{\fM'\fN'}^{\fA'}$ to be hyper-spin gauge invariant.  

In such a basic action, the relevant basic fields include the hyper-spinor field $\varPsi$, the hyper-spin gauge field $\cA_{\fM}^{\fA\fB}$, the gauge-type hyper-gravifield $\chi_{\fM}^{\; \fA}$, the conformal scaling gauge field $W_{\fM}$ and the scaling scalar field $\phi$. In general, Eq.(\ref{HUTactionQFT}) has a joined bimaximal local symmetry,
\be
G_{S} =  \mbox{GL(}D_h,\mbox{R)} \Join \mbox{SP(1},D_h\mbox{-1)}\times GS(1) \, ,
\ee
where the local symmetry GL($D_h$, R) emerges as a hidden general linear group symmetry. Such a hidden symmetry appears due to the fact that all the gauge field tensors in the basic action Eq.(\ref{HUTactionQFT}) are  antisymmetry tensors in hyper-spacetime. It involves no explicit interactions associated with the Christoffel symbols or Levi-Civita connection $\vGa_{\fM\fN}^{\fP}$ which is utilized to characterize the Riemannian geometry.  Such a property always allows us to choose the globally flat Minkowski hyper-spacetime as a base spacetime, so that the global Poincar\'e symmetry PO(1,$D_h$-1) becomes a basic symmetry. 

Essentially, the basic action of Eq.(\ref{HUTactionQFT}) is considered to be governed by the joined bimaximal global and local symmetry,
\be
G_{S} =  \mbox{PO(1},D_h\mbox{-1)} \times S(1) \Join \mbox{SP(1},D_h\mbox{-1)} \times \mbox{SG(1)}\, ,
\ee 
which enables us to establish hyperunified field theory within the framework of QFT.

The conformal scaling gauge symmetry SG(1) allows us to make a gauge fixing, so that the determinant of the hyper-gravifield $\chi_{\fM}^{\; \fA}$ can always be fixed to be unit by a specific conformal scaling gauge transformation $\xi_u^{D_h}(\hx)=\chi(\hx)$, i.e., 
\be
& & \chi_{\fM}^{\; \fA}(\hx) \to \chi_{\fM}^{(u)\fA}(\hx) = \xi_u^{-1}(\hx) \chi_{\fM}^{\; \fA}(\hx) \, , \nn \\
& & \chi^{(u)}= \det \chi_{\fM}^{(u)\fA} = 1 \, ,
\ee
which is referred to by a {\it unitary basis} for convenience of mention.

In such a {\it unitary basis}, we can rewritten the basic action Eq.(\ref{HUTactionQFT}) as follows:
\be \label{HUTactionQFTUB}
 I_H^{(u)} & \equiv &  \int [d\hx ] \; \kL ^{(u)} =    \int [d\hx ]  \phi^{D_h-4} \{ \,  \chih^{\fM\fN} \frac{1}{2}  \bar{\vPsi}  \varGamma_{\fA} \chi^{\; \fA}_{\fM} ( i \p_{\fN} + g_h\1 \cA_{\fN} )  \vPsi \nonumber \\
& - &   \frac{1}{4}( \tilde{\chi}^{\fM\fN\1 \fM'\fN'}_{\fA\fB\1 \fA'\fB' } \cF_{\fM\fN}^{\fA\fB}\cF_{\fM'\fN'}^{ \fA'\fB'}+ \chih^{\fM\fM'}\chih^{\fN\fN'} {\cal W}_{\fM\fN} {\cal W}_{\fM'\fN'} )  \nn \\
&  + & \alpha_E \phi^2 \frac{1}{4}\tilde{\chi}^{\fM\fN\fM'\fN'}_{\fA\fA'} \fG_{\fM\fN}^{\fA}\fG_{\fM'\fN'}^{\fA'} + \frac{1}{2}\hat{\chi}^{\fM\fN} d_{\fM} \phi d_{\fN}\phi - \beta_E\phi^4 \, \} \nn \\
&  + & 2\alpha_E g_h \p_{\fM}(\phi^{D_h-2} \cA_{\fN}^{\fN\fM} )   \, ,
\ee
where we have omitted for simplicity the superscript $(u)$ in the unitary basis. The basic action of Eq.(\ref{HUTactionQFTUB}) describes hyperunified field theory within the framework of QFT in the globally flat Minkowski hyper-spacetime. In such a framework, the dynamics of the gravitational interactions is truly characterized by the gauge-gravity theory through the conventional gauge interactions of the gauge-type hyper-gravifield $\chi_{\fM}^{\; \fA}$ in hyper-spacetime. 

Note that the last term in the basic action, Eq.(\ref{HUTactionQFTUB}) or Eq.(\ref{HUTactionQFT}), reflects the surface effect in $D_h$-dimensional hyper-spacetime. In the trivial case with the boundary conditions: $\phi(\hx)\to 0$ and/or $\cA_{\fN}^{\fN\fM}(\hx)\to 0$ as $\hx\to \infty$, the surface term can be ignored.

\subsection{Fundamental mass scale in hyperunified field theory with scaling gauge fixing}

The conformal scaling gauge symmetry in hyperunified field theory enables us to set an appropriate gauge fixing, so that we can always make a specific conformal scaling gauge transformation to transfer the scalar field into a constant. It motivates us to further postulate that {\it there should exist a fundamental mass scale in a conformal scaling gauge-invariant hyperunified field theory}. 

Let us now consider a typical gauge fixing via a specific conformal scaling gauge transformation $\xi_e(\hx)$,
\be
\chi_{\fM}^{\; \fA}(\hx) \to \chi_{\fM}^{(e)\fA}(\hx) = \xi_e^{-1}(\hx) \chi_{\fM}^{\; \fA}(\hx) \, , 
\ee
so that the {\it scaling scalar field} $\phi$ is transformed into a mass scale,
\be
\phi(\hx) \to \phi^{(e)}(\hx) = \xi_e(\hx)\phi(\hx) = M_S\, . 
\ee
We may refer to such a gauge fixing byan {\it Einstein-type basis}. The mass scale $M_S$ plays a role as a {\it fundamental mass scale} in hyperunified field theory.

In such an Einstein-type basis, the basic action Eq.(\ref{HUTactionQFT}) can be expressed as
\be \label{HUTactionQFTEB}
I_H^{(e)} & \equiv &  M_S^{D_h-4} \int [d\hx]  \chi\1 \kL^{(e)} \nn \\
& = &  M_S^{D_h-4} \int [d\hx] \chi\{ \chih^{\fM\fN} \frac{1}{2} \bar{\vPsi} \varGamma_{\fA} \chi^{\;\fA}_{\fM} ( i \p_{\fN} + g_h\1 \cA_{\fN} )  \vPsi \nonumber \\
&  - &  \frac{1}{4} ( \tilde{\chi}^{\fM\fN\1 \fM'\fN'}_{\fA\fB\1 \fA'\fB' } \cF_{\fM\fN}^{\fA\fB}\cF_{\fM'\fN'}^{ \fA'\fB'} + 
\chih^{\fM\fM'}\chih^{\fN\fN'} {\cal W}_{\fM\fN} {\cal W}_{\fM'\fN'}   ) \nn \\
&  + & \alpha_E M_S^2\frac{1}{4} \tilde{\chi}^{\fM\fN\fM'\fN'}_{\fA\fA'} \mG_{\fM\fN}^{\fA}\mG_{\fM'\fN'}^{\fA'} -\beta_E M_S^4     \nn \\
&  + & \frac{1}{2}g_w^2 M_S^{2} \chih^{\fM\fN} W_{\fM}W_{\fN} \}  + 2\alpha_E g_h M_S^{2}\p_{\fM} (\chi \cA_{\fN}^{\fN\fM}) ,
\ee
with 
\be
& & \mG_{\fM\fN}^{\fA} \equiv \p_{\fM} \chi_{\fN}^{\; \fA} - \p_{\fN} \chi_{\fM}^{\; \fA}\, .
\ee
We have omitted for simplicity the superscript $(e)$ for all basic fields. The field strengths $\mG_{\fM\fN}^{\fA}$ ($\fA=1,\ldots, D_h$) of the hyper-gravifield appear like the massless multi-Abelian gauge field strengths, but unlike ordinary multi-Abelian gauge fields, the dynamic term $\mG_{\fM\fN}^{\fA}\mG_{\fM'\fN'}^{\fA'}$ of the hyper-gravifield is associated with the {\it general hyper-spin gauge invariance tensor factor} $\tilde{\chi}^{\fM\fN\fM'\fN'}_{\fA\fA'}$, which concerns a highly nonlinear interaction of the hyper-gravifield.  In such a basis, the conformal scaling gauge field $W_{\fM}$ becomes gravitationally massive.

\subsection{ Gauge gravitational equations with conserved hyper-gravifield current and hyper-stress energy-momentum tensor in hyperunified field theory}

The gravitational gauge-geometry duality provides a useful tool to investigate the gravitational interactions in hyperunified field theory within the framework of QFT. Let us revisit the equation of motion for the gauge-type hyper-gravifield $\chi_{\fM}^{\;\; \fA}$ under the {\it general conformal scaling invariance condition} and the essential {\it gauge massless condition}. From the basic actions in both the Einstein-type basis and the general basis as shown in Eqs.(\ref{HUTactionQFTEB}) and (\ref{HUTactionQFT}), respectively, we obtain the following gravitational equations of motion for the gauge-type hyper-gravifield $\chi_{\fM}^{\;\; \fA}$ in two cases:
\be  \label{GEM}
& &    \p_{\fN}\1 \mG^{\; \fM\fN}_{\fA} = \mJ_{\fA}^{\; \;\fM} \, , \\
& &    \bar{\p}_{\fN}\1 \fG^{\; \fM\fN}_{\fA} = \fJ_{\fA}^{\;\; \fM} \, , \label{GEMG}
\ee
with $\bar{\p}_{\fN} =\p_{\fN} - \p_{\fN}\ln\phi$. We have introduced the definitions, 
\be \label{GEMT}
\mG^{\; \fM\fN}_{\fA} & \equiv & \alpha_E M_S^2 \chi    \tilde{\chi}^{[\fM\fN]\fM'\fN'}_{\fA\fA'}    \mG_{\fM'\fN' }^{\fA'} \, ,  \\
 \fG^{\; \fM\fN}_{\fA}  & \equiv & \alpha_E \phi^{D_h-2} \chi \tilde{\chi}^{[\fM\fN]\fM'\fN'}_{\fA\fA'}    \fG_{\fM'\fN' }^{\fA'} \, ,  
\ee
for the bicovariant tensors,  and 
\be \label{GEMC}
\mJ_{\fA}^{\;\; \fM} & = &  - \chi \hat{\chi}_{\fA}^{\;\;\fM} \kL^e + \chi \hat{\chi}_{\fA}^{\; \; \fP} [\, \frac{1}{2}   \hat{\chi}_{\fA''}^{\;\; \fM}  \bar{\vPsi} \varGamma^{\fA''} i {\mathcal D}_{\fP} \vPsi -  {\cal W}_{\fP\fQ} {\cal W}^{\fM\fQ}   \nonumber \\
& - & \tilde{\chi}^{[\fM\fQ]\fM'\fN'}_{\fA''\fB\; \fA'\fB'}  {\cal F}_{\fP\fQ}^{\fA''\fB} {\cal F}_{\fM'\fN'}^{\fA'\fB'}  + \alpha_E M_S^2\, \tilde{\chi}^{[\fM\fQ]\fM'\fN'}_{\fA''\fA'}  \mG_{\fP\fQ}^{\fA''} \mG_{\fM'\fN'}^{\fA'}  +  g_w^2M_S^2 W_{\fP}W^{\fM}   ] , \\  
 \fJ_{\fA}^{\;\; \fM} & = &  - \chi \hat{\chi}_{\fA}^{\;\;\fM} \kL +  \phi^{D_h-4} \chi \hat{\chi}_{\fA}^{\; \; \fP}  [\, \frac{1}{2}  \hat{\chi}_{\fA''}^{\;\; \fM}  \bar{\vPsi} \varGamma^{\fA''} i {\mathcal D}_{\fP} \vPsi -  {\cal W}_{\fP\fQ} {\cal W}^{\fM\fQ}   \nonumber \\
& - & \tilde{\chi}^{[\fM\fQ]\fM'\fN'}_{\fA''\fB\; \fA'\fB'}  {\cal F}_{\fP\fQ}^{\fA''\fB} {\cal F}_{\fM'\fN'}^{\fA'\fB'}  + \alpha_E\phi^2\, \tilde{\chi}^{[\fM\fQ]\fM'\fN'}_{\fA''\fA'}  \fG_{\fP\fQ}^{\fA''} \fG_{\fM'\fN'}^{\fA'}  + d_{\fP}\phi d^{\fM}\phi   ]  \, , \label{GEMCC}
\ee
for the bicovariant vector currents.  They may be called the {\it hyper-gravifield tensors} and the {\it hyper-gravifield currents}, respectively.

We shall refer to the gravitational equations presented in Eqs.(\ref{GEM}) and (\ref{GEMG}) with the definitions Eqs.(\ref{GEMT})-(\ref{GEMCC}) by the  {\it gauge gravitational equations}. For simplicity, we have omitted the superscript $(e)$ and adopted the same notations for the hyper-gravifield $\chi_{\fM}^{\; \fA}$ in two cases, but we shall keep in mind for their distinguishable features. The antisymmetric tensor factors are defined from Eq.(\ref{tensors}) as follows:
\be
& & \tilde{\chi}^{[\fM\fN]\fM'\fN'}_{\fA\fA'} \equiv \frac{1}{2} [ \tilde{\chi}^{\fM\fN\fM'\fN'}_{\fA\fA'} - \tilde{\chi}^{\fN\fM\fM'\fN'}_{\fA\fA'}], \nn \\
& & \tilde{\chi}^{[\fM\fN]\fM'\fN'}_{\fA\fB\fA'\fB'} \equiv \frac{1}{2} [ \tilde{\chi}^{\fM\fN\fM'\fN'}_{\fA\fB\fA'\fB'} - \tilde{\chi}^{\fN\fM\fM'\fN'}_{\fA\fB\fA'\fB'}] ,
\ee
with tensor factors $\tilde{\chi}^{\fM\fN\fM'\fN'}_{\fA\fB\fA'\fB'}$ and $\tilde{\chi}^{\fM\fN\fM'\fN'}_{\fA\fA'}$ defined in Eq.(\ref{tensors}).

The antisymmetry property of the {\it hyper-gravifield tensors} $ \mG^{\; \fM\fN}_{\fA}=-\mG^{\; \fN\fM}_{\fA} $ and $\fG^{\; \fM\fN}_{\fA}=-\fG^{\; \fN\fM}_{\fA}$ leads to the conserved {\it hyper-gravifield currents},
\be \label{CC}
&& \p_{\fM}\mJ_{\fA}^{\; \fM} = 0\, , \\
& &  \bar{\p}_{\fM}\fJ_{\fA}^{\; \fM} = 0\, ,
\ee
which implies that the hyper-gravifield does behave as a gauge field in the basic action of hyperunified field theory expressed within the framework of QFT.

The total {\it hyper-stress energy-momentum tensor} $\cT_{\fM}^{\; \fN}$ is related to the conserved hyper-gravifield currents as follows, 
\be
\cT_{\fM}^{\; \fN}   =   \fJ_{\fA}^{\; \fN} \chi_{\fM}^{\;\fA} =  M_S^{D_h-4} \mJ_{\fA}^{\; \fN} \chi_{\fM}^{\;\fA} =  M_S^{D_h-4} \cT_{\fM}^{(e)\fN} \, ,  
\ee
with the explicit forms 
\be
\cT_{\fM}^{(e)\fN}   & \equiv &  \mJ_{\fA}^{\; \fN} \chi_{\fM}^{\;\fA} = - \eta_{\fM}^{\;\fN} \chi  \kL^{(e)} + \chi [\, \frac{1}{2}  \hat{\chi}_{\fA}^{\;\; \fN}  \bar{\vPsi} \varGamma^{\fA} i {\mathcal D}_{\fM} \vPsi -  \tilde{\chi}^{[\fN\fQ]\fM'\fN'}_{\fA\fB\; \fA'\fB'}  {\cal F}_{\fM\fQ}^{\fA\fB} {\cal F}_{\fM'\fN'}^{\fA'\fB'}   \nonumber \\
&+ &  \alpha_E M_S^2\, \tilde{\chi}^{[\fN\fQ]\fM'\fN'}_{\fA\fA'}  \mG_{\fM\fQ}^{\fA} \mG_{\fM'\fN'}^{\fA'}  - {\cal W}_{\fM\fQ} {\cal W}^{\fN\fQ} + g_w^2 M_S^2 W_{\fM}W^{\fN} ] , \\
\cT_{\fM}^{\; \fN} & = &  - \eta_{\fM}^{\;\fN} \chi  \kL +  \phi^{D_h-4}\chi [\, \frac{1}{2}\hat{\chi}_{\fA}^{\;\; \fN}  \bar{\vPsi} \varGamma^{\fA} i {\mathcal D}_{\fM} \vPsi -  \tilde{\chi}^{[\fN\fQ]\fM'\fN'}_{\fA\fB\; \fA'\fB'}  {\cal F}_{\fM\fQ}^{\fA\fB} {\cal F}_{\fM'\fN'}^{\fA'\fB'}   \nonumber \\
&+ &  \alpha_E \phi^2\, \tilde{\chi}^{[\fN\fQ]\fM'\fN'}_{\fA\fA'}  \fG_{\fM\fQ}^{\fA} \fG_{\fM'\fN'}^{\fA'}  - {\cal W}_{\fM\fQ} {\cal W}^{\fN\fQ} + d_{\fM}\phi d^{\fN}\phi   ] , 
\ee
where $\cT_{\fM}^{\; \fN}$ is conformal scaling gauge invariant.

The conservation of the total hyper-stress energy-momentum tensor $\p_{\fN} \cT_{\fM}^{\; \fN} = 0 $ together with the conserved hyper-gravifield currents given in Eq.(\ref{CC}) leads to the relations,
\be
& & \vGa_{\fM\fN}^{\fP} \cT_{\fP}^{(e)\fM}  =   \vOm_{\fM\fB}^{\fA}  \mJ_{\fA}^{\; \fM} \chi_{\fN}^{\; \fB}, \\
& & \vGa_{\fM\fN}^{\fP} \cT_{\fP}^{\; \fM}  =  ( \vOm_{\fM\fB}^{\fA} - \p_{\fM}\ln\phi\1 \eta^{\fA}_{\;\fB} ) \fJ_{\fA}^{\; \fM} \chi_{\fN}^{\; \fB} \, ,  
\ee
where the left-hand side of the equations concerns the geometric quantities and the right-hand side involves the quantities of gauge field theory. 

In terms of the conserved hyper-stress energy-momentum tensor, the gauge gravitational equations of motion in two cases can be rewritten as follows
\be
& & \p_{\fP}\mG_{\fM}^{\fN\fP} -   \mG_{\fM}^{\fN}    =  \cT_{\fM}^{(e)\fN}  , \nn \\
& & \p_{\fP}\fG_{\fM}^{\fN\fP}  - \fG_{\fM}^{\fN}  =  \cT_{\fM}^{\; \fN}   , 
\ee
with the definitions
\be
 & & \mG_{\fM}^{\fN\fP}\equiv \chi_{\fM}^{\;\fA}\mG_{\fA}^{\fN\fP};\quad   \mG_{\fM}^{\fN}\equiv \mG_{\fA}^{\fN\fP} \p_{\fP} \chi_{\fM}^{\;\fA}  =  \chih_{\fA}^{\;\fQ} \p_{\fP} \chi_{\fM}^{\;\fA} \mG_{\fQ}^{\fN\fP} \, , \nn \\
 & &\fG_{\fM}^{\fN} \equiv \chi_{\fM}^{\;\fA}\fG_{\fA}^{\fN\fP} ; \quad \fG_{\fM}^{\fN} \equiv \fG_{\fA}^{\fN\fP} \hat{\p}_{\fP} \chi_{\fM}^{\;\fA} =  \chih_{\fA}^{\;\fQ} \hat{\p}_{\fP} \chi_{\fM}^{\;\fA} \fG_{\fQ}^{\fN\fP} \, ,
\ee
with $\hat{\p}_{\fP} \chi_{\fM}^{\;\fA} = (\p_{\fP} + \p_{\fP}\ln\phi ) \chi_{\fM}^{\;\fA}$.

\subsection{ Geometric gravitational equations of Einstein-like and beyond in hyper-spacetime }

To obtain the {\it geometric gravitational equation} in analogy to the Einstein equation in the general theory of relativity in four dimensional spacetime,  we shall apply an identity based on the {\it gravitational gauge-geometry duality} in hyper-spacetime,
\be
& &  \frac{1}{4} \chi \phi^{D_h-2} \tilde{\chi}^{\fM\fN\fM'\fN'}_{\fA\fA'} \fG_{\fM\fN}^{\fA}\fG_{\fM'\fN'}^{\fA'} - 2g_h  \p_{\fM}(\chi\phi^{D_h-2} \cA_{\fN}^{\fM\fN} )\nn \\
& & = \chi \phi^{D_h-2}( \cR - (D_h-1)(D_h-2)\p_{\fM}\ln\phi \p^{\fM}\ln\phi ) -  2g_h \p_{\fM} (\chi \phi^{D_h-2} \cH_{\fN}^{\fM\fN}), 
 \ee
which enables us to rewrite the basic action of Eq.(\ref{HUTactionQFT}) into an equivalent basic action in which the dynamics of the gravitational interactions is described by the conformal scaling gauge invariant Einstein-Hilbert type action in hyper-spacetime,
\be \label{HUTactionQFTEH}
 I_H & \equiv & \int \chi\1 \kL = \int \chi \phi^{D_h-4} \{ \chih^{\fM\fN} \frac{1}{2} \bar{\vPsi} \varGamma_{\fA} \chi^{\; \fA}_{\fM} ( i \p_{\fN} + g_h\1 \cA_{\fN} )  \vPsi \nonumber \\
& - &  \frac{1}{4}\1( \tilde{\chi}^{\fM\fN\1 \fM'\fN'}_{\fA\fB\1 \fA'\fB' } \cF_{\fM\fN}^{\fA\fB}\cF_{\fM'\fN'}^{ \fA'\fB'} +\chih^{\fM\fM'}\chih^{\fN\fN'} {\cal W}_{\fM\fN} {\cal W}_{\fM'\fN'}  ) \nn \\
 & + & \alpha_E  (\, \phi^2 \cR   - (D_h-1)(D_h-2) \chih^{\fM\fN}\p_{\fM}\phi \p_{\fN}\phi \, )   \nn \\
& + &  \frac{1}{2} \chih^{\fM\fN}d_{\fM} \phi d_{\fN}\phi - \beta_E\1\phi^4 \, \}  + 2 \alpha_Eg_h\p_{\fM} (\chi \phi^{D_h-2}\cH_{\fN}^{\fN\fM} ) .
\ee
In the Einstein-type basis, such a basic action gets the following form,  
\be \label{HUTactionQFTEB2}
I_H^{(e)} & \equiv & M_S^{D_h-4} \int [d\hx] \chi\1 \kL^{(e)}  \nn \\
&  = & M_S^{D_h-4}\int [d\hx] \chi\{ \chih^{\fM\fN} \frac{1}{2} \bar{\vPsi} \varGamma_{\fA} \chi^{\;\fA}_{\fM} ( i \p_{\fN} + g_h\1 \cA_{\fN} )  \vPsi \nonumber \\
&  - &  \frac{1}{4}\1 \tilde{\chi}^{\fM\fN\1 \fM'\fN'}_{\fA\fB\1 \fA'\fB' } \cF_{\fM\fN}^{\fA\fB}\cF_{\fM'\fN'}^{ \fA'\fB'}  + \alpha_E M_S^2 \cR  - \beta_E M_S^4 \nn \\
&  - & \frac{1}{4} \chih^{\fM\fM'}\chih^{\fN\fN'} {\cal W}_{\fM\fN} {\cal W}_{\fM'\fN'}  + \frac{1}{2}g_w^2 M_S^{2} \chih^{\fM\fN} W_{\fM}W_{\fN} \} ,
\ee
where we have ignored the surface time.  

From the above action, we are able to derive the {\it geometric gravitational equation} of the hyper-gravifield in terms of the Ricci curvature tensor $\cR_{\fM\fN}$ and the conserved hyper-gravifield current $\mJ_{\fM}^{\;\; \fA} $, 
\be
\alpha_E M_S^2 \cG_{\fA}^{\; \fM}  -2\alpha_EM_S^2\chi \cR_{\fM'\fN'}\chih_{\fA}^{\; \fM'}\chih^{\fN'\fM}  = \mJ_{\fA}^{\; \fM}   \, , 
\ee
with 
\be
\cG_{\fA}^{\; \fM}  \equiv \chi \chih_{\fA}^{\;\fP} \tilde{\chi}^{[\fM\fQ]\fM'\fN'}_{\fA''\fA'}  \mG_{\fP\fQ}^{\fA''} \mG_{\fM'\fN'}^{\fA'} \, .
\ee

When projecting the bicovariant vector tensors into the hyper-spacetime tensors, we arrive at two {\it geometric gravitational equations} corresponding to the symmetric and antisymmetric parts,
\be \label{EGE}
& & \mG_{\fM\fN}^{(e)} + \sT_{(\fM\fN)}^{(e)} =0 \, , \\
& & \sT_{[\fM\fN]}^{(e)} = 0\, ,  \label{BEGE}
\ee
with
\be 
\mG_{\fM\fN}^{(e)} & = & 2\alpha_EM_S^2 (\cR_{\fM\fN} - \frac{1}{2} \chi_{\fM\fN} \cR ) + \chi_{\fM\fN} \beta_E M_S^4 \, , 
\ee
and
\be
\sT_{(\fM\fN)}^{(e)} & = &  \frac{1}{4} (\bar{\vPsi}\chi_{\fN}^{\;\fA}  \varGamma_{\fA} i {\mathcal D}_{\fM} \vPsi  + \bar{\vPsi}\chi_{\fM}^{\;\fA}  \varGamma_{\fA} i {\mathcal D}_{\fN} \vPsi ) - \frac{1}{2} \chi_{\fM\fN} \bar{\vPsi}\chih^{\;\fP}_{\fA}  \varGamma^{\fA} i {\mathcal D}_{\fP} \vPsi   \nn \\
& - &  \tilde{\chi}^{[\fP\fQ]\fM'\fN'}_{\fA\fB\1 \fA'\fB' }[ \, \frac{1}{2}( \cF_{\fM\fQ}^{\fA\fB} \chi_{\fN\fP} + \cF_{\fN\fQ}^{\fA\fB} \chi_{\fM\fP} ) - \frac{1}{4} \chi_{\fM\fN} \cF_{\fP\fQ}^{\fA\fB} \, ] \cF_{\fM'\fN'}^{ \fA'\fB'}   \nn \\
& - & ( {\cal W}_{\fM\fP} {\cal W}_{\fN}^{\; \;\fP} - \frac{1}{4} \chi_{\fM\fN} {\cal W}_{\fP\fQ} {\cal W}^{\fP\fQ} ) + g_w^2 M_S^{2} (W_{\fM}W_{\fN} - \chi_{\fM\fN} W_{\fP}W^{\fP} )\, ,  \\
\sT_{[\fM\fN]}^{(e)} & = &  \frac{1}{4} (\bar{\vPsi}\chi_{\fN}^{\;\fA}  \varGamma_{\fA} i {\mathcal D}_{\fM} \vPsi  - \bar{\vPsi}\chi_{\fM}^{\;\fA}  \varGamma_{\fA} i {\mathcal D}_{\fN} \vPsi )  \nn \\
& - &   \tilde{\chi}^{[\fP\fQ]\fM'\fN'}_{\fA\fB\1 \fA'\fB' } \frac{1}{2}(\, \cF_{\fM\fQ}^{\fA\fB} \chi_{\fN\fP} - \cF_{\fN\fQ}^{\fA\fB} \chi_{\fM\fP} \, ) \cF_{\fM'\fN'}^{ \fA'\fB'} \, ,
\ee
where $ \mG_{\fM\fN}^{(e)}$ is the Einstein-type gravitational curvature tensor with the {\it cosmological constant} in hyper-spacetime, $\sT_{(\fM\fN)}^{(e)} $ is the Einstein-type symmetric hyper-stress energy-momentum tensor and $\sT_{[\fM\fN]}^{(e)} $ is the antisymmetric hyper-stress energy-momentum tensor in hyper-spacetime.

The geometric gravitational equation (\ref{EGE}) with the symmetric tensors in hyper-spacetime is analogous to Einstein equation of the general theory of relativity in four-dimensional spacetime, while the geometric gravitational equation (\ref{BEGE}) with the antisymmetric  tensor represents a new feature of the gravitational interactions in hyperunified field theory.

\section{Conclusions and remarks}

We have performed a general analysis and a detailed construction of hyperunified field theory. The postulates of gauge invariance and coordinate independence are proposed to be more general and fundamental than the postulate of general covariance of coordinates to describe the laws of nature. All the spin-like charges of elementary particles have been considered as the essential quantum numbers of the basic building blocks of nature to determine the hyper-spin symmetry. All the elementary particles have been merged into a single hyper-spinor field in the spinor representation of the hyper-spin symmetry, and all the basic forces have been unified into a fundamental interaction governed by the hyper-spin gauge symmetry SP(1,$D_h$-1). The hyper-spin gauge field $\cA_{\fM}^{\fA\fB}$ associated with the bicovariant vector hyper-gravifield $\chi_{\fM}^{\;\;\fA}$ is taken as the basic force field to realize the hyper-spin gauge symmetry SP(1,$D_h$-1). The hyper-spin charge of the hyper-spinor field is conjectured to relate coherently with the dimension of hyper-spacetime. The hyper-gravifield fiber bundle structure of biframe hyper-spacetime has been shown to be natural and crucial in the construction of hyperunified field theory. To unify all the elementary particles and basic forces in the standard model, we have built the general action of hyperunified field theory with a minimal dimension $D_h=19$ of hyper-spacetime, which possesses the joined bimaximal global and local symmetry $G_S$ = PO(1,18)$\times$S(1)$\Join $SP(1,18)$\times$SG(1). 

The gauge symmetry in hyperunified field theory has been shown to be characterized by the gauge-type Goldstone-like hyper-gravifield $\chi_{\fM}^{\;\;\fA}$, which reveals the gravitational origin of gauge symmetry. Such a Goldstone-like hyper-gravifield allows us to transmute between the coordinate and non-coordinate systems of hyper-spacetime and to express hyperunified field theory in various equivalent formalisms, which enables us to demonstrate explicitly both the gauge-gravity and gravity-geometry correspondences and to corroborate the gravitational gauge-geometry duality.

In the hidden coordinate formalism of hyperunified field theory, it has been shown that all the basic forces between elementary particles are described by the fundamental gauge interaction, which is governed by the hyper-spin gauge symmetry SP(1, $D_h$-1) of the single hyper-spinor field $\varPsi$ and characterized by the hyper-spin gauge field $\cA_{\fC}^{\fA\fB} = \vOm_{\fC}^{\fA\fB} + \cH_{\fC}^{\fA\fB}$. The nonhomogeneous hyper-spin gauge transformations are attributed to the hyper-spin gravigauge field $\vOm_{\fC}^{\fA\fB}$ which is determined by the Goldstone-like hyper-gravifield $\chi_{\fM}^{\;\;\fA}$. The dynamics of the hyper-gravifield  $\chi_{\fM}^{\;\;\fA}$ is described by the field strength $\cR_{\fC\fD}^{\; \fA\fB}$ characterized by the hyper-spin gravigauge field $\vOm_{\fC}^{\fA\fB}$, which reveals the {\it gauge-gravity correspondence} in the locally flat hyper-gravifield spacetime ${\bf G}_h$. Such a non-coordinate spacetime ${\bf G}_h$ spanned by the hyper-gravifield basis $\{\chi^{\fA}\}$ reflects a non-commutative geometry, which is viewed as a dynamically emerged hyper-spacetime characterized by the hyper-spin gravigauge field $\vOm_{\fC}^{\fA\fB}$ as exhibited in Eq.(\ref{NCG}).

In the hidden gauge formalism of hyperunified field theory with gauge fixing to the unitary gauge, the basic gauge fields consist of the symmetric gauge-type hyper-gravifield $\chi_{\fM\fA}(\hx)= \chi_{\fA\fM}(\hx)$ (or  the nonlinearly realized symmetric Goldstone-like hyper-gravifield $G_{\fM\fA}(\hx)=G_{\fA\fM}(\hx)$) and the antisymmetric hyper-spacetime homogauge field $\cH_{\fM}^{\fP\fQ}(\hx)$. So the gravitational interactions for both hyper-spinor and boson fields are described by the gauge fixing symmetric Goldstone-like hyper-gravifield $\chi_{\fM\fA}(\hx)$. The geometry of hyper-spacetime is characterized by the hyper-gravimetric field  $\chi_{\fM\fN}(\hx)$ that is determined by the symmetric Goldstone-like hyper-gravifield $\chi_{\fM\fA}(\hx) $, i.e., $\chi_{\fM\fN}(\hx) = (\chi_{\fM\fA}(\hx))^2 $. The dynamics of gravitational interactions is described by the conformal scaling gauge invariant Einstein-Hillbert type action, which is characterized by the hyper-spacetime gravigauge field $\vGa_{\fM\fN}^{\fP}(\hx)$ as the Christoffel symbols. The general conformal scaling invariance condition as shown in Eq.(\ref{CCrelation1}) has been found to eliminate all the terms in quadratic Riemann and Ricci tensors, so that there exists no unitarity problem involving at a higher derivative gravity. Such a formalism affirms the {\it gravity-geometry correspondence} in hyper-spacetime. 

From the hyper-gravifield fiber bundle structure of biframe hyper-spacetime, the gravitational interaction has been formulated as the gauge interaction of the hyper-gravifield  $\chi_{\fM}^{\;\fA}$ based on the general hyper-spin gauge invariance with the general gauge massless condition shown in Eq.(\ref{CCrelation2}). Such a gauge gravitational interaction has been demonstrated to be dual to the geometrical gravitational interaction described solely by the conformal scaling gauge invariant Einstein-Hilbert type action in hyper-spacetime, which corroborates explicitly the gravitational gauge geometry duality.

We have focused in this paper mainly on the building of hyperunified field theory from the bottom-up approach. The basic properties of hyperunified field theory and the issue on the fundamental mass scale have been discussed within the framework of QFT, which allows us to derive the gauge gravitational equation with the conserved hyper-gravifield current and deduce the geometric gravitational equations of Einstein-like and beyond corresponding to the symmetric and antisymmetric hyper-stress energy-momentum tensors in hyper-spacetime. Such a hyperunified field theory is conjectured to hold at a fundamental mass scale $M_S$ based on the conformal scaling gauge symmetry in hyper-spacetime. 

Once a hyperunified field theory is established, a more sophisticated task is to figure out a realistic model to describe the real world from the top-down approach. It is inevitable to carry out the basic issues, such as: how to realize an appropriate symmetry breaking mechanism and a reliable dimension reduction to reach the observable four-dimensional spacetime of the real world; how to reproduce the standard model with three families of quarks and leptons and explain the observed matter-antimatter asymmetry and the dark matter component in the present universe; how to reveal a potential inflationary period of early universe and understand the observed accelerating expansion of present universe with the dominant dark energy component.  Making those issues is beyond the scope of the present paper. Nevertheless, it is clear that, unlike the usual unified theories and the extra-dimensional models, either the symmetry breaking or the dimension reduction in hyperunified field theory should be no longer independent and isolated, they must be correlated and associated with each other as shown in Ref.\cite{YLWU3} for a geometric symmetry breaking mechanism. This is because the dimensions of hyper-spacetime are coherently related with the basic quantum numbers of  the hyper-spinor field, which determines the basic symmetry of hyperunified field theory. 

We hope that the present hyperunified field theory has provided us a new insight for the unity of all the basic forces and elementary particles. To test whether such a theory is the true choice of nature, there remain more theoretical work and experimental efforts to be made exploring the subject.

\centerline{{\bf Acknowledgement}}

The author is grateful to many colleagues for useful discussion and conversation during the International Symposium on Gravitational Waves, May 25-29, 2017 at ICTP-AP/UCAS. This work was supported in part by the National Science Foundation of China (NSFC) under Grant Nos. 11690022 and 11475237, and by the Strategic Priority Research Program of the Chinese Academy of Sciences (CAS), Grant No. XDB23030100, Key Research program QYZDY-SSW-SYS007, and by the CAS Center for Excellence in Particle Physics (CCEPP).

\end{document}